\documentclass[12pt]{article}
\usepackage{amsmath}
\usepackage{amsfonts}
\usepackage{graphicx,psfrag,epsf}
\usepackage{subfigure}
\usepackage{enumerate}
\usepackage{xcolor}
\usepackage{bm}
\usepackage{natbib}
\usepackage{diagbox}
\usepackage{bbm}
\usepackage{url}
\newcommand{\blind}{0}
\usepackage{algorithm}
\usepackage{algpseudocode}
\usepackage{multirow}
\usepackage{hyperref}
\usepackage{rotating}  

\algrenewcommand\algorithmicrequire{\textbf{Input:}}
\algrenewcommand\algorithmicensure{\textbf{Output:}}

\DeclareMathOperator{\Diag}{Diag}
\DeclareMathOperator{\vect}{vec}

\DeclareMathOperator{\Prob}{Pr}
\def\spacingset#1{\renewcommand{\baselinestretch}%
{#1}\small\normalsize} \spacingset{1.0}

\begin{document}

\if0\blind
{
  \title{\bf Bayesian Signal Matching for Transfer Learning in ERP-Based Brain Computer Interface}
  \author{Tianwen Ma\\
 Department of Biostatistics and Bioinformatics, Emory University,\\
    Jane E. Huggins\\
    Department of Physical Medicine and Rehabilitation and\\  Department of Biomedical Engineering, University of Michigan,\\
     and \\
     Jian Kang\thanks{
    To whom correspondence should be addressed: jiankang@umich.edu}\\
     Department of Biostatistics, University of Michigan\\
    } 
  \date{}
  \maketitle

} \fi

\if1\blind
{
  \bigskip
  \bigskip
  \bigskip
  \begin{center}
    {\LARGE\bf Bayesian Signal Matching for Transfer Learning in ERP-Based Brain Computer Interface}
\end{center}
  \medskip
} \fi

\bigskip
\begin{abstract}
An Event-Related Potential (ERP)-based Brain-Computer Interface (BCI) Speller System assists people with disabilities to communicate by decoding electroencephalogram (EEG) signals. A P300-ERP embedded in EEG signals arises in response to a rare, but relevant event (target) among a series of irrelevant events (non-target). Different machine learning methods have constructed binary classifiers to detect target events, known as calibration. The existing calibration strategy uses data from participants themselves with lengthy training time. Participants feel bored and distracted, which causes biased P300 estimation and decreased prediction accuracy. To resolve this issue, we propose a Bayesian signal matching (BSM) framework to calibrate EEG signals from a new participant using data from source participants. BSM specifies the joint distribution of stimulus-specific EEG signals among source participants via a Bayesian hierarchical mixture model. We apply the inference strategy. If source and new participants are similar, they share the same set of model parameters; otherwise, they keep their own sets of model parameters; we predict on the testing data using parameters of the baseline cluster directly. Our hierarchical framework can be generalized to other base classifiers with parametric forms. We demonstrate the advantages of BSM using simulations and focus on the real data analysis among participants with neuro-degenerative diseases. 
\end{abstract}

\noindent%

\newpage
\spacingset{1.8}
\section{Introduction}
\label{sec:intro}

\subsection{Background}
\label{subsec:background}

A Brain-Computer Interface (BCI) is a device that interprets brain activity to assist people with severe neuro-muscular diseases with normal communications. An electroencephalogram (EEG)-based BCI speller system is one of the most popular BCI applications, which enables a person to ``type'' words without using a physical keyboard by decoding the recorded EEG signals. The EEG brain activity has the advantages of non-invasiveness, low cost, and high temporal resolution \citep{niedermeyer2005electroencephalography}.
The P300 event-related potential BCI design, known as the P300 ERP-BCI design \citep{donchin2000mental}, is one of the most widely used BCI frameworks and the focus of this work.

An ERP is an event potential in the EEG signals in response to an external event. The P300 ERP is a particular ERP that occurs in response to a rare, but relevant event (i.e., highlighting a group of characters on the screen) among a group of frequent, but irrelevant ones. The relevant (target) P300 ERP usually has a \textit{positive} deflection in voltage with a latency around 300 ms. Here, the latency refers to the time intervals between the onset of the event and the first response peak. Although P300 ERP is the most useful ERP, other types of ERPs, i.e., N200, exist and help interpret and classify brain activity.
During the experiment, participants were asked to wear an EEG cap with multiple electrodes, sit next to a virtual screen, and focus on a specific character to spell, denoted as the ``target key.'' The study applied the row-and-column paradigm (RCP) developed by \citep{farwell1988talking}, which is a 6 $\times$ 6 grid with 36 keys. Each event is either a row stimulus or a column stimulus. The order of rows and columns is random, but it loops through all rows and columns every 12 stimuli, known as a sequence. Participants were asked to mentally count whenever they saw a row or column stimulus containing the target key, and do nothing, otherwise. Thus, each sequence always contained exactly two stimuli that elicited target P300 ERPs. 

\begin{figure}[htbp]
\begin{subfigure}
     \centering
     \includegraphics[width=0.49\textwidth]{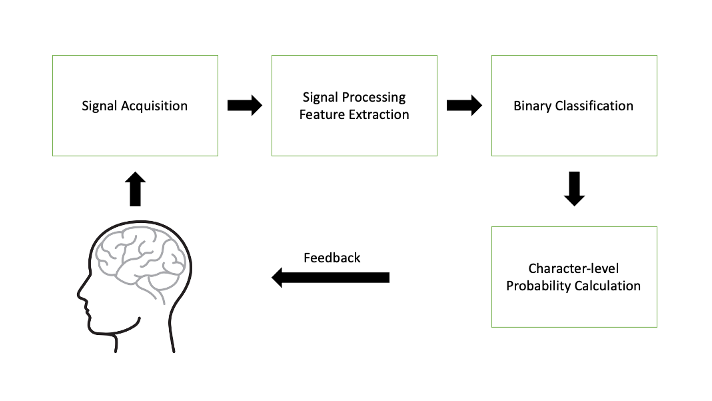}
\end{subfigure}
\hfill
\begin{subfigure}
     \centering
     \includegraphics[width=0.49\textwidth]{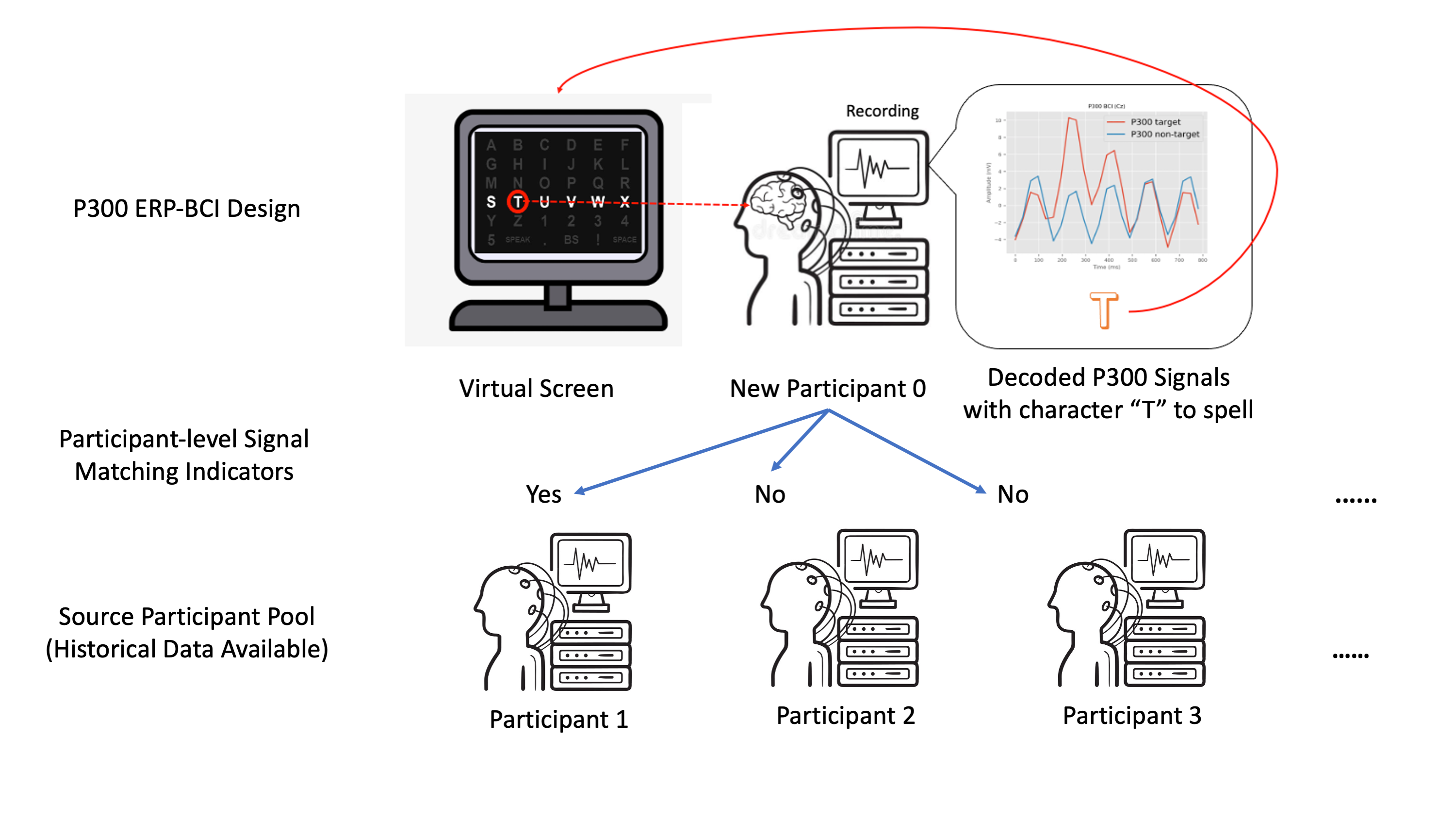}
\end{subfigure}
    \caption{\small \textbf{Left}: The conventional offline framework for P300 ERP spelling includes signal acquisition, pre-processing and  feature extraction, binary classification, character-level probability calculation, and final result demonstration. \textbf{Right}: The proposed data integration framework leverages data of source participants and small data from the new participant to construct the BSM binary classifier with participant-specific selection indicators.}
    \label{fig:bsm_flowchart_illustration}
\end{figure}

The left panel of Figure \ref{fig:bsm_flowchart_illustration} demonstrates the fundamental mechanism of the P300 ERP-BCI design. The virtual screen presents a sequence of stimuli; Participants respond to the stimuli with EEG signals recorded; The machine truncates the EEG signals in a fixed time response window after each stimulus and generates a EEG feature vector. For raw data with multiple electrodes, the ultimate truncated signal segment is achieved by concatenating multi-channel EEG signals. The machine then makes a \textit{binary} decision whether the participant's brain elicits a target P300 ERP response. After stimulus-specific binary classifier scores are generated, they are converted into character-level probabilities. Finally, the character with the highest probability is selected and shown on the virtual screen. For offline spelling, the final decision per character is usually made after multiple sequence replications.
Many machine learning (ML) methods have successfully constructed such binary classifiers for P300 ERPs, including stepwise linear discriminant analysis (swLDA) \citep{donchin2000mental}, \citep{krusienski2008toward}, support vector machine (SVM) \citep{kaper2004bci}, independent component analysis (ICA) \citep{xu2004bci},  linear discriminant analysis (LDA) with xDAWN filter \citep{rivet2009xdawn}, convolutional neural network (CNN) \citep{cecotti2010convolutional}, and Riemannian geometry (RG) \citep{barachant2011multiclass}.
Recently, \cite{ma2022bayesian} attempted to improve prediction accuracy by identifying spatial-temporal discrepancies between target and non-target P300 responses and explicitly addressed the overlapping ERP issue.

\subsection{Challenges and Existing Work}
\label{subsec:existing_work}

The conventional framework for P300 ERP spelling task requires training and testing, known as calibration and free-typing, as well. Typically, participants were asked to copy a multi-character phrase to build a participant-specific training profile. Due to the low signal-to-noise ratio in EEG, current calibration strategy relies on collecting data from participants themselves with a fixed but large amount of sequence replications. However, participants tended to feel bored and tired after repetitions, leading to biased target P300 ERPs and inefficient calibration process.Therefore, it is important to identify efficient calibration methods while maintaining decent prediction accuracy for free-typing. Existing work has tackled this problem by applying transfer learning \citep{bozinovski1976influence}. In statistical fields, we denoted this concept as data integration \citep{lenzerini2002data}. To avoid confusion, we do not distinguish among ``transfer learning,'' ``data integration,'' or ``borrowing'' in this work. We briefly introduce existing works that have applied transfer learning under different domains with applications to P300 ERP-BCIs. A recent review \citep{wu2020transfer} on transfer learning summarized methods between 2016 and 2020 with six paradigms and applications. For P300-ERP BCI paradigm, they considered cross-session/participant and cross-device transfer learning scenarios and produced a nice table for alignment-related approaches including offline/online status, known/unknown labels, alignment framework, referencing objects, base classifiers, and computational costs. We continue introducing the work that were not covered previously. 


Ensemble Learning Generic Information (ELGI) \citep{xu2015inter} examined whether inter-participant information is beneficial to ERP classification with simple base classifiers. They applied swLDA to each participant's dataset as base classifier and combine them to construct the final classifier. Later, they introduced the \textit{Weighted Ensemble Learning Generic Information} (WELGI) \citep{xu2016incorporation} by adding weights to each base classifier. Similarly,  \citeauthor{an2020weighted}in 2020 proposed a weighted participant-semi-independent classification method (WSSICM). They applied SVM and fit each base classifier by combining the entire data of each source participant and a small portion of data of the new participant. Weights for base classifiers were determined by an ad-hoc approach. Finally, \citeauthor{adair2017evolving} proposed the \textit{Evolved Ensemble Learning Generic Information} (eELGI) that developed a data-driven algorithm to determine the grouping criterion for training sets among source participants and applied swLDA to form base classifiers.

Transfer learning based on Riemannian geometry (RG) has gained increasing attention recently due to its fast speed to converge and a natural framework to leverage information from source participants. The input data for RG were the sample covariance matrices, and the distance-based algorithm based on RG was called Minimum Distance to Mean (MDM) \citep{barachant2011multiclass}. 
\citeauthor{congedo2013new} and \citeauthor{barachant2014plug} extended the original RG-based classifier to the P300 ERP design by augmenting the covariance matrix with cross-covariance between observed signals and reference target signals.
\citeauthor{rodrigues2018riemannian} applied the Riemannian procrustes analysis (RPA) to address the heterogeneity of EEG signals across different sessions or participants. Before the authors applied the MDM classifier, they applied certain affine transformations 
to raw participant-level covariance matrices such that the resulting covariance matrices were less heterogeneous across sessions or participants while their Riemannian distances were preserved. 
\citeauthor{li2020transfer} standardized the covariance matrices across participants by applying the affine transformation with the participant-specific Riemannian geometric mean covariance matrix. Finally, \citeauthor{khazem2021minimizing} proposed Minimum Distance to Weighted Mean (MDWM) by combining estimated mean covariance matrices from source participants and the new participant via the Riemannian distance with a trade-off hyper-parameter. 


\subsection{Our Contributions}
\label{subsec:novelty}

To reduce the calibration time while maintaining similar prediction accuracy, we proposed a Bayesian Signal Matching (BSM) method to build a participant-dependent, calibration-less framework. 
BSM reduced the calibration time of a new participant by borrowing data from pre-existing source participants’ pool. We performed an intuitive clustering approach by simply borrowing data from eligible participants without data transformation and homogenization. 
BSM specified the joint distribution of stimulus-specific EEG signals from all participants via a Bayesian hierarchical mixture model. Our method has several advantages: (i). From the perspective of user experience, fewer training data required for calibration under this framework increased higher spelling efficiency and placed less burden on participants. (ii). From the perspective of inference, we performed pairwise selection between each source participant and the new participant and specified the baseline cluster corresponding to the new participant. The selection indicators represented how close source calibration data of each source participant resembled the data of the new participant. If the selection indicator was 1, we merged source participant's data with the new participant's data, otherwise, we kept the source participant's data as a separate cluster. This approach did not focus on the similarity evaluation among source participants, and thereby, avoided the label-switching issue. (iii). From the perspective of prediction, we used the baseline cluster for \textit{directly} prediction without refitting the model with the augmented data. The applied Bayesian framework for parameter estimation also  ``leveraged'' the shared information from source participants implicitly and was robust to noises.
Our proposed hierarchical framework is also flexible to generalize to other classifiers with parametric likelihood functions. (iv). From the perspective of computation, we performed the data analysis on the transformed ERP responses after xDAWN filter \citep{rivet2009xdawn}, The xDAWN filter was a technique for dimensionality reduction. This pre-processing technique increased the signal-to-noise-ratio and matching probability, and improved the modeling efficiency compared to modeling the raw EEG signals or joint modeling transformed EEG signals.

The paper is organized as follows: Sections \ref{sec:method} and \ref{sec:posterior} present the framework for the Bayesian hierarchical mixture model and posterior inference, respectively. Sections \ref{sec:simulation} and \ref{sec:real_data_analysis} present the numerical results for simulation and real data analysis, respectively. Section \ref{sec:discussion} concludes the paper with a discussion.

\section{Methods}
\label{sec:method}

\subsection{Basic Notations and Problem Setup}
\label{subsec:problem_setup}

Let $\mathcal{MVN}(\bm{\mu},\Sigma)$ be a multivariate normal distribution with mean and covariance matrix parameters $\bm{\mu}$ and $\Sigma$. Let $\mathcal{MN}(M, U, V)$ be a matrix normal distribution with location matrix $M$ and two scale matrix parameters $U$ and $V$ \citep{dawid1981some}. Let $\mathrm{Diag}(\cdot)$ be a diagonal matrix notation.
Let $\mathcal{LN}(\mu,\sigma)$ be a Log-Normal distribution with mean and scale parameters $\mu$ and $\sigma$. Let $\mathcal{HC}(x_0,\sigma)$ be a Half-Cauchy distribution with location and scale parameters $x_0$ and $\sigma$. Let $\mathcal{U}(a,b)$ be a Uniform distribution with lower and upper bounds $a$ and $b$. 

Let $n=0,\ldots,N$ be the participant index, where $n=0$ refers to the new participant and $n=1,\ldots,N$ refer to source participants. Let $l=1,\ldots,L_n$ and $i=1,\ldots, I_n$ be the character index and sequence index for participant $n$, respectively. We follow the conventional RCP design such that each sequence contains $J(J=12)$ stimuli, including six row stimuli from top to bottom $(1,\ldots,6)$ and six column stimuli from left to right $(7,\ldots,12)$ on the $6\times 6$ virtual keyboard (See the virtual screen in Figure \ref{fig:bsm_flowchart_illustration}). For the $i$th sequence, $l$th target character, and $n$th participant, let $\bm{W}_{n,l,i}=(W_{n,l,i,1},\ldots, W_{n,l,i,12})^{\top}$ be a stimulus code indicator that takes values from the permutation of $\{1,\ldots,12\}$. Under the RCP design, given the target character and the stimulus-code indicators, there are exactly two target stimuli and ten non-target stimuli within each sequence. We define $\bm{Y}_{n,l,i}=(Y_{n,l,i,1},\ldots, Y_{n,l,i,12})^{\top}$ as the stimulus-type indicator for the $i$th sequence, $l$th target character, and $n$th participant, where $Y_{n,l,i,j}\in\{0,1\}$. For example, given a target character ``T'' and a stimulus code indicator $\bm{W}_{n,l,i}=(7,  9, 10,  5,  1, 2,  8, 11,  6,  4,  3, 12)^{\top}$, we obtain its stimulus type indicator $\bm{Y}_{n,l,i}=(0,0,0,0,0,0,1,0,0,1,0,0)^{\top}$, where the indices of $1$s correspond to the indices of the $4$th row and $2$nd column (denoted by $8$) in $\bm{W}_{n,l,i}$. Let $k=0,\ldots, K-1$ be the cluster index, where $k=0$ is the cluster that matches the new participant and $k=1,\ldots,K-1$ are the clusters within source participants. Finally, we incorporate $E$ channels of EEG signals and let $e(e=1,\ldots,E)$ be the channel index. We extract channel-specific EEG segments from the onset of each stimulus with a long EEG response window length (with $T_0$ time points). We denote $\bm{X}_{n,l,i,j,e}(s)$ as the extracted EEG signal segment of the $j$th stimulus, $i$th sequence, $l$th target character, and $n$th participant from channel $e$ at time $s\in[0,s_{T_0}]$. 

\subsection{The Framework of Bayesian Signal Matching}
\label{subsec:bayesian_split_and_merge_clustering}

We make three assumptions as follows: (i). We assume that P300 ERPs are the same regardless of target character to spell. To simplify, we let all source participants spell the same character $\omega$ with the same sequence replication size $I$, so we drop the target character index $l$ and the source participant index $n$ from $I_n$. For example, $Y_{n,l,i,j}$ is reduced to $Y_{n,i,j}$. The biological mechanism behind the speller system is the response to an unexpected stimulus. Under the RCP design, there were always two target stimuli supposed to elicit target P300 ERPs, and this rule applied regardless of the spelling characters. 
(ii). We assumed that target and non-target ERP functions shared the same covariance matrix within the same cluster. We simplified the temporal covariance matrix as a scaled AR(1) correlation matrix. Linear discriminant analysis (LDA) with its variations have been shown to be effective binary classifiers for this problem, we argued that the difference on the mean level had sufficient separation effect, and sharing the same covariance matrix reduced the number of parameters. Empirical correlation matrix also suggested that the correlation declined fast and they were similar for target and non-target stimuli, even after the xDAWN filter. The shared and simplified covariance matrix also reduced the number of parameters and made the proposed model parsimonious, especially when we considered borrowing data from source participants. (iii). We performed selection with respect to the target ERP data and we excluded non-target ERP data from the source participants' pool. Despite a larger sample size of non-target ERP data, unfortunately, their signals showed fewer unique patterns and would add more noises to the model fitting. The right panel of Figure \ref{fig:bsm_flowchart_illustration} provides a detailed illustration of the data integration framework of proposed Bayesian signal matching method. Details of three assumptions can be found in the supplementary material Section S1.

For the new participant, i.e., when $n=0$, we assume
\begin{equation}\label{eq:multi_new_participant}
\bm{X}_{0,i,j} = \bm{B}_{0,1}Y_{0,i,j} + \bm{B}_{0,0} (1-Y_{0,i,j}) + \bm{\epsilon}_{0,i,j},
\end{equation}
where $\bm{B}_{0,1},\bm{B}_{0,0}$, and $\bm{\epsilon}_{0,i,j}$ are target ERP matrix, non-target ERP matrix, and random noise matrix of the new participant, of the same size $T_0\times E$, respectively. For source participant $n, n>0$, we introduce a participant-specific binary indicator $Z_n\in\{0,1\}$ and assume that $\Prob(Z_n=1) =\pi_n$, where $\pi_n$ is the probability whether source participant $n$ matches the new participant. With $\{Z_n\}$, we consider a signal matching model for the EEG signal matrix for source participant $n$ as follows:
\begin{equation}
\label{eq:multi_source_participant}
\begin{aligned}
\bm{X}_{n,i,j} &=\bm{B}_{n,1}Y_{n,i,j} + \bm{B}_{n,0}(1-Y_{n,i,j})+\bm{\epsilon}_{n,i,j},\\
\bm{B}_{n,1}&=\bm{B}_{0,1}Z_n + \tilde{\bm{B}}_{n,1}(1-Z_n).\\
\end{aligned}
\end{equation}
where $\bm{B}_{n,1}, \tilde{\bm{B}}_{n,1},\bm{B}_{n,0}$, and $\bm{\epsilon}_{n,i,j}$ are target ERP matrix after matching, target ERP matrix before matching, non-target ERP matrix, and stimulus-specific random error matrix of source participant $n$, respectively. Equation ~\eqref{eq:multi_source_participant} assumes that each source participant has either has the new participant's parameter set with a successful matching or their own parameter set. Note that we do not borrow non-target EEG data of source participants. The quantities $\bm{X}_{n,i,j}, \tilde{\bm{B}}_{\cdot},\bm{B}_{\cdot}$, and $\bm{\epsilon}_{n,i,j}$ are matrices with the same size of $T_0$ by $E$.

For random error matrix $\bm{\epsilon}_{n,i,j}$, where $n\in\{0,1,\cdots, N\}$, we assume an additive relationship to characterize its spatial-temporal error structure.
\begin{equation}
\begin{aligned}
    \bm{\epsilon}_{n,i,j} &= \bm{\xi}_{n,i,j} +
    \bm{\varepsilon}_{n,i,j},\\
    \bm{\xi}_{n,i,j} &= (\bm{\xi}_{n,i,j,1},\cdots, \bm{\xi}_{n,i,j,T_0})^{\top}, \quad 
    \bm{\varepsilon}_{n,i,j} = (\bm{\varepsilon}_{n,i,j,1}, \cdots, \bm{\varepsilon}_{n,i,j,E}),\\
\end{aligned}
\end{equation}
where $\bm{\xi}_{n,i,j,t}, (t=1,\cdots,T_0)$ is a vector of spatial random effects for $E$ channels at time point $t$, and $\bm{\varepsilon}_{n,i,j,e}, (e=1,\cdots,E)$ is a vector of temporal random effects for $T_0$ time points of channel $e$. Let $\bm{\Sigma}_n^s=\bm{V}_n\bm{R}_n^s\bm{V}_n^{\top}$, $\bm{R}_n^s$, and $\bm{R}_n^t$ be the spatial covariance matrix, spatial correlation matrix, and temporal correlation matrix for participant $n,n\in\{0,...,N\}$, respectively. $\bm{R}_n^s$ is a correlation matrix with the structure of compound symmetry characterized by a scale parameter $\eta_n\in(0,1)$; $\bm{V}_k$ is a diagonal matrix characterized by $(\sigma_{n,1}, \cdots, \sigma_{\sigma_{n,E}})$, and $\bm{R}_k^t$ is a correlation matrix with the structure of exponential decay characterized by a scalar parameter $\rho_n\in(0,1)$. 
Given $Z_n=1$, where $n\in\{1,\cdots, N\}$, we assume that $\bm{\xi}_{n,i,j,t}$ and $\bm{\varepsilon}_{n,i,j,e}$ are noise vectors generated by the new participant's parameters; Given $Z_n=0$, we assume that $\bm{\xi}_{n,i,j,t}$ and $\bm{\varepsilon}_{n,i,j,e}$ are noise vectors generated by participant $n$'s parameters.
\begin{equation}
\begin{aligned}
(\bm{\xi}_{n,i,j,t}\mid Z_n) & \sim \mathcal{MVN}\{\bm{0}, \bm{\Sigma}_0^sZ_n+
\bm{\Sigma}_n^s(1-Z_n)\},\\ 
(\bm{\varepsilon}_{n,i,j,e} \mid Z_n) & \sim \mathcal{MVN}\{\bm{0},\bm{R}_0^tZ_n + \bm{R}_n^t(1-Z_n)\}.\\
\end{aligned}
\end{equation}

Let $\bm{\Theta}$ be a collection of all unknown parameters. Our primary parameters of interest are those associated with the new participant, i.e., $\bm{B}_{0,1},\bm{B}_{0,0}, \sigma_{0,e}^2, \eta_0$, and $\rho_0$, and posterior probability of matching indicators, i.e., $\Prob(Z_n=1\mid \bm{X},\bm{Y}), n=1,\cdots, N$. The remaining parameters are considered as nuisance parameters.

\section{Posterior Inferences}
\label{sec:posterior}

\subsection{Model Reparametrization and Prior Specifications}
\label{subsec:represent_prior}

We rewrite the model by applying the notation of matrix normal distribution
\begin{equation}
\begin{aligned}
\bm{X}_{0,i,j} \mid Y_{0,i,j} = 1; \bm{\Theta} & \sim \mathcal{MN}(\bm{B}_{0,1}, \bm{\Sigma}_0^s, \bm{R}_0^t), \\
\bm{X}_{0,i,j} \mid Y_{0,i,j}=0 ;\bm{\Theta} & \sim \mathcal{MN}(\bm{B}_{0,0}, \bm{\Sigma}_0^s, \bm{R}_0^t),\\
\bm{X}_{n,i,j} \mid Z_n, Y_{n,i,j}=1; \bm{\Theta} & \sim \mathcal{MN}\{\bm{B}_{0, 1}Z_n + \bm{B}_{n,1}(1-Z_n), \\
& \quad \quad \quad \quad \bm{\Sigma}_0^sZ_n + \bm{\Sigma}_n^s (1-Z_n), 
\bm{R}_0^tZ_n + \bm{R}_n^t(1-Z_n)\},\\
\end{aligned}
\label{eq:matrix_normal_model}
\end{equation}
where $\bm{B}_{n,1}=(\bm{\beta}_{n,1,1}^{\top},\cdots,\bm{\beta}_{n,1,E}^{\top}) (n>0)$ and $\bm{B}_{0,0}=(\bm{\beta}_{0,0,1}^{\top},\cdots,\bm{\beta}_{0,0,E}^{\top})$. 

To better characterize the multi-channel ERP response functions $\{\bm{B}_{n,1}\}_{n=0}^N$ and $\bm{B}_{0,0}$, we frame them using the notation of a multivariate Gaussian process. First, we adopt the definition \citep{dixon2018multivariate} to define a multivariate Gaussian processes as follows: Let $\bm{\mu}:\mathbb{R}\rightarrow \mathbb{R}^E$, $\kappa: \mathbb{R}\times \mathbb{R} \rightarrow \mathbb{R}$, $\bm{\Omega}\in\mathbb{R}^{E\times E}$, and $\{s_t\}_{t=1}^{T_0}$ be a vector-valued mean function, a kernel on the temporal domain, a positive definite parameter covariance matrix on the spatial domain, and input points, respectively. Here, the general multidimensional input vector is reduced to the scalar indexed by time only. $\bm{f}$ is a multivariate Gaussian process on $\mathbb{R}$ with vector-valued mean function $\bm{\mu}$, kernel $\kappa$, and row covariance matrix $\bm{\Omega}$ if the vectorization of any finite collection of vector-valued variables have a joint multivariate Gaussian distribution, i.e.,
\begin{equation}
\begin{aligned}
  \vect\left\{\left[\hat{\bm{f}}(s_1),\cdots, \hat{\bm{f}}(s_{T_0})\right]\right\}\sim \mathcal{MVN}\left(\vect\left\{\left[\hat{\bm{\mu}}(s_1),\cdots, \hat{\bm{\mu}}(s_{T_0})\right]
    \right\}, \bm{\Sigma}\otimes \bm{\Omega}\right),\\ 
\end{aligned}
\end{equation}
where $\vect$ was the vectorization operator; $\hat{\bm{f}}(s), \hat{\bm{\mu}}(s)\in\mathbb{R}^E$ were column vectors that evaluated the functions $\bm{f}$ and $\bm{\mu}$ at the index point $s$. Furthermore, $\bm{\Sigma}\in\mathbb{R}^{T_0\times T_0}$ with $\bm{\Sigma}_{t_1,t_2}=k(s_{t_1},s_{t_2})$, where $t_1,t_2\in\mathbb{Z}$ and $0< t_1,t_2\leq T_0$, and $\otimes$ is the Kronecker product operator. We denote $\bm{f}\sim \mathcal{MGP}(\bm{\mu}, \kappa, \bm{\Omega})$.

In our case, we let $\bm{b}_{n,1}$ be an $E$-dim target ERP function of participant $n$ characterized by $\bm{\mu}_{n,1}, \kappa_1, \bm{\Sigma}_n^s$, $n=0,\cdots, N$, and $\bm{b}_{0,0}$ be another $E$-dim non-target ERP function of participant $0$ characterized by $\bm{\mu}_{0,0},\kappa_0, \bm{\Sigma}_0^s$. To simplify, we let $\bm{\mu}_{n,1}=\bm{\mu}_{0,0}=\bm{0}$ and apply a $\gamma$-exponential kernel function to specify target kernel $\kappa_{1}$ and non-target kernel $\kappa_{0}$ as follows:
\begin{equation}
\begin{aligned}
k(z_i,z_j) = \psi_0 \exp \left\{-\left(\frac{||z_i-z_j||_2^2}{s_0}\right)^{\gamma_0}\right\},
\end{aligned}
\label{eq:gamma_exp_kernel}
\end{equation}
where $0\leq \gamma_0<2,s_0>0$. In practice, $\gamma_0$ and $s_0$ are treated as hyper-parameters and selected among a group of values and we do not further differentiate among participants. 
For participant-specific kernel variance parameters $\psi_{n,1}$ and $\psi_{0,0}$, we assign Log-Normal priors with the a mean zero and a scale one. 
In addition, we define a function $\phi(\cdot)$ as an eigenfunction of kernel $k$ with eigenvalue $\lambda$ if it satisfies the following integral: $\int k(s,s')\phi(s)ds=\lambda\phi(s').$
In general, the number of eigenfunctions is infinite, i.e., $\{\phi_l(s)\}_{l=1}^{+\infty}$, and we rank them by the descending order of their corresponding eigenvalues $\{\lambda_l\}_{l=1}^{+\infty}$. The Mercer's Theorem suggests that the kernel function $k$ can be decomposed as a weighted summation of normalized eigenfunctions $\phi_l(s)$ with positive eigenvalues $\lambda_l$ and approximated with a sufficient large $L$ subgroup as follows: $k(s,s')=\sum_{l=1}^{+\infty}\lambda_l\phi_l(s)\phi_l(s')\approx \sum_{l=1}^{L}\lambda_l\phi_l(s)\phi_l(s')$.
Thus, a random function $\bm{f}$ can be represented as a linear combination of the above eigenfunctions and approximated with a sufficient large $L$ subgroup evaluated at $s$  as follows:
\begin{equation*}
\begin{aligned}
    \bm{f}(s)& =\begin{pmatrix}
        f_1(s)\\
        \vdots \\
        f_E(s)\\
    \end{pmatrix}=
    \begin{pmatrix}
        \sum_{l=1}^{\infty} a_{1,l}\phi_l(s)\\
        \vdots\\
        \sum_{l=1}^{\infty} a_{E,l}\phi_l(s)\\
    \end{pmatrix}\approx \begin{pmatrix}
        \sum_{l=1}^L a_{1,l}\phi_l(s)\\
        \vdots\\
        \sum_{l=1}^L a_{E,l}\phi_l(s)\\
    \end{pmatrix},\\
\end{aligned}
\end{equation*}
where $(a_{1,l},\cdots,a_{E,l})^{\top} \sim \mathcal{MVN}(\bm{0}, \bm{\Omega}), l>0$.
For participant-and-channel-specific noise variance parameters $\sigma_{n,e}^2$, we assign Half-Cauchy priors with a mean zero and a scale five. For temporal and spatial parameters $\rho_n$ and $\eta_n$, we assign discrete uniform priors among a grid of candidate values between 0 and 1. For $n=0,\cdots, N, e=1,\cdots, E$,
\begin{equation}
\begin{aligned}
& \bm{b}_{n,1}(s) \sim \mathcal{MGP}(\bm{0}, \psi_{n,1}\kappa_{1}, \bm{\Sigma}_n^s),\quad \bm{b}_{0,0}(s)\sim \mathcal{MGP}(\bm{0}, \psi_{0,0}\kappa_{0}, \bm{\Sigma}_0^s),\\
& \psi_{n,1}, \psi_{0,0} \overset{\mathrm{indep.}}{\sim}  \mathcal{LN}(0, 1), \quad 
\sigma_{n,e} \sim \mathcal{HC}(0, 5.0), \quad \rho_n \sim \mathcal{U}(0, 1), \quad \eta_n \sim \mathcal{U}(0, 1), \\
\end{aligned}
\label{eq:prior_specifications}
\end{equation}

\subsection{Markov Chain Monte Carlo}
\label{subsec:mcmc}

We apply the Gibbs Sampling method to estimate random functions $\bm{b}_{n,1}, \bm{b}_{0,0}$
and covariance-related parameters
$\sigma_{n,e}^2,\rho_n$, and $\eta_n$. By Mercer's representation theorem, we obtain that
\begin{equation}
\begin{aligned}
\bm{b}_{n,1}(s)& =\psi_{n,1}\sum_{l=1}^{+\infty}\Diag (\bm{a}_{n,1,l}) \phi_{1,l}(s)\approx \psi_{n,1}\sum_{l=1}^{L_1}\Diag (\bm{a}_{n,1,l}) \phi_{1,l}(s),\quad n=0,\cdots, N,\\
\bm{b}_{0,0}(s)& =\psi_{0,0}\sum_{l=1}^{+\infty}\Diag (\bm{a}_{0,0,l}) \phi_{0,l}(s)\approx\psi_{0,0} \sum_{l=1}^{L_0}\Diag (\bm{a}_{0,0,l}) \phi_{0,l}(s),\\
\end{aligned}
\label{eq:gp_b_function}
\end{equation}
where $\{\bm{a}_{n,1,l}\}_{n=0}^N$ and $\bm{a}_{0,0,l}$ are column vectors of coefficient parameters associated with eigenfunctions $\phi_{1,l}$ and $\phi_{0,l}$, respectively.
We evaluate $\bm{b}_{n,1}(s)$ and $\bm{b}_{0,0}(s)$ over a group of input time points $\{s_t\}_{t=1}^{T_0}$ and define them $\bm{B}_{n,1}$ and $\bm{B}_{0,0}$, respectively. Then, we adopt Eq \eqref{eq:gamma_exp_kernel} to calculate the kernel covariances $\bm{\Psi}_1$ and $\bm{\Psi}_0$ associated with kernel function $\kappa_1$ and $\kappa_0$, respectively. $\bm{B}_{n,1}$ and $\bm{B}_{0,0}$ can be approximated by
\begin{equation}
\begin{aligned}
\bm{B}_{n,1} &\approx \psi_{n,1} \cdot \bm{A}_{n,1}\bm{\Psi}_1^{1:L_1},\quad
\bm{A}_{n,1} = \left(
    \bm{a}_{n,1,1} \cdots  \bm{a}_{n,1,l} \cdots \bm{a}_{n,1,L_1}
\right), \quad n\in\{0,\cdots, N\}, \\ 
\bm{B}_{0,0} & \approx \psi_{0,0} \cdot \bm{A}_{0,0}\bm{\Psi}_0^{1:L_0},\quad 
\bm{A}_{0,0}=\left(
    \bm{a}_{0,0,1} \cdots  \bm{a}_{0,0,l}  \cdots  \bm{a}_{0,0,L_0}
\right),\\
\bm{A}_{n,1} & \sim \mathcal{MN}\left(\bm{0}, \bm{\Sigma}_n^s, \bm{V}_{\bm{\Psi}_1}^{1:L_1}\right), \quad \bm{A}_{0,0}\sim \mathcal{MN}\left(\bm{0}, \bm{\Sigma}_0^s, \bm{V}_{\bm{\Psi}_0}^{1:L_0}\right),\\
\end{aligned}
\end{equation}
where $\bm{A}_{n,1}$ and $\bm{A}_{0,0}$ follow matrix normal prior distributions with closed posterior forms
\citep{zhang2021high}, and $\bm{\Psi}_1^{1:L_1},\bm{\Psi}_0^{1:L_0},\bm{V}_{\bm{\Psi}_1}^{1:L_1},$ and $\bm{V}_{\bm{\Psi}_0}^{1:L_0}$ are the first $L_1$ rows of $\bm{\Psi}_1$, the first $L_0$ rows of $\bm{\Psi}_0$, the diagonal matrix of the first $L_1$ eigenvalues associated with $\bm{\Psi}_1$, and the diagonal matrix of the first $L_0$ eigenvalues associated with $\bm{\Psi}_0$, respectively. In practice, $L_1$ and $L_0$ are determined by the minimum integer of which cumulative sum of eigenvalues divided by the total sum of eigenvalues is over 95\% after reordering them by the descending order. For the remaining parameters, we apply Metropolis-Hastings algorithm to draw posterior samples. For the convergence check, we run multiple chains with different seed values and evaluate the convergence by Gelman-Rubin statistic \citep{gelman1992inference}. In addition, we examine the chain-specific posterior means of $\{Z_n\}$ as a post-hoc way to demonstrate the convergence. Details of MCMC implementations can be found in the supplementary material.

\begin{figure}[htbp]
     \centering
      \includegraphics[width=0.95\textwidth]{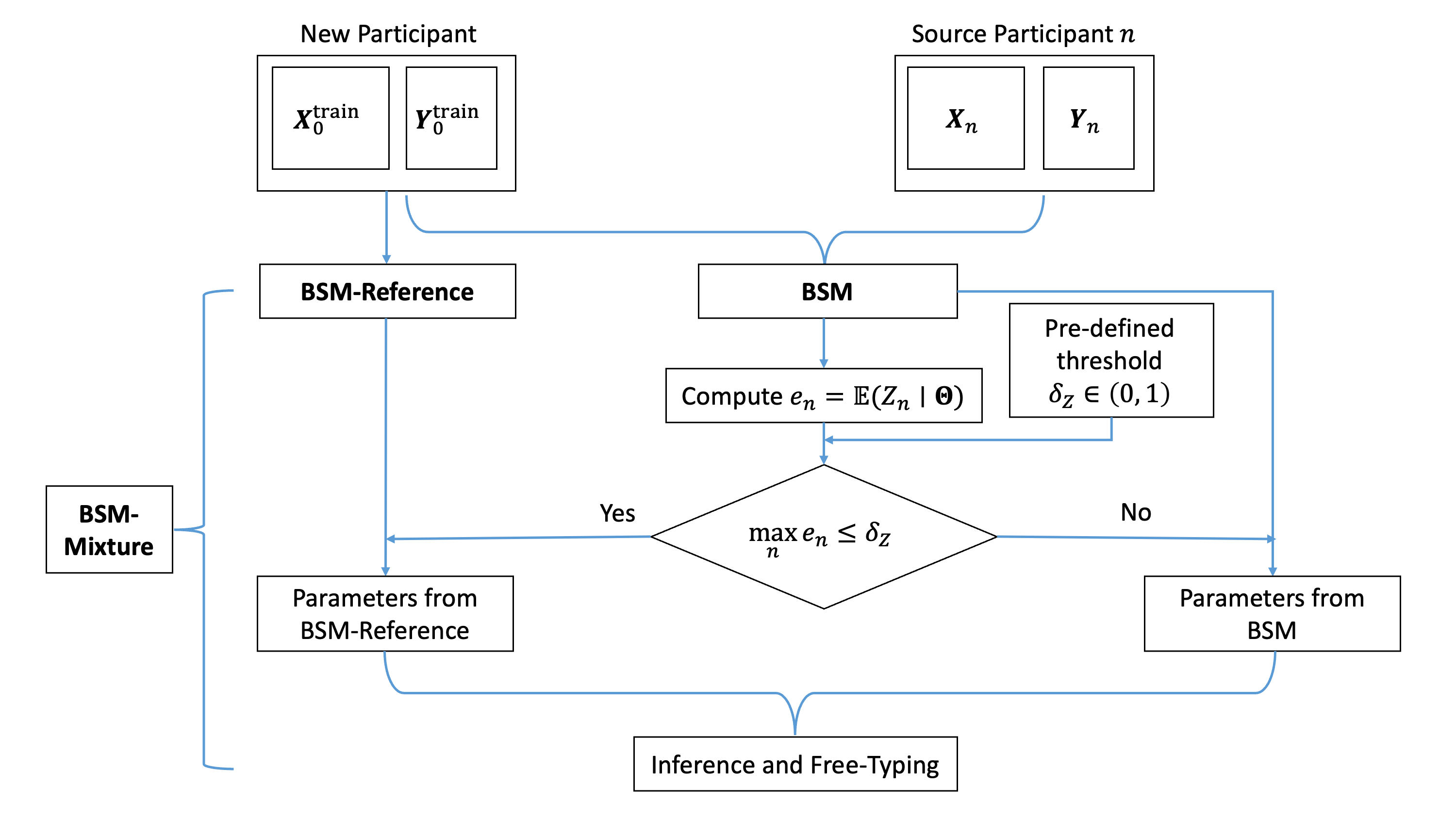}
    \caption{\small An illustration of the BSM-Mixture algorithm. First, we pre-define a threshold $\delta_Z\in(0,1)$. Second, we fit BSM model and obtain posterior samples of $\{Z_n\}$ to compute $\mathbb{E}(Z_n\mid\bm{\Theta})$. If none of the posterior expectations exceed $\delta_Z$, we apply BSM-Reference, otherwise, we apply BSM model for prediction. Note that both BSM-Reference and BSM models can be fitted simultaneously, and it will not have extra computational burden to the total calibration effort.}
    \label{fig:bsm_mixture_flowchart}
\end{figure}

\subsection{Hybrid Model Selection Strategy: BSM-Mixture}
\label{subsec:bsm_mixture}
Since our primary goal is to use small amount of training data from the new participant for calibration data, it may be challenging for our BSM model to identify source participants when no source participants are available for matching. In practice, we propose the hybrid model selection strategy, namely \textbf{BSM-Mixture}, such that if no source participants were selected satisfying certain threshold, we would use BSM-Reference for prediction. We define the hybrid model selection strategy as follows: First, we pre-define a threshold $\delta_Z \in (0, 1)$. Second, we fit BSM model and obtain the posterior samples of $\{Z_n\}$ to compute $\mathbb{E}(Z_n\mid\bm{\Theta})$. If none of $\mathbb{E}(Z_n\mid \bm{\Theta})$ exceed $\delta_Z$, we apply BSM-Reference, otherwise, we apply BSM model for prediction. Note that BSM-Reference and BSM model can be fitted simultaneously, it does not add much computational burden to the total calibration effort.
In addition, since the current data have been collected from the single lab source, we do not expect to see significant impact of machine effect on the parameter estimationa and prediction accuracy.
Figure \ref{fig:bsm_mixture_flowchart} provides an illustration of the hybrid model selection strategy on BSM-Mixture.

\subsection{Posterior Character-Level Prediction}
\label{subsec:post_char_prediction}

One of the advantages of our method is to predict on testing data directly without refitting the data. Under the RCP design, the character-level prediction depends on selecting the correct row and column within each sequence. Let $\bm{W}, \bm{Y}$, and $\bm{X}$ be existing stimulus-code indicators, stimulus-type indicators, and matrix-wise EEG signals, respectively. Let $\bm{W}_0^*, \bm{Y}_0^*, \bm{X}_0^*$, and $\bm{\Theta}_0$ be additional one sequence of stimulus-code indicators, stimulus-type indicators, matrix-wise EEG signals, and the parameter set, respectively, associated with the new participant. To simplify the notation, we assume that the new participant is spelling the same target character $\omega$. Based on the property described in Section \ref{subsec:problem_setup}, let $\bm{y}^*_{\omega}$ be the possible values of stimulus-type indicators given $\bm{W}_0^*$ and the target character $\omega$.
\begin{equation}
\begin{aligned}
\Prob(\bm{Y}_0^* = \bm{y}^*_{\omega} \mid \bm{X}_0^*, \bm{W}_0^*, \bm{X}, \bm{W}, \bm{Y}) & = \int \Prob (\bm{Y}_0^* = \bm{y}^*_{\omega} \mid \bm{\Theta}_0; \bm{X}_0^*, \bm{W}_0^*)\Prob (\bm{\Theta}_0 \mid \bm{X}, \bm{W}, \bm{Y})d\bm{\Theta}_0,\\
\Prob(\bm{Y}_0^* = \bm{y}^*_{\omega} \mid \bm{\Theta}_0; \bm{X}_0^*, \bm{W}_0^*)
& \propto \underbrace{\Prob(\bm{X}_{0}^*\mid \bm{\Theta}_0; \bm{W}_0^*, \omega)}_{\text{Equation } \eqref{eq:matrix_normal_model}} \cdot \Prob(\omega \text{ is target}),\\
\end{aligned}
\label{eq:char_prediction}
\end{equation}
Here, $\Prob(\omega \text{ is target}) = 1/36$ is the predictive prior on each candidate character with non-informative priors under the RCP design. When we need multiple sequence replications to select the target character, we modify the above formula by multiplying the sequence-specific posterior conditional likelihood within each cluster.

\section{Simulation Studies}
\label{sec:simulation}
We conduct extensive simulation studies to demonstrate the advantage of our proposed framework. We show the results of two simulation scenarios: a naive multi-channel setting and a real data based setting.

\subsection{Naive Multi-channel Simulation Scenario}
\label{subsec:simulation_naive_multi}
\noindent \textbf{Setup} \quad We consider the scenario with $N=7$ and $K=3$. The simulated EEG signals are generated from multivariate normal distributions  with a response window of 35 time points per channel, i.e., $T_0=35$. We design three groups of parameters including two-dimensional pre-specify mean ERP functions and variance nuisance parameters, where the shape and magnitude parameters are based on real participants from the database \citep{thompson2014plug}. For simulated data generated from group 0, we create a typical P300 pattern for the first channel where the target ERP function reach its positive peak around 10th time point post stimulus and reverse the sign for both ERP functions for the second channel; for simulated data generated from group 1, we create the same target ERP function as group 0 for the first channel and reduce the peak magnitude for the second channel; for simulated data generated from group 2, we reduce the peak magnitude for the first channel and create the same ERP functions as group 1 for the second channel. 
We consider an autoregressive temporal structure of order 1 (i.e., AR(1)), compound symmetry spatial structure, and channel-specific variances for background noises, where the true parameters for groups 0, 1, and 2 are $(\rho_0 = 0.7, \eta_0=0.6, \sigma_{0,1} =8.0,\sigma_{0,2}= 8.0)$, $(\rho_1 = 0.7, \eta_1=0.4, \sigma_{1,1} =8.0,\sigma_{1,2}= 6.0)$, and $(\rho_2 = 0.5, \eta_2=0.4, \sigma_{2,1} =2.0,\sigma_{2,2}= 2.0)$, respectively.
We design two cases for this scenario, a case without matched data among source participants (Case 1) and a case with matched data (Case 2). For Case 1, the group labels for source participants 1-6 are 1, 1, 1, 2, 2, and 2, respectively. For Case 2, the group labels for source participants 1-6 were 0, 0, 1, 1, 2, and 2, respectively. We perform 100 replications for each case. Within each replication, we assume that each participant is spelling the character ``T'' three times with ten sequence replications per character for training and generate additional testing data of the same size as the testing data of the single-channel scenario. 

\noindent \textbf{Settings and Diagnostics} \quad All simulated datasets are fitted with equation \eqref{eq:matrix_normal_model}. We also apply two covariance kernels $\kappa_1$ and $\kappa_0$ for target and non-target ERP functions, respectively. The two kernels are characterized by $\gamma$-exponential kernels with length-scale and gamma hyper-parameters as $(0.2, 1.2)$ and $(0.3, 1.2)$, respectively. The pre-specified threshold $\delta_Z$ is $0.5$. We run the MCMC with three chains, with each chain containing 5,000 burn-ins and 3,000 MCMC samples.

\begin{table}[htbp]
\centering
\caption{\small The \textbf{upper} and \textbf{lower} panels show summary of means and standard deviations of $Z_n=1$ for Cases 1 and 2 across 100 replications by BSM, respectively. The numerical values are multiplied by 100 for convenience and were reported with respect to the training sequence replications. 
}
\resizebox{0.80\columnwidth}{!}{
\centering
\begin{tabular}{c|cccccccccc} 
\hline
\textbf{Participant ID} & \multicolumn{10}{c}{\textbf{Sequence Size}} \\ \cline{2-11} 
  & 1 & 2 & 3 & 4 & 5 & 6 & 7 & 8 & 9 & 10 \\ \hline
1 & 31.7, 25.6 & 14.3, 18.5 & 8.2, 13.6 & 9.1, 14.8 & 8.7, 14.6 & 7.2, 13.7 & 7.9, 14.0 & 8.8, 15.4 & 8.2, 15.1 & 7.8, 13.2 \\
2 & 28.5, 26.3 & 15.3, 19.8 & 9.8, 16.6 & 9.5, 17.0 & 9.8, 16.6 & 7.5, 15.4 & 8.1, 13.6 & 6.8, 12.5 & 8.8, 14.0 & 9.2, 14.9 \\
3 & 32.8, 26.2 & 15.3, 19.3 & 9.2, 15.0 & 11.4, 17.2 & 8.9, 14.7 & 9, 15.7 & 11.1, 17.1 & 7.2, 13.7 & 6.5, 12.3 & 9.4, 15.8 \\
4 & 1.3, 3.3 & 1.6, 4.5 & 1.1, 0.3 & 1.4, 3.3 & 0.9, 0.3 & 0.9, 0.3 & 1.3, 3.3 & 1.0, 0.3 & 1.0, 0.3 & 1.7, 4.6 \\
5 & 2.3, 6.4 & 1.4, 3.3 & 1.7, 4.6 & 2.3, 6.4 & 1.0, 0.3 & 1.0, 0.3 & 1.0, 0.3 & 1.6, 4.6 & 2.0, 5.7 & 1.3, 3.2 \\
6 & 1.0, 0.4 & 1.0, 0.3 & 1.7, 4.6 & 1.0, 0.3 & 1.7, 4.6 & 1.3, 3.3 & 1.3, 3.2 & 1.7, 4.6 & 1.3, 3.3 & 2.0, 5.6 \\\hline
\hline
\textbf{Participant ID} & \multicolumn{10}{c}{\textbf{Sequence Size}} \\ \cline{2-11} 
 & 1 & 2 & 3 & 4 & 5 & 6 & 7 & 8 & 9 & 10 \\ \hline
1 & 75.9, 22.9 & 83.0, 20.0 & 89.0, 16.0 & 87.8, 17.9 & 92.3, 13.3 & 92.5, 14.1 & 92.2, 15.0 & 92.9, 13.2 & 93.3, 11.4 & 93.2, 13.4 \\
2 & 76.5, 22.0 & 81.0, 21.4 & 87.9, 17.6 & 87.3, 18.7 & 92.6, 14.2 & 91.2, 16.5 & 92.2, 14.1 & 91.8, 14.8 & 91.6, 15.2 & 91.2, 14.1 \\
3 & 15.5, 19.7 & 10.0, 14.5 & 12.4, 17.5 & 8.2, 15.1 & 6.6, 12.3 & 10.0, 15.4 & 6.9, 13.4 & 4.3, 10.9 & 6.8, 14.1 & 7.9, 14.1 \\
4 & 12.1, 18.0 & 11.7, 15.4 & 12.1, 18.6 & 9.2, 16.5 & 9.1, 16.3 & 7.5, 13.1 & 9.0, 15.5 & 8.2, 14.3 & 7.8, 14.8 & 6.5, 12.3 \\
5 & 1.7, 4.5 & 1.0, 0.3 & 1.0, 0.4 & 1.3, 3.3 & 1.0, 0.3 & 1.0, 0.3 & 1.6, 4.6 & 2.0, 5.6 & 1.3, 3.3 & 1.0, 0.3 \\
6 & 1.4, 3.3 & 1.0, 0.3 & 1.3, 3.3 & 1.3, 3.2 & 1.0, 0.3 & 1.1, 0.3 & 1.0, 0.3 & 0.9, 0.3 & 1.0, 0.3 & 1.0, 0.4 \\\hline
\end{tabular}}
\label{tab:simulation_multi_channel_cluster}
\end{table}

\noindent \textbf{Criteria} \quad We evaluate our method by matching and prediction. For matching, we report the proportion that each source participant is matched to the new participant and produce the ERP function estimates with 95\% credible bands with respect to the training sequence size. For prediction, we report the character-level prediction accuracy of the testing data, using Bayesian signal matching with source participants' data (BSM), Bayesian generative methods with the new participant's data only (BSM-Reference), the hybrid model selection strategy (BSM-Mixture), an existing classification method with data borrowing using Riemannian Geometry (MWDM), and swLDA with new participant's data only (swLDA). We do not include SMGP for comparison in the simulation studies because SMGP assumes that data are generated under sequences of stimuli, while our data generative mechanism is based on stimuli. We does not include other conventional ML methods because the prediction accuracy of swLDA is representative of those types of methods.

\noindent \textbf{Matching Results} \quad The upper and lower panels of Table \ref{tab:simulation_multi_channel_cluster} show the means and standard deviations of $Z_n=1$ across 100 replications for cases without and with matched data, respectively, by BSM. The numerical values are multiplied by 100 for convenient reading and are reported with respect to the training sequence size of the new participant. For Case 1, all the values quickly drop below 10.0 with 3 training sequence replications; for Case 2, values associated with Participants 1-2 are above 75.0 across all the training sequences and values associated with Participants 3-6 quickly drop below 10.0 within 4 sequence replications. The results indicate that our method correctly perform the participant selection.

\begin{sidewaysfigure}[htbp]
     \centering
     \begin{subfigure}
     \centering
      \includegraphics[width=0.45\textwidth]{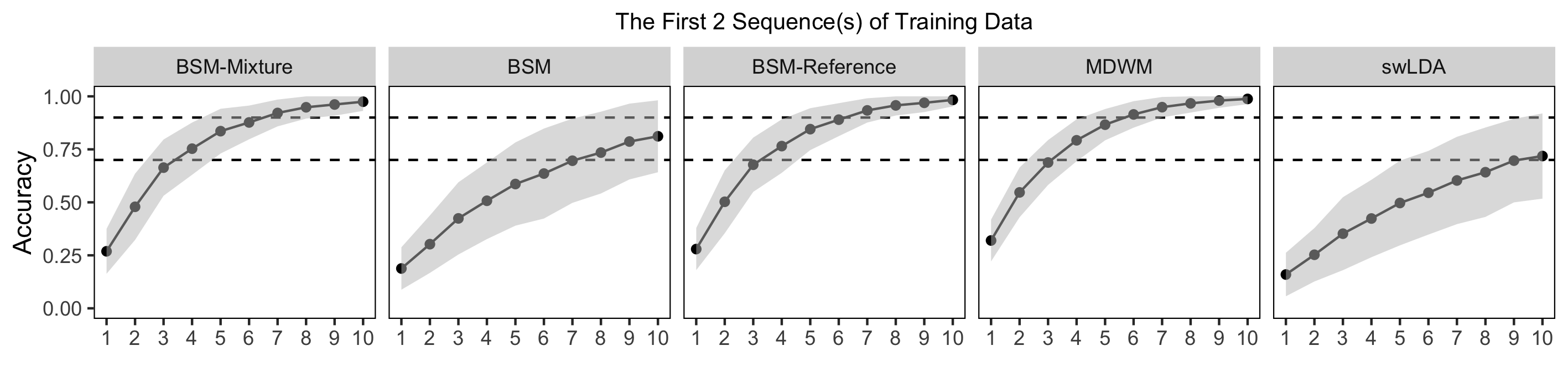}
     \end{subfigure}
     \hfill
     \begin{subfigure}
     \centering
      \includegraphics[width=0.45\textwidth]{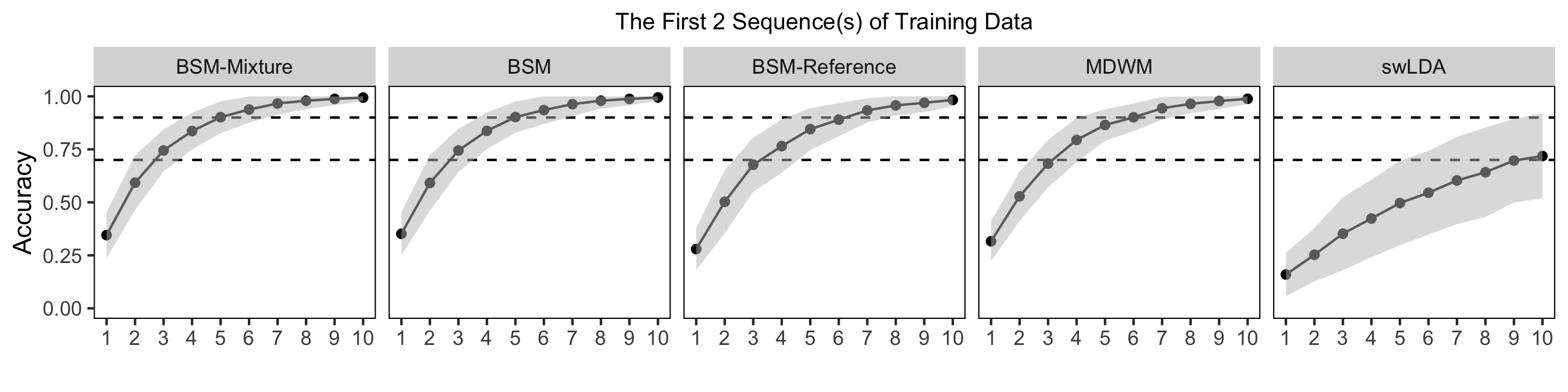}
     \end{subfigure}
     \vfill
     \begin{subfigure}
     \centering
      \includegraphics[width=0.45\textwidth]{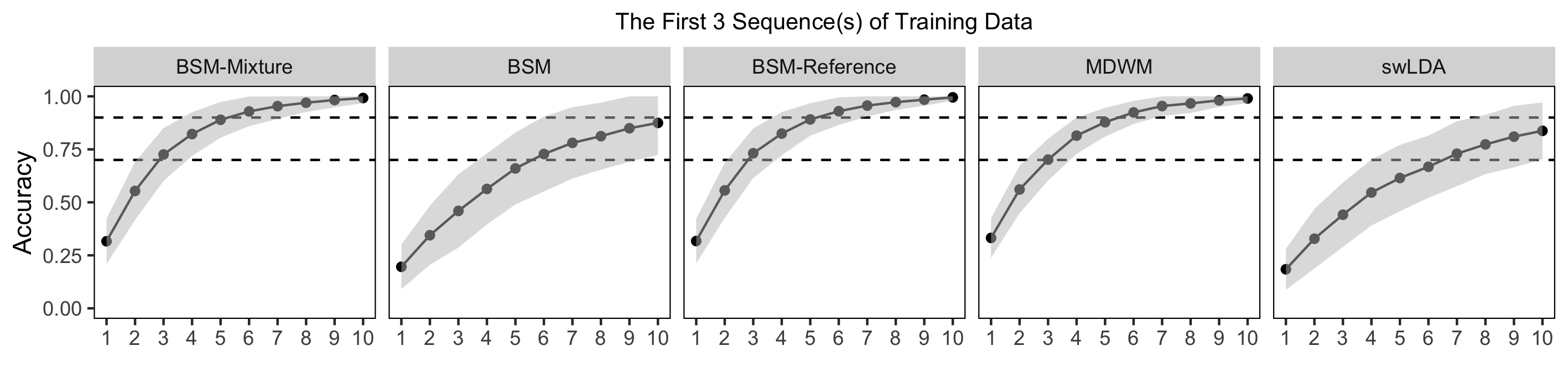}
     \end{subfigure}
     \hfill
      \begin{subfigure}
     \centering
      \includegraphics[width=0.45\textwidth]{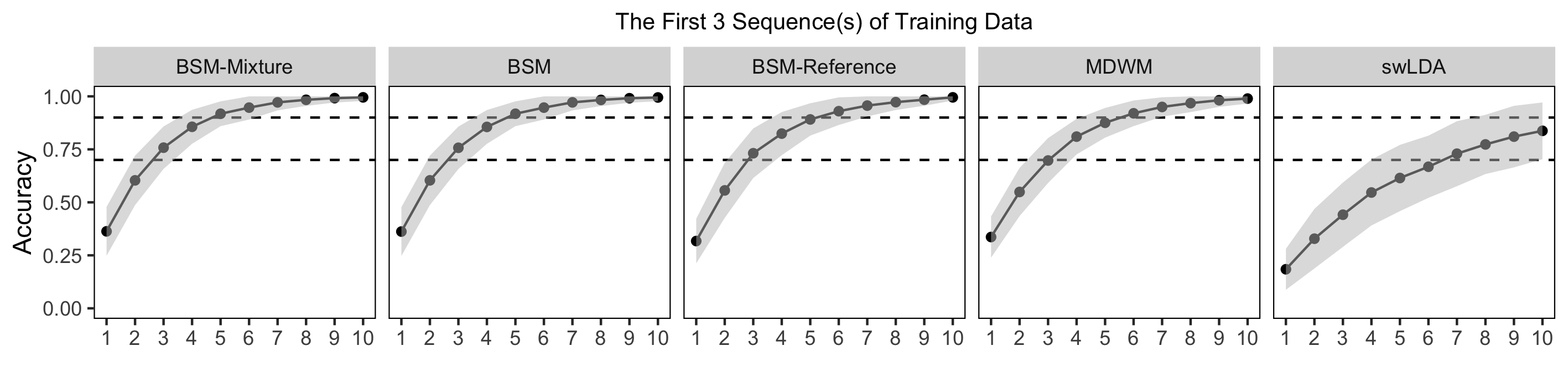}
     \end{subfigure}
     \vfill
     \begin{subfigure}
     \centering
      \includegraphics[width=0.45\textwidth]{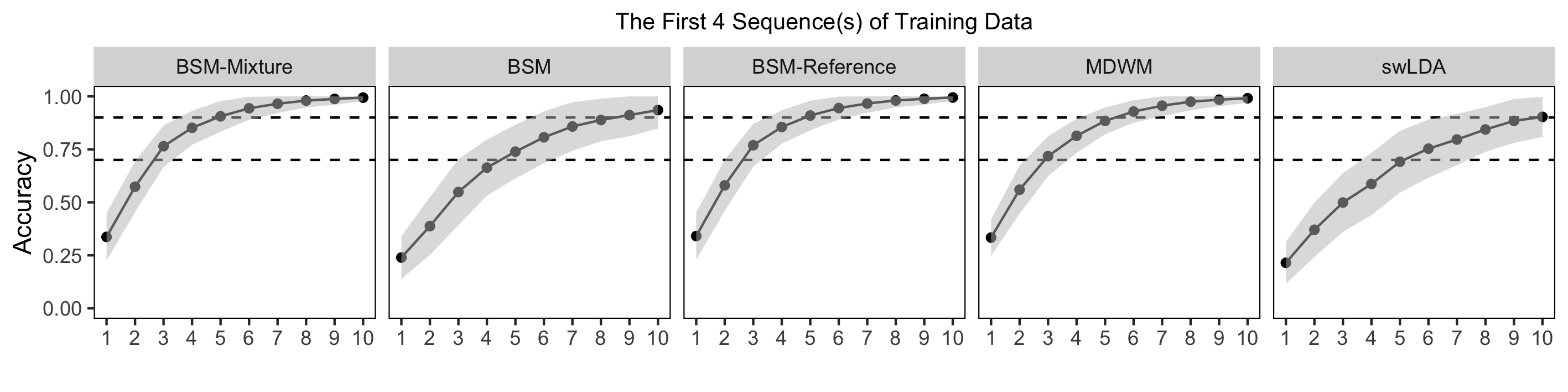}
     \end{subfigure}
     \hfill
     \begin{subfigure}
     \centering
      \includegraphics[width=0.45\textwidth]{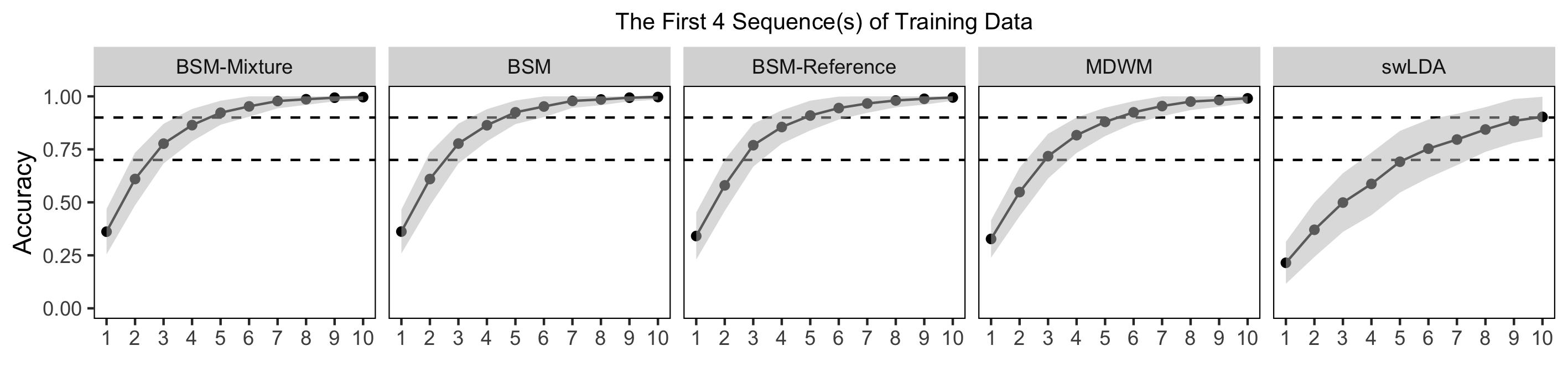}
     \end{subfigure}
     \vfill
    \caption{\small The \textbf{left} and \textbf{right} panels show the means and standard errors of testing prediction accuracy of Cases 1 (Unmatched) and 2 (Matched) by BSM-Mixture, BSM, BSM-Reference, MDWM, and swLDA across 100 replications, respectively. The upper, middle, and lower panels show prediction accuracy with the first 2, 3, and 4 training sequence replications, respectively. For Unmatched scenarios, BSM-Mixture was robust to source participants by adopting BSM-Reference's results, while for Matched scenario, BSM-Mixture required one testing sequence fewer than BSM-Reference to achieve the 90\% threshold with narrower error bars.}
    \label{fig:simulation_multi_channel_prediction}
\end{sidewaysfigure}

\noindent \textbf{Prediction Results} \quad The left and right panels of Figure \ref{fig:simulation_multi_channel_prediction} show the means and standard errors of testing prediction accuracy of Case 1 (No Matched) and Case 2 (Matched) by BSM-Mixture, BSM, BSM-Reference, MDWM, and swLDA, respectively. The testing prediction accuracy is further stratified by training the first 2, 3, and 4 sequences of the new participant. For Case 1, within the same row, BSM-Mixture and BSM-Reference have similar performances to MDWM. The poor performance of BSM is due to the uncertainty among participant selection in presence of no source participant and small training sample size. Therefore, the hybrid model selection strategy solves this issue. For Case 2, within the same row, BSM-Mixture performs the best, followed by MDWM, suggesting that our method successfully identifies the matched participants and leverages their information correctly. Within the same column, four methods benefit from an increasing number of training data from the new participant.

\subsection{Real Data Based Simulation Scenario}
\label{subsec:simulation_real_data_based}

We design a real-data based multi-channel simulation scenario. True parameters and sample sizes of training sets are copied from the real data analysis. We follow the tradition that participant 0 refers to the new participant, while participants 1 to 23 refer to the source participants. The participant-specific mean ERP functions and variance nuisance parameters are obtained from posterior means of each participant in the real data analysis. We multiply the variance estimates by 2 in this simulation scenario. The simulated EEG signals are also generated from multivariate normal distributions with a response window of 25 time points per channel, i.e., $T_0=25$. We perform 100 replications for this scenario. Within each replication, we assume that each participant is spelling the character ``T'' eight times with ten sequence replications per character for training and generate additional testing data of the same size as the testing data of the naive multi-channel scenario.
All simulated datasets are fitted with equation \eqref{eq:matrix_normal_model}. We used the same kernels, hyper-parameters, and the threshold $\delta_Z$ as the real data analysis. We run the MCMC with three chains, with each chain containing 5,000 burn-ins and 3,000 MCMC samples. Both matching and prediction results are reported, and we follow the same criteria as mentioned in Section \ref{subsec:simulation_naive_multi}.

\begin{figure}[ht]
     \centering
     \begin{subfigure}
     \centering
      \includegraphics[width=0.95\textwidth]{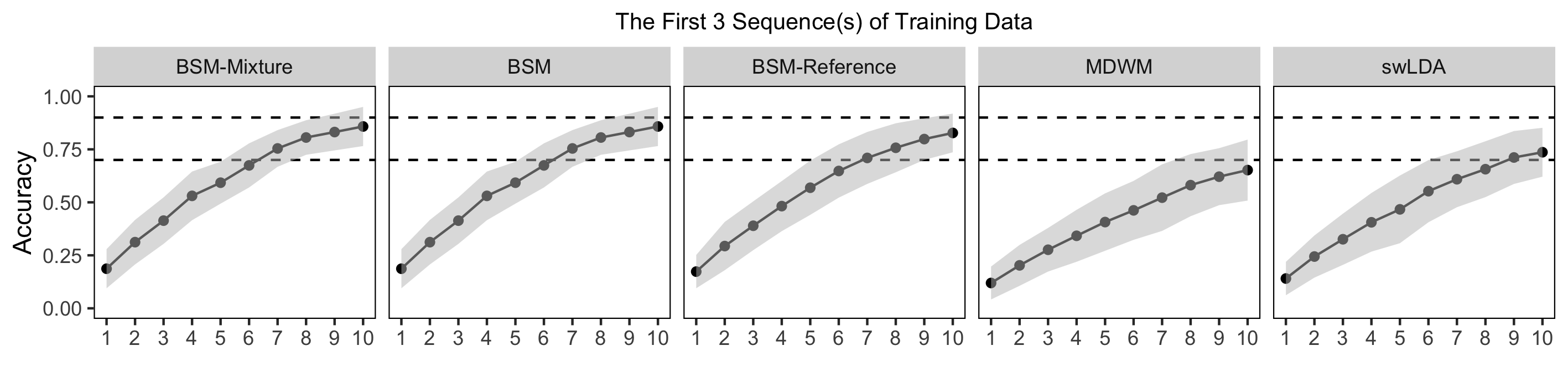}
     \end{subfigure}
     \vfill
     \begin{subfigure}
     \centering
      \includegraphics[width=0.95\textwidth]{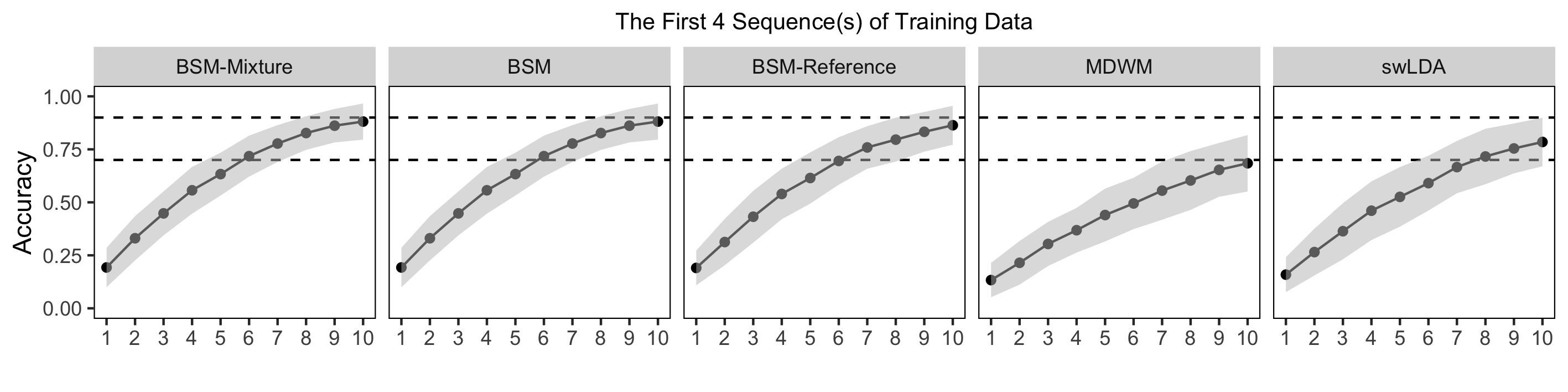}
     \end{subfigure}
     \vfill
     \begin{subfigure}
     \centering
      \includegraphics[width=0.95\textwidth]{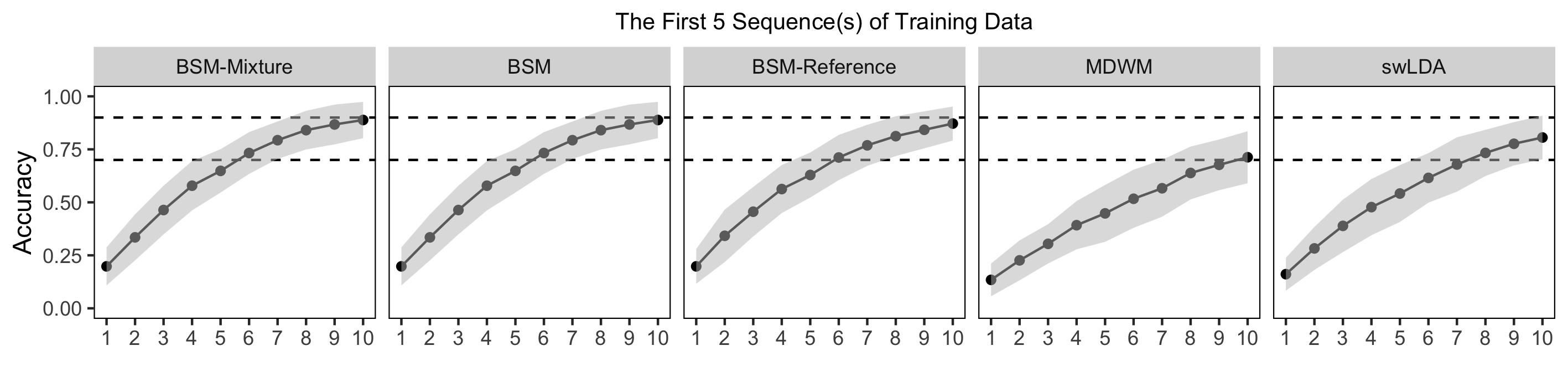}
     \end{subfigure}
     \vfill
    \caption{\small A graphic of the testing prediction accuracy. The upper, middle, and lower panel corresponds to the prediction accuracy with the first 3, 4, and 5 sequences of training data, respectively. All results are evaluated across 100 replications.}
    \label{fig:simulation_multi_channel_real_data_based_prediction}
\end{figure}

\noindent \textbf{Results} \quad
For matching, we only observe that participant 1 has a posterior mean of being merged around 0.2, while the posterior means of other source participants are below 0.1. The finding is consistent with the matching from the real data analysis given the same participant ordering and supports our argument that we typically expect one or two matching based on BSM classifier. Details of selection probability can be found in Section S4.2 in the supplementary.
For prediction, Figure \ref{fig:simulation_multi_channel_real_data_based_prediction} shows the testing prediction accuracy by the same five methods across 100 replications. The upper, middle, and lower panels are the prediction accuracy by training the first 2, 3, and 4 training sequences of the new participant. Overall, we observe that BSM-Mixture performs better than MDWM and swLDA. The prediction curves of BSM-Mixture and BSM are almost identical, suggesting that the algorithm identifies at least one source participant with the posterior mean of being merged exceeds the threshold $\delta_Z=0.1$, which is also consistent with the matching result. The prediction accuracy of BSM is slightly higher and narrower than that of BSM-Reference, possibly because the borrowing information may be limited when only one participant has reached the selection threshold. Finally, when we increase the sample size of training data, the prediction accuracy improves accordingly. 

\section{Analysis of BCI data from Real Participants}
\label{sec:real_data_analysis}

We examine the performance of our BSM framework on real-participant data, where the real data were originally collected from the \if0\blind University of Michigan Direct Brain Interface Laboratory (UMDBI)\fi
\if1\blind XXX Lab\fi. A total of 41 participants were recruited to accomplish the experiment, and we demonstrated the results of nine participants with brain damages or neuro-degenerative diseases.
Each participant completed one calibration (training) session (TRN) and up to three free-typing sessions (FRT). For the calibration session, each participant was asked to copy a 19-character phrase ``THE\_QUICK\_BROWN\_FOX'' including three spaces with 15 sequences, while wearing an EEG cap with 16 channels and sitting close to a monitor with a virtual keyboard with a $6\times 6$ grid. Details of the experimental setup can be found \cite{thompson2014plug}. 

The steps of the real data analysis are outlined as follows: First, we performed pre-processing techniques on raw EEG signals. Next, we identified a proper source participant pool based on testing swLDA accuracy. We trained swLDA and SMGP as two reference methods using the calibration data from the new participant only and trained BSM-Mixture and MDWM as two methods with data borrowing. The swLDA method was an existing efficient classifier. The MDWM method simply identified common features between the new participant and each source participant and did not perform explicit participant selection. Note that the SMGP method primarily focused on detecting the spatial-temporal discrepancies on brain activity between target and non-target ERP functions under the sequence-based analysis framework, while the other three methods treated stimulus-specific EEG data as feature vectors for classification.
We did not include other conventional ML methods because they did not have the borrowing version.
We demonstrated the detailed results of Participant K151, a senior male diagnosed with amyotrophic lateral sclerosis (ALS), and showed the results of remaining participants in the supplementary materials.

\subsection{Data Pre-processing Procedure and Model Fitting}
\label{subsec:pre_processing}

We applied the band-pass filter between 0.5 Hz and 6 Hz and down-sampled the EEG signals with a decimation factor of 8. Next, we extracted a fixed response window of approximately 800 ms after each event such that the resulting EEG matrix contained 15 sequence segments, and each sequence contained 12 extracted event-related EEG signal segments. The resulting event-related EEG signal segment had around 800 ms, i.e., 25 sampling points per channel. 
Since the SNR of EEG signals is typically low, we further applied the xDAWN filter to enhance the evoked potentials \citep{rivet2009xdawn}. The xDAWN method was an unsupervised algorithm that decomposed the raw EEG signals into relevant P300 ERPs and background noises and aimed to find a robust spatial filter that projected raw EEG signal segments onto the estimated P300 subspace by \textit{directly} modeling noise in the objective function.
We used all 16 channels as the signal input and selected the first two major components. We will denote raw EEG signal inputs and ERP functions after the xDAWN algorithm as transformed EEG signals and transformed ERP functions, respectively.

To emphasize the calibration-less point, we used a small portion of K151's training set and abundant training set from source participants. 
We identified 23 participants whose prediction accuracy of swLDA on FRT files exceeded 50\% and used them as the source participant pool:
For each source participant, we included the first 10 sequence replications of the first five characters: ``THE\_Q'', and the testing prediction accuracy was obtained with all FRT files (up to three sessions per participant). 
For K151, we included the first five sequence replications of the same five characters. 
Therefore, the resulting EEG feature matrices for the new participant and source participants had sizes of (300, 2, 25) and (600, 2, 25), respectively. Here, 300, 600, 2, and 25 were the number of  stimuli for new participant, the number of stimuli for source participants, the number of transformed ERP components and the length of EEG response window, respectively.

The datasets were fitted with the equation \eqref{eq:matrix_normal_model}. We applied two covariance kernels $\kappa_1$ and $\kappa_0$ for target and non-target ERP functions, respectively. Two kernels were characterized by $\gamma$-exponential kernels with length-scale and gamma hyper-parameters as $(0.2, 1.2)$ and $(0.3, 1.2)$, respectively. We ran the MCMC with three chains, with each chain containing 8,000 burn-ins and 1,000 MCMC samples. The Gelman-Rubin statistics were computed to evaluate the convergence status of each model. For matching, we reported the selection indicator $Z_n$. Due to the data heterogeneity, we expect to see partial matching. The largest posterior mean of $\{Z_n\}$ is around 0.2, the $\delta_Z$ should not exceed 0.2. In addition, most posterior means are around 4\%, and we expect to see two or three source participants to match with the new participant. We set the $\delta_Z=0.1$, i.e., the mean of $2/24$ and $3/24$. We also produced the transformed EEG signal function estimates by BSM and compared them to the results by BSM-Reference. 
For prediction, we reported the prediction accuracy on FRT files by BSM-Mixture, MDWM, swLDA, and SMGP up to five testing sequences. We only reported BSM-Mixture to utilize both advantages of BSM and BSM-Reference. 

\subsection{Inference and Prediction Results}
\label{subsec:real_data_K151}

\begin{figure}[ht]
    \centering
    \includegraphics[width=0.9\textwidth]{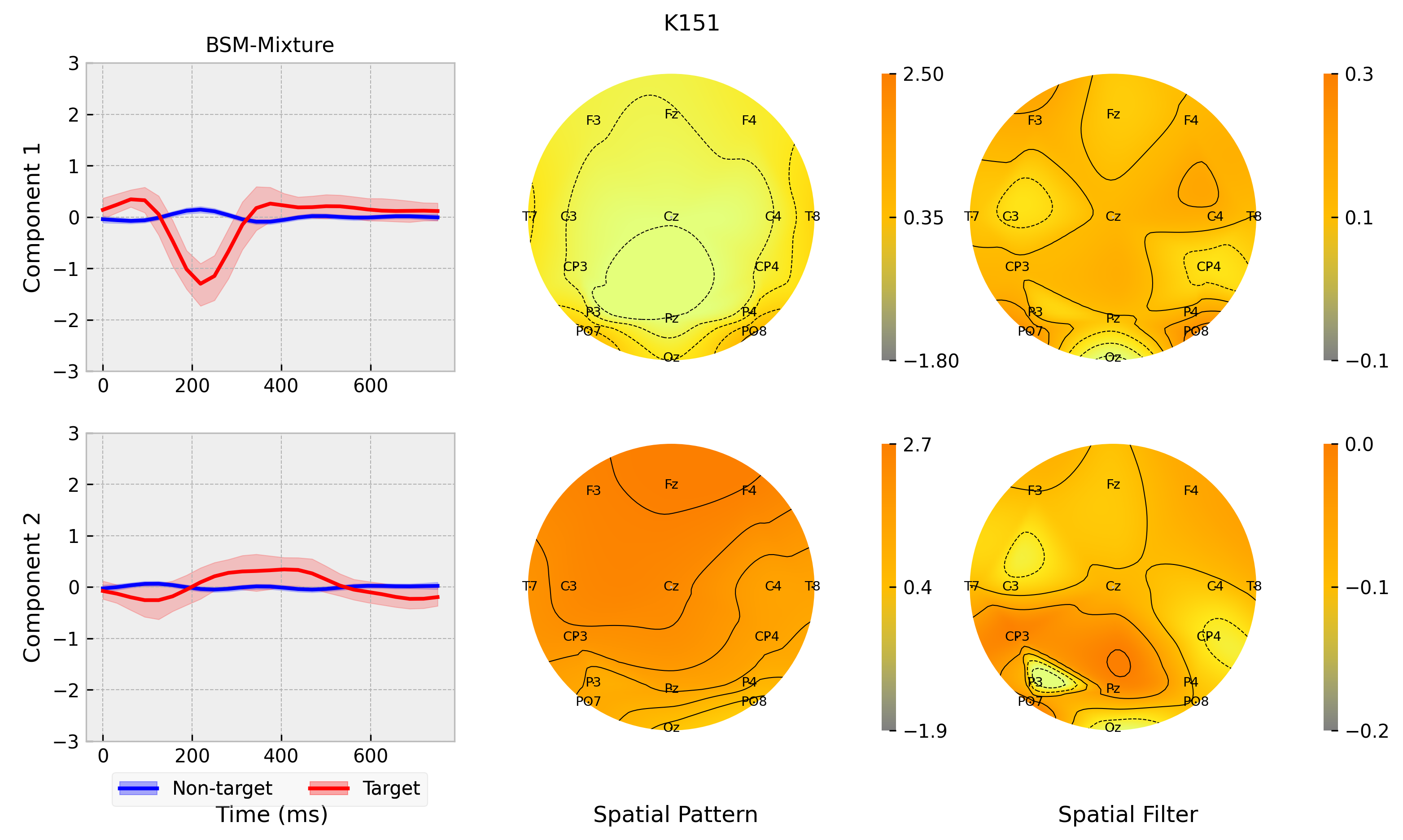}
    \caption{\small The first column shows the mean estimates with 95\% credible bands of the first two transformed ERP functions by BSM-Mixture. Only target ERP information is borrowed. The middle and right columns show the spatial patterns and spatial filters of the new participant's partial training data by xDAWN, respectively. 
    }
    \label{fig:real_data_K151}
\end{figure}

For the estimation of mean-level transformed signals, Figure \ref{fig:real_data_K151} shows the inference results of K151. The first columns show the mean estimates and their 95\% credible bands of the first two transformed ERP functions by BSM-Mixture. Only target ERP information was borrowed. The middle and right columns show the spatial pattern and spatial filter of the new participant's partial training data by xDAWN, respectively. 

For Component 1, we observed a sinusoidal pattern such that the transformed ERP response achieved a minor positive peak around 50 ms post stimulus and a major negative peak around 200 ms post stimulus. The signal achieved its second positive peak around 350 ms post stimulus and gradually decreased to zero. Spatial pattern and filter plots indicated that the major negative peak was primarily attributed to parietal-occipital and occipital regions (channels PO7, PO8, and Oz). For Component 2, we observed a Mexican-hat pattern with a double-camel-hump feature. The transformed target ERP response achieved its first negative peak around 100 ms post stimulus and its two positive peaks around 250 ms and 450 ms post stimulus before it gradually decreased to zero. Spatial pattern and filter plots indicated that such double-positive peak might be associated with the signals in central and parietal regions. The credible bands of BSM-Mixture were generally wider than BSM-Reference due to the uncertainty of pairwise comparisons among source participants.

For matching results, based on the P300 ERP estimation from BSM-Reference, we expect one or two source participants to match with the new participant, so we set the prior probability of $Z_n=1$ as $\Pr\{Z_n=1\}=2/24\approx 8\%$. And we set the $\delta_Z$ to be slightly higher than the prior probability, i.e., 10\%. Among 23 source participants, five source participants had posterior probability of being merged with K151 greater than or equal to 10\%. Of all, K111 showed the greatest similarity to K151, with a posterior probability of 25\%, followed by K158 (16\%), K183 (15\%), and K172 (13\%). The mean and median values of posterior probability $\Pr\{Z_n=1\}$ were 6.4\% and 4.0\%, respectively. Details of the matching results can be found in the Section S5.2 in the Supplementary material. Given the general low signal-to-noise ratio in EEG data, our borrowing results based on mean-level separation effects were not dominated by any particular source participant.

\begin{table}[htbp]
\centering
\caption{Prediction accuracy (percentage) of all FRT files with up to six testing sequences by BSM-Mixture, MDWM, swLDA, and SMGP with five training sequences.}
\resizebox{0.75\textwidth}{!}{%
\begin{tabular}{c|cc|cc} \hline 
 \textbf{Testing} & \multicolumn{2}{c|}{\textbf{Borrowing Methods}}  & \multicolumn{2}{c}{\textbf{Reference Methods}}        \\\cline{2-5}
   \begin{tabular}[c]{@{}c@{}} \textbf{Sequence Size}\end{tabular} & \textbf{BSM-Mixture} & \textbf{MDWM} & \textbf{swLDA} &
  \textbf{SMGP} \\\hline
  1 & 54.1\% & 47.3\% & 37.8\% & 16.2\%\\
  2 & 82.4\% & 71.6\% & 50.0\% & 23.0\% \\
  3 & 83.8\% & 75.7\% & 67.6\% & 40.5\% \\
  4 & 89.2\% & 81.1\% & 75.7\% & 39.2\% \\
  5 & 90.5\% & 90.5\% & 81.1\% & 51.4\% \\
  6 & 94.6\% & 95.9\% & 83.8\% & 59.8\% \\\hline
\end{tabular}}
\label{tab:real_data_K151}
\end{table}

Table \ref{tab:real_data_K151} shows the total prediction accuracy of all FRT files with respect to the testing sequences up to six by BSM-Mixture, MDWM, swLDA, and SMGP with five training sequences for K151. Overall, BSM-Mixture performed better than two reference methods. BSM-Mixture achieves an 85.1\% accuracy within 2 testing sequences, which is slightly faster than the existing method MDWM. Unfortunately, the SMGP method performs poorly in this case due to different assumptions in the analysis framework. The SMGP method treats one sequence of EEG data as an analysis unit, while our BSM framework treats one stimulus of EEG data as an analysis unit. Therefore, under the RCP design, the sample size for SMGP is 1/12 of that for BSM-Mixture, which leads to inaccurate estimation of P300 response functions and poor prediction on FRT files. 

In general, the advantage of BSM-Mixture is reflected in later testing sequences when participants are more likely to get distracted and have a higher heterogeneity level of EEG signals over time. Certain temporal intervals may become less separable and the attenuated amplitude estimates on these potentially confusing regions by BSM-Mixture actually reduce the negative effect of growing signal heterogeneity on prediction accuracy. For K151, temporal regions such as [500 ms, 800 ms] of Component 1, the double peaks of Component 2 by BSM-Mixture are both attenuated than those by BSM-Reference. However, that the initial 100 ms and intervals around 350 ms of Component 1 becomes even more separable by BSM-Mixture may suggest that those intervals are robust over time and essential to maintain high prediction accuracy during FRT tasks.

\section{Discussion}
\label{sec:discussion}

In this article, we propose a hierarchical Bayesian Signal Matching (BSM) framework to build a participant-semi-dependent, calibration-less framework with applications to P300 ERP-based Brain-Computer Interfaces. The BSM framework reduces the sample size for calibration of the new participant by borrowing data from pre-existing source participants’ pool at the participant level. The BSM framework specifies the joint distribution of stimulus-related EEG signal inputs via a Bayesian hierarchical mixture model. Unlike conventional clustering approaches, our method specifies the baseline cluster associated with the new participant and conducts pairwise comparisons between each source participant and the new participant by the criterion of log-likelihood. In this way, our method avoids the cluster label switching issue and can pre-compute model parameters from source participants to speed up the calibration. In addition, since our method intrinsically borrows the information from source participants via the binary selection indicators, we can use parameters from the baseline cluster to test directly without refitting the model with the augmented data. Finally, our hierarchical framework is base-model-free, such that it can be extended to any other classifiers with clear likelihood functions. We demonstrate the advantages of our method using extensive single- and multi-channel simulation studies and focus on the cohort with brain damages.

In the real data analysis, we find that first transformed ERP response is associated with the effect of visual cortex (higher positive values within parietal-occipital and occipital regions, while close to zero elsewhere) and that the second transformed ERP response is associated with the typical P300 responses (higher positive values within central and parietal regions). The resulting pattern with double-camel humps may be due to the effect of latency jitter \citep{guy2021peak} or simply the superimposed effect of various signal contributions from central and parietal electrodes by xDAWN. The cross-participant results also suggest that major negative peaks before 200 ms and positive peaks around 350 ms post stimulus are preserved in the first two transformed ERP responses. The above scientific findings are consistent with the finding from our previous study \cite{ma2022bayesian} that the performance of the P300 speller greatly depends on the effect of the visual cortex, in addition to electrodes located in central and parietal regions.

In addition, compared to the simulation studies, the borrowing results of the real data analysis show partial matching. Several reasons accounted for the results: First, the data generation mechanism of our method assumed that each observed EEG signal input is generated independently across stimuli, while the real data were extracted with overlapping EEG components between adjacent stimuli, and the overlapping temporal features still exist after applying the xDAWN filter. It is a potential model mis-specification to apply our method to the real data, which might make it difficult to find a very good match on the participant level. 
Second, our similarity criterion is based on calculating the log-likelihood of the target transformed ERP functions using the generative base classifier, which may be more sensitive to noise than discriminant classifiers. In simulation studies, the noises are clearly defined and well controlled. In the real data analysis, the noises may not be normally distributed, and the heterogeneity level of data increases not only across participants but also within the same participant over time, and the growing variability of data tends to compromise the separation effects of transformed ERPs. Fortunately, since BSM models the selection indicator in a soft-threshold manner, it has a flavor of Bayesian averaging that generally attenuates the magnitudes of peaks. Such diminishing differences between target and non-target transformed ERP function estimates suppress the weights of ``ambiguous'' temporal intervals and reduce the negative impact of false positives. 

\begin{figure}[ht]
    \centering
    \includegraphics[width=0.9\textwidth]{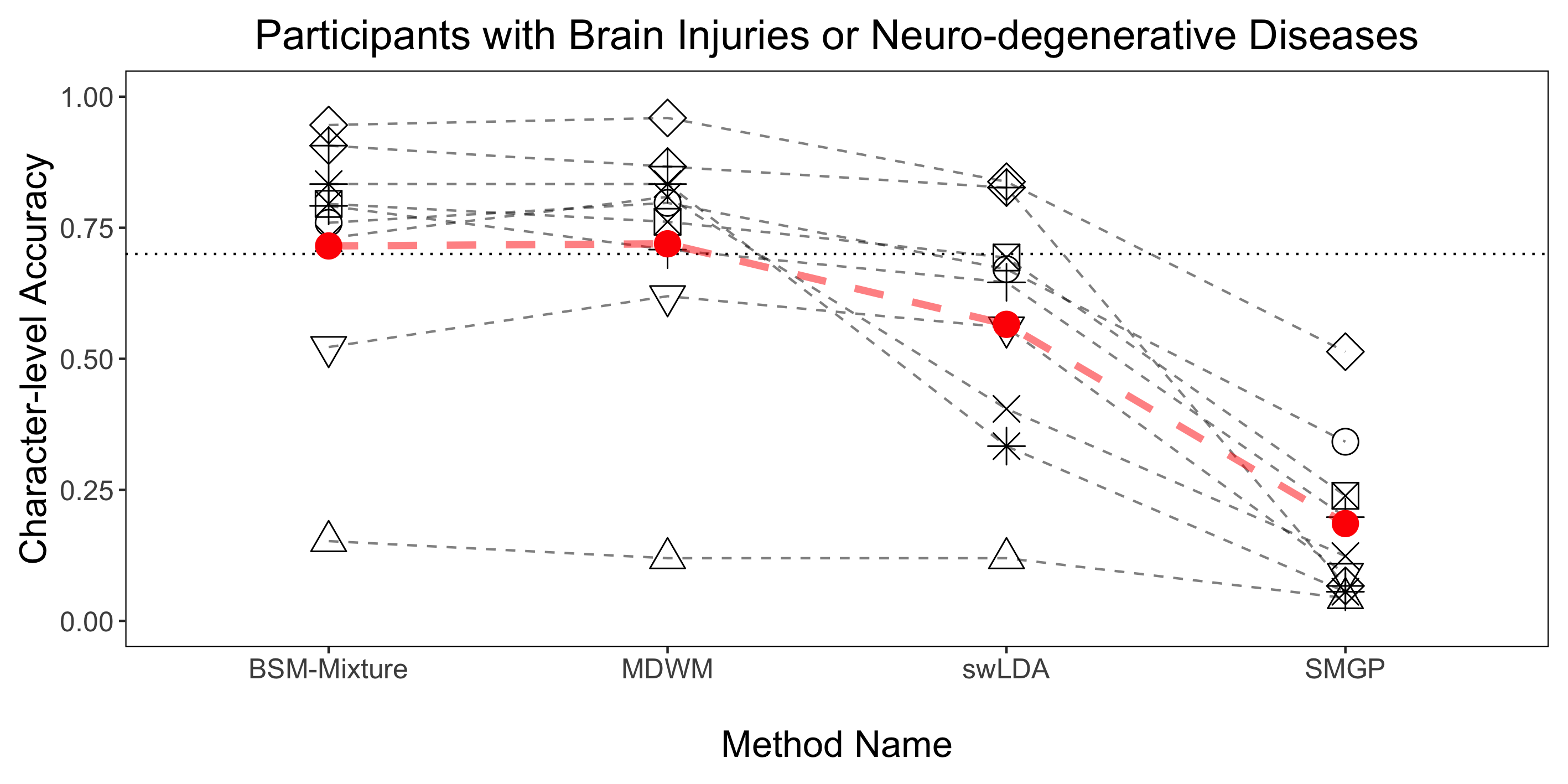}
    \caption{\small A Spaghetti plot of character-level testing accuracy by BSM-Mixture, MDWM,, swLDA, and SMGP for 9 participants with five training sequences. Means of each method are shown in red. A minimum accuracy of 0.7 for practical usage is shown.}
    \label{fig:real_data_cross_participant_summary}
\end{figure}

For cross-participant analysis, although we acknowledge the difference among the large and diverse source population, the features of interests for task-based, P300 ERP speller system are more consistent and easier to identify, compared to resting-state EEG data. We argue that the similarity exists among the participants who use this application. The transformed ERP function estimates of BSM are consistently attenuated compared to BSM-Reference, and the patterns of first two components are similar across nine participants.
Figure \ref{fig:real_data_cross_participant_summary} demonstrates the prediction accuracy of our target population cohort by BSM-Mixture, MDWM, swLDA, and SMGP. BSM-Mixture has similar prediction accuracy to MDWM, and they perform better than two reference methods. The minimal useful accuracy for communication is considered to be 70\% \citep{kubler2005brain}. The improvement in BSM-Mixture shows that people with brain injuries or neuro-degenerative diseases benefit from data borrowing.

There are some limitations to our study. First, the generative base classifier we used is still sensitive to background noise compared to other discriminant classifiers. It might overly extract information from source participants to overfit the model. Second, the generative classifier might require tuning hyper-parameters such as the kernel hyper-parameter selection of Gaussian process priors although our sensitivity analysis suggests that the inference and prediction are robust with respect to moderate changes of kernel hyper-parameters. The current strategy is to determine the values with heuristics, which might potentially affect the inference results and the prediction accuracy. Finally, although selection at the participant level is an intuitive idea, the characteristic of heterogeneity over time within the same participant suggested that we should have secondary sequence-level binary selection indicators. However, a side analysis of using sequence-level BSM-Mixture surprisingly performs worse than the one using participant-level selection indicators. It might be due to the fact that such refined selection indicators paired with generative base classifiers might overfit the training data and therefore has less power to generalize to the FRT data over time. Nevertheless, our main contribution is to demonstrate the advantages of the proposed hierarchical signal-matching framework, and the current framework could definitely be expanded and improved.

For future work, first, we would replace the generative base classifier with other robust discriminant classifiers, e.g., Riemannian geometry. The entire framework would remain unchanged except that we would specify the likelihood function by introducing Riemannian Gaussian distributions \citep{zanini2016parameters}, \citep{said2017riemannian}. 
If we followed the generative pathway for classifiers, it could be further improved byadopting the split-and-merge concept to explicitly control the discrepancies between target and non-target (transformed) EEG signals.
Finally, we would incorporate our framework into calibration-free methods \citep{verhoeven2017improving}. In addition to borrowing the training data from source participants and testing separately, we would directly start testing on the new participant and dynamically update the parameters of the baseline group as the data arrive.

\textbf{Data Available Statement}
Since the participants in the original study did not consent to release their data in public directly, even for de-identified version. However, the de-identified data are available upon request for other researchers. An outgoing data user agreement (DUA) will be required to initiate the data transfer. Please contact the corresponding author for details.

\section{Acknowledgments}

The authors would like to thank the Editor Professor Michael Stein, the Associate Editor and reviewers for their helpful comments and constructive suggestions, which led to this much-improved manuscript. The authors would also like to acknowledge the participants in our BCI experiments. The content is solely the responsibility of the authors and does not represent the official viewpoints of NICHD, NIH, NIDRR, or the Department of Education.

\section{Disclosures}
This work was partially supported by grants NIH R01DA048993 (Kang and Johnson), NIH R01MH105561 (Guo and Kang), and NSF IIS2123777 (Kang). The authors have no conflicts of interests to declare.

\bibliographystyle{apalike}
\spacingset{1.8}
\bibliography{main}
\end{document}


\maketitle

This supplementary document includes elaboration on model assumptions in Section \ref{supsec:three_assumptions}, the MCMC algorithm for single-channel scenario in Section \ref{supsec:mcmc_single}, the MCMC algorithm for multi-channel scenario in Section \ref{supsec:mcmc_multi}, results of single-channel simulation studies in Section \ref{supsec:simulation}, and remaining results of real data analysis and sensitivity analysis in Section \ref{supsec:real_data_analysis}. In addition, you can check the relevant code on the GitHub repository. \\
\url{https://github.com/NiubilityDiu/Bayesian-Signal-Matching-JASA}.

\section{Rationale behind Model Assumptions}
\label{supsec:three_assumptions}

We made three assumptions in this study.

\begin{figure}[ht]
\centering
\includegraphics[width=0.50\linewidth]{first_round_review/Figures/oddball_paradigm_illustration.png}
\caption{\small An illustration of P300 oddball paradigm.}
\label{fig:sup_oddball}
\end{figure}

\textit{First, we assume that the P300 ERPs were the same regardless of target character $\omega$ to spell}. The biological mechanism behind the speller system is the oddball paradigm that examines how the human brain responds to an unexpected stimulus \citep{luck2014introduction}. The human brain naturally elicits target P300 ERPs in response to a rare deviant stimulus among a series of repeated standard stimuli. Figure \ref{fig:sup_oddball} provides an illustration of such oddball paradigm for P300 ERPs. In our row-and-column (RCP) design, given a specific character $\omega$ to spell, only two target stimuli (row and column ones containing $\omega$) are supposed to elicit target P300 ERPs, while the remaining ten stimuli are non-target P300 ERPs. This rule applies regardless of the detailed spelling characters. Although we admit such character-level heterogeneity may exist per user, they are usually trivial (no significant differences of ERP estimation stratified by characters) and tend to cancel out after replications.

\begin{figure}[ht]
\centering
\includegraphics[width=0.33\linewidth]{first_round_review/Figures/heatmap_cov_mat_xdawn_component_1_target.png}
\hfill
\includegraphics[width=0.33\linewidth]{first_round_review/Figures/heatmap_cov_mat_xdawn_component_1.png}\hfill
\includegraphics[width=0.33\linewidth]{first_round_review/Figures/heatmap_cov_mat_xdawn_component_1_non_target.png}

\caption{\small The heatmap plots of temporal covariance matrix of the first component after xDAWN. The left, middle, and right correspond to target, non-target, and all filtered EEG features, respectively. Data are used from K151 in the real data analysis.}
\label{fig:sup_response_three_heatmaps}
\end{figure}

\textit{Second, we assume that the target and non-target ERP functions share the same covariance matrix within the same cluster, and we simplify the temporal covariance matrix as a scaled AR(1) correlation matrix.} 
This is supported because of the \textbf{special} way of data preprocessing. In practice, raw EEG recordings are typically stored in long time series ($E\times T_0$, where $E$ and $T_0$ refer to the channel and the streaming time length, respectively). One of the preprocessing steps is to truncate the raw EEG recordings into small EEG signal segments. Given a particular channel, the stimulus-specific EEG signal segment is a fixed-length vector that starts when the stimulus is highlighted and lasts about 800 ms thereafter. However, since the time interval between adjacent stimuli is about 150 ms, the resulting truncated EEG signal segment contains a large amount of overlap. The overlap universally exists for all truncated EEG segments, regardless of their stimulus types. The shared structure is preserved after the xDAWN preprocessing.

In addition, linear discriminant analysis (LDA) and its variations have been shown to be effective binary classifiers to decode the type of data \citep{krusienski2006comparison}. They assume that mean-level features have sufficient separation effect, which also makes us comfortable with the assumption of the shared covariance matrix. Finally, the shared and simplified covariance matrix also reduce the number of parameters and make the proposed model parsimonious, especially when we need to consider additional data from source participants.
Figure \ref{fig:sup_response_three_heatmaps} visualizes the covariance matrices for target, non-target, and all filtered EEG features. Despite the oscillation pattern in the left-sided plot, we argue that it is reasonable assume that target and non-target ERPs share the same covariance matrix as a scaled AR(1) correlation matrix given the above three arguments.

\begin{figure}[ht]
\centering
\includegraphics[width=0.45\linewidth]{first_round_review/Figures/P300_Cz.png}
\hfill
\includegraphics[width=0.45\linewidth]{first_round_review/Figures/P300_BSM_component_1.png}
\\
\caption{\small \textbf{Left}: Empirical target and non-target P300 ERPs of channel Cz using grand averaging (descriptive technique). \textbf{Right}: Fitted target and non-target transformed P300 ERP by BSM-Mixture.}
\label{fig:sup_response_two_P300}
\end{figure}

\textit{Finally, we perform selection with respect to the target ERP data and we exclud non-target ERP data from the source participants' pool.} We do it for two reasons: First, the typical target P300 ERPs have more distinct features compared to non-target P300 ERPs. 
The left and right panels of Figure \ref{fig:sup_response_two_P300} show a typical target and non-target P300 ERPs before and after xDAWN filters, respectively. Both figures show that target ERPs tend to have larger amplitude, clearer definition of latency, and a more recognizable shape. Those features will provide more useful information for signal matching. Second, the special rule of RCP design leads to the fact that the ratio of target and non-target ERPs is 1:5 (Among a $6\times 6$ grid, one target row and column versus five remaining rows and five remaining columns). By only borrowing the target ERPs, we add a moderate amount of training data to our model, allowing faster mixing and better parameter estimation.

\section{MCMC Algorithm for Single-channel Scenario}
\label{supsec:mcmc_single}

We present the MCMC algorithm to update parameters $\{\bm{\alpha}_{n,1}, \psi_{n,1}, \sigma_n, \rho_n\}_{n=0}^N$, $\bm{\alpha}_{0,0}$, and $\psi_{0,0}$ conditional on the stimulus-type indicators $\{\bm{Y}_{n,i,j}\}$ and extracted EEG signals $\{\bm{X}_{n,i,j}\}$. To clarify, we use $\tilde{(\cdot)}_n$ to denote the parameters estimated using source participant $n$ only, and we use $(\cdot)_n$ to denote the parameters for source participant $n$ under the framework of our approach. We use ``rest'' to denote the remaining parameters for simplicity. In practice, in the ``no-matching'' step, we import a random sample from existing MCMC outputs.

\subsection{Joint Distribution and Prior Specifications}
\label{supsubsec:post_single}

The joint distribution is 
\begin{equation*}
\begin{aligned}
& \mathrm{Pr}(\{\bm{X}_{n,i,j}\},\bm{\Theta},\{Z_n\} \mid \{Y_{n,i,j}\};\{\pi_n\})\\
& = \underbrace{\left(\prod_{n=0}^{N} \mathrm{Pr}(\bm{\Theta}_n)\right)}_{\mathrm{Prior \, Distribution}} \cdot 
\underbrace{\left(\prod_{n=1}^N \prod_{Y_{n,i,j}=1}\mathrm{Pr}(\{\bm{X}_{n,i,j}\}\mid Z_n \bm{\Theta}_0 + (1-Z_n) \bm{\Theta}_n; \{Y_{n,i,j}\}) \cdot \mathrm{Pr}(Z_n \mid \pi_n)\right)}_{\mathrm{Source \, Participants' \, Likelihood}} \\
& \cdot 
\underbrace{\prod_{i,j}\mathrm{Pr}\left(\{\bm{X}_{0,i,j}\}\mid  \bm{\Theta}_0; \{Y_{0,i,j}\}\right)}_{\mathrm{New \, Participant's \, Likelihood}},\\
\end{aligned}
\end{equation*}
Recall that within participant-specific parameter set $\bm{\Theta}_n,\: n=0,\ldots,N$,
\begin{equation*}
\begin{aligned}
\beta_{n,1}(t) & \sim \mathcal{GP}(0, \psi_{n,1}\kappa_{1}), \quad \beta_{0,0}(t) \sim \mathcal{GP}(0, \psi_{0,0}\kappa_{0}), \quad \rho_n\sim \mathcal{U}(0, 1), \quad \sigma_n \sim \mathcal{HC}(0, 5.0).\\
\end{aligned}
\end{equation*}

\noindent By Mercer's Theorem, we further obtain that
\begin{equation*}
\begin{aligned}
\bm{\beta}_{n,1} & = \psi_{n,1} \bm{\Psi}_1 \bm{\alpha}_{n,1}, \quad \bm{\alpha}_{n,1} \sim \mathcal{MVN}(\bm{0}, \Diag(\bm{\lambda}_1)), \quad \psi_{n,1} \sim \mathcal{LN}(0, 1), \quad n\geq0,\\
\bm{\beta}_{0,0} & = \psi_{0,0} \bm{\Psi}_0 \bm{\alpha}_{0,0}, \quad \bm{\alpha}_{0,0} \sim \mathcal{MVN}(\bm{0}, \Diag(\bm{\lambda}_0)), \quad \psi_{0,0} \sim \mathcal{LN}(0, 1), \\
\end{aligned}
\end{equation*}
where $\bm{\Psi}_1,\bm{\lambda}_1$ and $\bm{\Psi}_0,\bm{\lambda}_0$ are eigen-functions and eigen-vectors for target and non-target ERP response functions, respectively. To simplify, we only assume two kernels based on the stimulus-type indicators.

\subsection{Markov Chain Monte Carlo}
\label{sup_subsec:mcmc_single}

\subsubsection{Update $\{\bm{\alpha}_{n,1}\}_{n=0}^N$ and $\bm{\alpha}_{0,0}$}
\label{supsub2:mcmc_single_alpha}

For $n=0$,
\begin{equation*}
\begin{aligned}
\bm{\alpha}_{0,1} \mid \mathrm{rest} & \sim \mathcal{MVN}  (\bm{\Lambda}^{-1}_{\alpha_{0,1}}\bm{\eta}_{\alpha_{0,1}}, \bm{\Lambda}^{-1}_{\alpha_{0,1}}),\\
\bm{\Lambda}_{\alpha_{0,1}} & = \sum_{\{n:Z_n=1\}\cup\{0\}} \sum_{Y_{n,i,j}=1} \left(\psi_{0,1}\bm{\Psi}_1\right)^{\top}\left(\sigma_0^2 \bm{R}_0^t\right)^{-1} \left(\psi_{0,1}\bm{\Psi}_1\right) + \Diag(\bm{\lambda}_1^{-1}),\\
\bm{\eta}_{\alpha_{0,1}} & = \sum_{\{n:Z_n=1\}\cup\{0\}}\sum_{Y_{n,i,j}=1}\left(\psi_{0,1}\bm{\Psi}_1\right)^{\top} \left(\sigma_0^2 \bm{R}_0^t\right)^{-1} \bm{X}_{n,i,j},\\
\bm{\alpha}_{0,0} \mid \mathrm{rest} & \sim \mathcal{MVN} (\bm{\Lambda}^{-1}_{\alpha_{0,0}}\bm{\eta}_{\alpha_{0,0}}, \bm{\Lambda}^{-1}_{\alpha_{0,0}}),\\
\bm{\Lambda}_{\alpha_{0,0}} & = \sum_{Y_{0,i,j}\neq 1} \left(\psi_{0,0}\bm{\Psi}_0\right)^{\top}\left(\sigma_0^2 \bm{R}_0^t\right)^{-1} \left(\psi_{0,0}\bm{\Psi}_0\right) + \mathrm{Diag}(\bm{\lambda}_0^{-1}),\\
\bm{\eta}_{\alpha_{0,0}} & = \sum_{Y_{0,i,j}\neq1}\left(\psi_{0,0}\bm{\Psi}_0\right)^{\top} \left(\sigma_0^2 \bm{R}_0^t\right)^{-1} \bm{X}_{0,i,j},\\
\end{aligned}
\end{equation*}
For $n>0$, 
\begin{equation*}
\begin{aligned}
\bm{\alpha}_{n,1} \mid \mathrm{rest} &\sim  \left\{
\begin{array}{ll}
\mathcal{MVN}   (\bm{0}, \Diag(\bm{\lambda}_1)),   &  Z_n=1,\\
\mathcal{MVN}   (\bm{\Lambda}^{-1}_{\alpha_{n,1}} \bm{\eta}_{\alpha_{n,1}}, \bm{\Lambda}^{-1}_{\alpha_{n,1}}),  &  Z_n =0.\\
\end{array}\right.\\
\bm{\Lambda}_{\alpha_{n,1}} & =  \sum_{Y_{n,i,j}=1} \left(\psi_{n,1}\bm{\Psi}_1\right)^{\top}\left(\sigma_n^2 \bm{R}_n^t)\right)^{-1} \left(\psi_{n,1}\bm{\Psi}_1\right) + \Diag(\bm{\lambda}_1^{-1}),\\
\bm{\eta}_{\alpha_{n,1}} & =  \sum_{Y_{n,i,j}=1}\left(\psi_{n,1}\bm{\Psi}_1\right)^{\top} \left(\sigma_n^2 \bm{R}_n^t)\right)^{-1} \bm{X}_{n,i,j}.\\
\end{aligned}
\end{equation*}

\subsubsection{Update $\{\psi_{n,1}\}_{n=0}^N$ and $\psi_{0,0}$}
\label{supsub2:mcmc_single_psi}

Let $f_{\mathcal{LN}}(x;0,1)$ be the density function for the Log-Normal prior.
For $n=0$, 
\begin{equation*}
\begin{aligned}
f(\psi_{0,1}\mid \mathrm{rest}) & \propto \prod_{\{n:Z_n=1\}\cup\{0\}} \prod_{Y_{n,i,j}=1}\bm{\phi}\left(\bm{X}_{n,i,j};\psi_{0,1}\bm{\Psi}_1\bm{\alpha}_{0,1},\sigma_0^2 \bm{R}_0^t\right) \cdot f_{\mathcal{LN}}(\psi_{0,1};0, 1), \\
f(\psi_{0,0}\mid \mathrm{rest}) & \propto \prod_{Y_{0,i,j}\neq1}\bm{\phi}\left(\bm{X}_{0,i,j};\psi_{0,0}\bm{\Psi}_0\bm{\alpha}_{0,0},\sigma_0^2 \bm{R}_0^t\right) \cdot f_{\mathcal{LN}}(\psi_{0,0};0, 1).\\
\end{aligned}
\end{equation*}
For $n>0$, 
\begin{equation*}
\begin{aligned}
f(\psi_{n,1}\mid \mathrm{rest}) & \propto
\left\{
\begin{array}{cl}
  f_{\mathcal{LN}}(\psi_{n,1};0, 1),  & Z_n=1,  \\
 \prod_{Y_{n,i,j}=1}\bm{\phi}\left(\bm{X}_{n,i,j};\psi_{n,1}\bm{\Psi}_1\bm{\alpha}_{n,1},\sigma_n^2 \bm{R}_n^t\right) \cdot f_{\mathcal{LN}}(\psi_{n,1};0, 1),   & Z_n=0.\\ 
\end{array}
\right.\\
\end{aligned}
\end{equation*}

\subsubsection{Update $\{\sigma_n\}_{n=0}^N$}
\label{supsub2:mcmc_single_sigma} 

Let $f_{\mathcal{HC}}(x;0,5)$ be the density function for the Half-Cauchy prior. For $n=0$,
\begin{equation*}
\begin{aligned}
f(\sigma_0 \mid \mathrm{rest}) & \propto  \prod_{\{n:Z_n=1\}\cup\{0\}}\prod_{Y_{n,i,j}=1}\bm{\phi}\left(\bm{X}_{n,i,j}; \psi_{0,1}\bm{\Psi}_1\bm{\alpha}_{0,1}, \sigma_0^2 \bm{R}_0^t\right) \\
& \quad \quad \quad \cdot \prod_{Y_{0,i,j}\neq1} \bm{\phi}\left(\bm{X}_{0,i,j};\psi_{0,0}\bm{\Psi}_0\bm{\alpha}_{0,0},\sigma_0^2\bm{R}_0^t \right) \cdot f_{\mathcal{HC}}(\sigma_0;0,5).\\
\end{aligned}
\end{equation*}
For $n>0$,
\begin{equation*}
\begin{aligned}
f(\sigma_n \mid \mathrm{rest}) & \propto \left\{
\begin{array}{lc}
  f_{\mathcal{HC}}(\sigma_n;0,5),   & Z_n = 1,  \\
 \prod_{Y_{n,i,j}=1}\bm{\phi}\left(\bm{X}_{n,i,j}; \psi_{n,1}\bm{\Psi}_1\bm{\alpha}_{n,1}, \sigma_n^2\bm{R}_n^t\right)  \cdot  f_{\mathcal{HC}}(\sigma_n;0,5),     & Z_n=0.\\
\end{array}
\right.\\
\end{aligned}
\end{equation*}

\subsubsection{Update $\{\rho_n\}_{n=0}^N$}
\label{supsub2:mcmc_single_rho}

For $\rho_n$, we construct the exponential-decay correlation matrix as
\begin{equation*}
\begin{aligned}
\bm{R}_n^t & = \begin{pmatrix}
1 & \rho_n & \rho_n^2 & \cdots & \rho_n^{T_0-1}\\
\rho_n & 1 & \rho_n & \cdots & \rho_n^{T_0-2}\\
\vdots & \ddots & \vdots & \ddots & \vdots \\
\rho_n^{T_0-1} & \cdots & \rho_n^2 & \rho_n & 1 \\
\end{pmatrix},\\
\end{aligned}
\end{equation*}
where $T_0$ is the EEG response window length. Then, for $n=0$,
\begin{equation*}
\begin{aligned}
f(\rho_0 \mid \mathrm{rest}) & \propto  \prod_{\{n:Z_n=1\}\cup\{0\}}\prod_{Y_{n,i,j}=1}\bm{\phi}\left(\bm{X}_{n,i,j}; \psi_{0,1}\bm{\Psi}_1\bm{\alpha}_{0,1}, \sigma_0^2 \bm{R}_0^t\right) \cdot \prod_{Y_{0,i,j}\neq1}\bm{\phi}\left(\bm{X}_{0,i,j};\psi_{0,0}\bm{\Psi}_0\bm{\alpha}_{0,0},\sigma_0^2\bm{R}_0^t\right).\\
\end{aligned}
\end{equation*}
For $n>0$, 
\begin{equation*}
\begin{aligned}
f(\rho_n \mid \mathrm{rest}) & \propto 
\left\{
\begin{array}{lc}
    \mathcal{DU}(N_{\rho}), & Z_n=1,  \\
    \prod_{Y_{n,i,j}=1}\bm{\phi}\left(\bm{X}_{n,i,j}; \psi_{n,1}\bm{\Psi}_1\bm{\alpha}_{n,1}, \sigma_n^2 \bm{R}_n^t\right), & Z_n=0.\\ 
\end{array}
\right.\\
\end{aligned}
\end{equation*}

\subsubsection{Update $\{Z_n\}_{n=1}^N$}
\label{supsub2:mcmc_single_zn}

For $Z_n$, since we only borrow target EEG signals from source participants, we obtain that
\begin{equation*}
\begin{aligned}
\Prob(Z_n=1 \mid \mathrm{rest}) & = \frac{\prod_{Y_{n,i,j}=1}\bm{\phi}(\bm{X}_{n,i,j}; \bm{\beta}_{0,1}, \sigma_0^2 \bm{R}_0^t)}{\prod_{Y_{n,i,j}=1}\bm{\phi}(\bm{X}_{n,i,j}; \bm{\beta}_{0,1}, \sigma_0^2 \bm{R}_0^t) + \prod_{Y_{n,i,j}=1}\bm{\phi}(\bm{X}_{n,i,j};\bm{\beta}_{n,1}, \sigma_n^2 \bm{R}_n^t)}.\\
\end{aligned}
\end{equation*}

\section{MCMC Algorithms for Multi-channel Scenario}
\label{supsec:mcmc_multi}

We present the MCMC algorithm to update parameters $\{\bm{A}_{n,1}, \psi_{n,1}, \{\sigma_{n,e}\}_{e=1}^E, \rho_n, \eta_n\}_{n=0}^N$, $\bm{A}_{0,0}$, and $\psi_{0,0}$ conditional on the stimulus-type indicators $\{\bm{Y}_{n,i,j}\}$ and extracted EEG signal matrices $\{\bm{X}_{n,i,j,e}\}$. 
Here, we use $(\cdot)_n$ and $(\tilde{\cdot})_n$ to denote parameters associated with source participant $n$ after matching and before matching, respectively.
We denote $\bar{\bm{\phi}}(\cdot)$ as the density function of a matrix normal distribution. We use ``rest'' to denote the remaining parameters for simplicity. 

\subsection{Joint Distribution and Prior Specifications}
\label{supsubsec:post_multi}

The joint distribution is 
\begin{equation*}
\begin{aligned}
& \mathrm{Pr}(\{\bm{X}_{n,i,j}\},\bm{\Theta},\{Z_n\} \mid \{Y_{n,i,j}\}; \{\pi_n\})\\
& = \underbrace{\left(\prod_{n=0}^{N} \mathrm{Pr}(\bm{\Theta}_n)\right)}_{\mathrm{Prior \, Distribution}} \cdot 
\underbrace{\left(\prod_{n=1}^N \prod_{Y_{n,i,j}=1}\mathrm{Pr}\left(\{\bm{X}_{n,i,j}\}\mid Z_n\bm{\Theta}_0+(1-Z_n)\bm{\Theta}_n; \{Y_{n,i,j}\}\right) \cdot \mathrm{Pr}(Z_n \mid \pi_n)\right)}_{\mathrm{Source \, Participants' \, Likelihood \, (Target \, Only)}} \\
& \cdot 
\underbrace{\prod_{i,j}\mathrm{Pr}\left(\{\bm{X}_{0,i,j}\}\mid \bm{\Theta}_0; \{Y_{0,i,j}\}\right)}_{\mathrm{New \, Participant's \, Likelihood}},\\
\end{aligned}
\end{equation*}

\noindent To apply the conjugate property, we assume that $\bm{A}_{n,1}$ and $\bm{A}_{0,0}$ share the same spatial covariance prior as $\bm{X}_{n,i,j}$, i.e.,
\begin{equation*}
\begin{aligned}
\bm{B}_{n,1} & = \bm{A}_{n,1} \bm{\Psi}_1, \quad \bm{A}_{n,1} \sim \mathcal{MN}(\bm{0}_{E,N_{\psi_1}}, \bm{\Sigma}_n^s, \Diag(\bm{\lambda}_1)), \quad \psi_{n,1} \sim \mathcal{LN}(0,1),\\
\bm{B}_{0,0} & = \bm{A}_{0,0} \bm{\Psi}_0, \quad \bm{A}_{0,0}  \sim \mathcal{MN}(\bm{0}_{E,N_{\psi_0}}, \bm{\Sigma}_n^s, \Diag(\bm{\lambda}_0)), \quad \psi_{0,0} \sim \mathcal{LN}(0,1),\\
\end{aligned}
\end{equation*}
where $\bm{\Psi}_1,\bm{\lambda}_1$ and $\bm{\Psi}_0,\bm{\lambda}_0$ are eigen-functions and eigen-vectors for target and non-target ERP response functions, respectively. $N_{\psi_1}$ and $N_{\psi_0}$ are numbers of eigen-values for eigen-functions $\bm{\Psi}_1$ and $\bm{\Psi}_0$, respectively. To simplify, we have two kernels based on stimulus-type indicators.

\subsection{Markov Chain Monte Carlo}
\label{supsubsec:mcmc_multi}

\subsubsection{Update $\{\bm{A}_{n,1}\}_{n=0}^N$ and $\bm{A}_{0,0}$}
\label{supsub2:mcmc_multi_alpha}

We adopt the similar method \citep{zhang2021high} to update $\bm{A}_{n,1}$ and $\bm{A}_{0,0}$. For $n=0$,
\begin{equation*}
\begin{aligned}
\bm{A}_{0,1} \mid \mathrm{rest} & \sim \mathcal{MN}(\bm{U}_{0,1}, \bm{R}_0^s, \bm{\Lambda}_{0,1}^{t}),\\
\bm{\Lambda}_{0,1}^t & = \left\{ \psi_{0,1}^2 \sum_{\{n:Z_n=1\}\cup\{0\}}
 \sum_{Y_{n,i,j}=1}\bm{\Psi}_1 (\bm{R}_0^t)^{-1}\bm{\Psi}_1^{\top}+\Diag(\bm{\lambda}_1)\right\}^{-1},\\
\bm{U}_{0,1}^{\top} & = \bm{\Lambda}_{0,1}^t \cdot \left(\psi_{0,1} \bm{\Psi}_1 (\bm{R}_0^t)^{-1} \sum_{\{n:Z_n=1\}\cup\{0\}}\sum_{Y_{n,i,j}=1}\bm{X}_{n,i,j}^{\top}\right).\\
\bm{A}_{0,0} \mid \mathrm{rest} & \sim \mathcal{MN}(\bm{U}_{0,0}, \bm{\Sigma}_0^s, \bm{\Lambda}_{0,0}^{t}),\\
\bm{\Lambda}_{0,0}^t & = \left\{\psi_{0,0}^2 
 \sum_{Y_{0,i,j}\neq1}\bm{\Psi}_0 (\bm{R}_0^t)^{-1}\bm{\Psi}_0^{\top}+\Diag(\bm{\lambda}_0)\right\}^{-1},\\
\bm{U}_{0,0}^{\top} & = \bm{\Lambda}_{0,0}^t \cdot \left(\psi_{0,0} \bm{\Psi}_0 (\bm{R}_0^t)^{-1} \sum_{Y_{0,i,j}\neq1}\bm{X}_{0,i,j}^{\top}\right).\\
\end{aligned}
\end{equation*}

\noindent For $n>0$,
\begin{equation*}
\begin{aligned}
&
\left\{
\begin{array}{lc}
\bm{A}_{n,1} \mid \mathrm{rest}=\bm{A}_{0,1}
,     &  Z_n=1, \\
\bm{A}_{n,1}=\tilde{\bm{A}}_{n,1}, \tilde{\bm{A}}_{n,1}\sim \mathcal{MN}(\bm{U}_{n,1}, \bm{\Sigma}_n^s, \bm{\Lambda}_{n,1}^{t}),     & Z_n=0.\\ 
\end{array}
\right.\\
\bm{\Lambda}_{n,1}^t & = \left\{\psi_{n,1}^2 
 \sum_{Y_{n,i,j}=1}\bm{\Psi}_1 (\bm{R}_n^t)^{-1}\bm{\Psi}_1^{\top}+\Diag(\bm{\lambda}_1)\right\}^{-1},\\
\bm{U}_{n,1}^{\top} & = \bm{\Lambda}_{n,1}^t \cdot \left(\psi_{n,1} \bm{\Psi}_1 (\bm{R}_n^t)^{-1} \sum_{Y_{n,i,j}=1}\bm{X}_{n,i,j}^{\top}\right).\\
\end{aligned}
\end{equation*}

\subsubsection{Update $\{\psi_{n,1}\}_{n=0}^N$ and $\psi_{0,0}$}
\label{supsub2:mcmc_multi_psi}

\noindent Let $f_{\mathcal{LN}}(x;0,1)$ be the density function for the Log-Normal prior. For $n=0$, 
\begin{equation*}
\begin{aligned}
f(\psi_{0,1}\mid \mathrm{rest}) & \propto \prod_{\{n:Z_n=1\}\cup\{0\}} \prod_{Y_{n,i,j}=1}\bar{\bm{\phi}}\left(\bm{X}_{n,i,j};\psi_{0,1}\bm{A}_{0,1}\bm{\Psi}_1,\bm{\Sigma}_0^s,\bm{R}_0^t\right) \cdot f_{\mathcal{LN}}(\psi_{0,1};0, 1), \\
f(\psi_{0,0}\mid \mathrm{rest}) & \propto \prod_{Y_{0,i,j}\neq1}\bar{\bm{\phi}}\left(\bm{X}_{0,i,j};\psi_{0,0}\bm{A}_{0,0}\bm{\Psi}_0,\bm{\Sigma}_0^s, \bm{R}_0^t\right) \cdot f_{\mathcal{LN}}(\psi_{0,0};0, 1),\\
\end{aligned}
\end{equation*}

\noindent For $n>0$, 
\begin{equation*}
\begin{aligned}
 & \left\{
\begin{array}{lc} 
\psi_{n,1}=\psi_{0,1},
 &  Z_n=1,\\
 \psi_{n,1}=\tilde{\psi}_{n,1}, &  Z_n=0.\\
f(\tilde{\psi}_{n,1}\mid \mathrm{rest})\propto\prod_{Y_{n,i,j}=1}\bar{\bm{\phi}}\left(\bm{X}_{n,i,j};\tilde{\psi}_{n,1}\bm{A}_{n,1}
\bm{\Psi}_1,\bm{\Sigma}_n^s,\bm{R}_n^t\right) \\
\quad \cdot f_{\mathcal{LN}}(\tilde{\psi}_{n,1};0, 1).\\
\end{array}
\right.\\
\end{aligned}
\end{equation*}

\subsubsection{Update $\{\{\sigma_{n,e}\}_{e=1}^E\}_{n=0}^N$}
\label{supsub2:mcmc_multi_sigma} 

Let $f_{\mathcal{HC}}(x;0,5)$ be the density function for the Half-Cauchy prior. We assume channel-specific variance parameters, i.e., $\sigma_{n,e}$ is the $(e,e)$-th entry of the diagonal matrix $\bm{V}_n$. We let $\bm{\Sigma}_n^s = \bm{V}_n\bm{R}_n^s\bm{V}_n^{\top}$. For $n=0$, we loop through channel $e$,
\begin{equation*}
\begin{aligned}
f(\sigma_{0,e} \mid \mathrm{rest}) & \propto  \prod_{\{n:Z_n=1\}\cup\{0\}}\prod_{Y_{n,i,j}=1}\bar{\bm{\phi}}\left(\bm{X}_{n,i,j}; \bm{B}_{0,1}, \bm{V}_0\bm{R}_0^s\bm{V}_0^{\top}, \bm{R}_0^t\right) \\
& \quad \quad \quad \cdot \prod_{Y_{0,i,j}\neq1} \bar{\bm{\phi}}\left(\bm{X}_{0,i,j};\bm{B}_{0,0},\bm{V}_0\bm{R}_0^s\bm{V}_0^{\top},\bm{R}_0^t \right) \cdot f_{\mathcal{HC}}(\sigma_
{0,e};0,5).\\
\end{aligned}
\end{equation*}

\noindent For $n>0$, we loop through each channel $e$,
\begin{equation*}
\begin{aligned}
 & 
\left\{
\begin{array}{lc}
\sigma_{n,e}=\sigma_{0,e},
&  Z_n=1,\\
\sigma_{n,e} = \tilde{\sigma}_{n,e}, & Z_n=0.\\
f(\tilde{\sigma}_{n,e}\mid \mathrm{rest})\propto\prod_{Y_{n,i,j}=1}\bar{\bm{\phi}}\left(\bm{X}_{n,i,j};\bm{B}_{n,1},\bm{V}_n\bm{R}_n^s\bm{V}_n^{\top}, \bm{R}_n^t\right)\\
\quad \cdot  f_{\mathcal{HC}}(\tilde{\sigma}_{n,e};0,5).\\ 
\end{array}
\right.\\
\end{aligned}
\end{equation*}

\subsubsection{Update $\{\eta_n\}_{n=0}^N$}
\label{supsub2:mcmc_multi_eta}

We construct the spatial correlation matrix $\bm{R}_n^s$ using the compound symmetry structure characterized by $\eta_n$. Let $g(x)$ be the prior density function for $\eta_n$. For $n=0$,
\begin{equation*}
\begin{aligned}
f(\eta_0 \mid \mathrm{rest}) & \propto  \prod_{\{n:Z_n=1\}\cup\{0\}}\prod_{Y_{n,i,j}=1} \bar{\bm{\phi}}\left(\bm{X}_{n,i,j}; \bm{B}_{0,1}, \bm{V}_0\bm{R}_0^s\bm{V}_0^{\top}, \bm{R}_0^t\right) \\
& \quad \quad \quad \cdot \prod_{Y_{0,i,j}\neq1}\bar{\bm{\phi}}\left(\bm{X}_{0,i,j};\bm{B}_{0,0},\bm{V}_0\bm{R}_0^s\bm{V}_0^{\top},\bm{R}_0^t\right) \cdot g(\eta_0).\\
\end{aligned}
\end{equation*}

\noindent For $n>0$,
\begin{equation*}
\begin{aligned}
\left\{
\begin{array}{lc}
\eta_n=\eta_0,     &  Z_n=1, \\
\eta_n = \tilde{\eta}_n, & Z_n=0.\\
\quad f(\tilde{\eta}_n\mid \mathrm{rest}) \propto \prod_{Y_{n,i,j}=1} \bar{\bm{\phi}}\left(\bm{X}_{n,i,j}; \bm{B}_{n,1}, \bm{V}_n \bm{R}_n^s \bm{V}_n^{\top}, \bm{R}_n^t\right) \cdot g(\tilde{\eta}_n).\\
\end{array}
\right.\\
\end{aligned}
\end{equation*}

\subsubsection{Update $\{\rho_n\}_{n=0}^N$}
\label{supsub2:mcmc_multi_rho}

We construct the same exponential-decay correlation matrix as in Section \ref{supsub2:mcmc_single_rho}. Let $h(x)$ be the prior density function for $\rho_n$. For $n=0$,
\begin{equation*}
\begin{aligned}
f(\rho_0 \mid \mathrm{rest}) & \propto  \prod_{\{n:Z_n=1\}\cup\{0\}}\prod_{Y_{n,i,j}=1}
\bar{\bm{\phi}}\left(\bm{X}_{n,i,j}; \bm{B}_{0,1}, \bm{V}_0 \bm{R}_0^s \bm{V}_0^{\top}, \bm{R}_0^t\right) \\
& \quad \quad \quad \cdot \prod_{Y_{0,i,j}\neq1}\bar{\bm{\phi}}\left(\bm{X}_{0,i,j};\bm{B}_{0,0},\bm{V}_0\bm{R}_0^s\bm{V}_0^{\top},\bm{R}_0^t\right) \cdot h(\rho_0)\\
\end{aligned}
\end{equation*}

\noindent For $n>0$,
\begin{equation*}
\begin{aligned}
\left\{
\begin{array}{lc}
    \rho_n=\rho_0, & Z_n=1,  \\
    \rho_n = \tilde{\rho}_n, & Z_n=0.\\
    f(\tilde{\rho}_n\mid \mathrm{rest})\propto\prod_{Y_{n,i,j}=1} \bar{\bm{\phi}}\left(\bm{X}_{n,i,j}; \bm{B}_{n,1}, \bm{V}_n \bm{R}_n^s \bm{V}_n^{\top}, \bm{R}_n^t\right)\cdot h(\tilde{\rho}_n).\\ 
\end{array}
\right.\\
\end{aligned}
\end{equation*}

\subsubsection{Update $\{Z_n\}_{n=1}^N$}
\label{supsub2:mcmc_multi_zn}

For $Z_n$, given the prior $\Prob(Z_n=1\mid \theta),\theta\in(0,1)$, we obtain that 
\begin{equation*}
\begin{aligned}
\Prob(Z_n=1 \mid \mathrm{rest}) & \propto\prod_{(i,j):Y_{n,i,j}=1}\bar{\bm{\phi}}(\bm{X}_{n,i,j}; \bm{B}_{0,1}, \bm{\Sigma}_0^s, \bm{R}_0^t) \Prob(Z_n=1\mid \theta).\\
\end{aligned}
\end{equation*}

\section{Additional Simulation Studies}
\label{supsec:simulation}

\subsection{Single-channel  Simulation Scenario}
\label{supsubsec:single_channel_sim}

\begin{figure}
    \centering
    \includegraphics[width=0.90\textwidth]{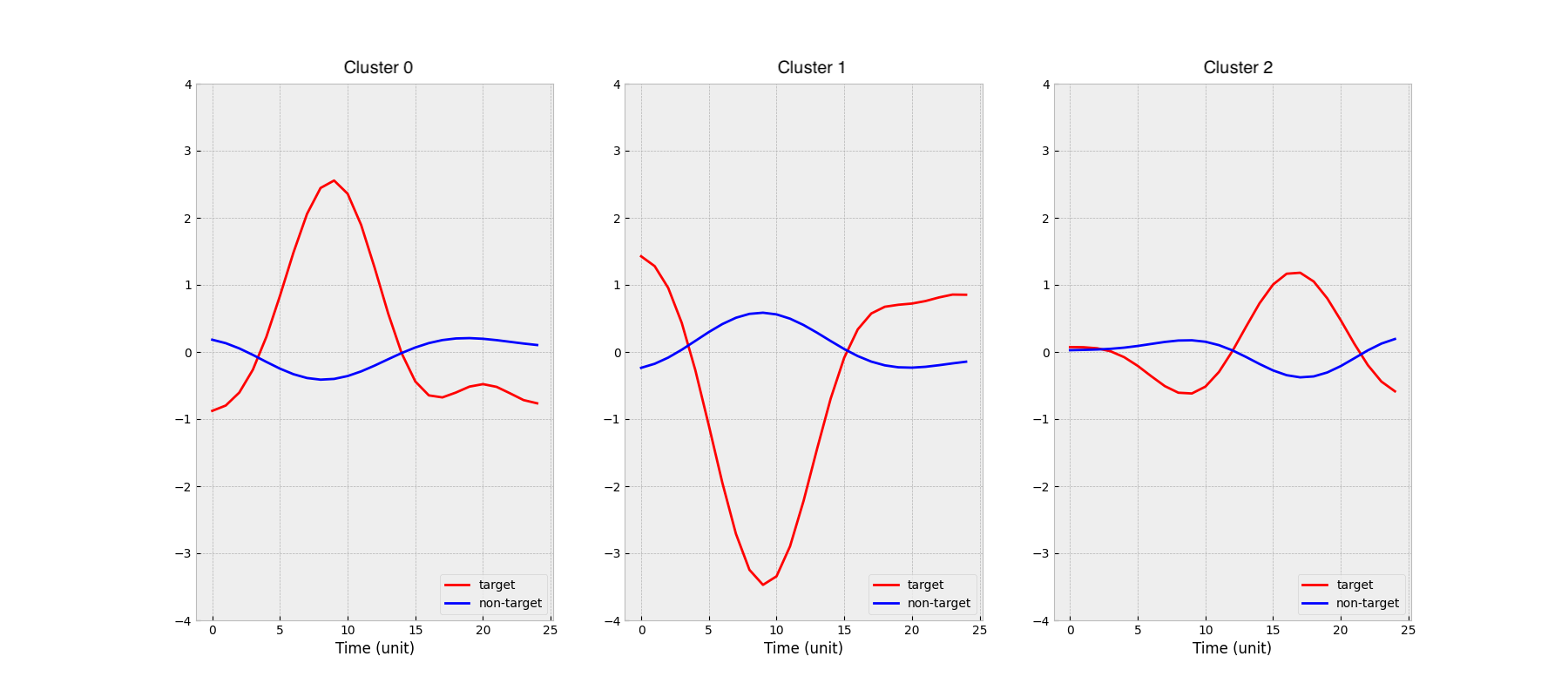}
    \caption{\small The one-dimensional true ERP function design for single-channel simulation scenario. The shape and magnitude were based on \citep{thompson2014plug}. The response window contained 35 time points. For cluster 0, we create a typical P300 pattern with a positive peak around 10th time point post stimulus; for cluster 1, we multiplied the typical P300 pattern by -1; for cluster 2, we created a different target ERP function with a late peak around 25th time point post stimulus.}
    \label{fig:sup_simulation_single_ERP_function}
\end{figure}

%
\noindent \textbf{Setup} \quad We considered the same scenario as the multi-channel scenario reported in the main text, i.e., $N=7$ and $K=3$. Figure \ref{fig:sup_simulation_single_ERP_function} demonstrated three pre-specified mean ERP functions, the shape and magnitude of which were based on a real participant \citep{thompson2014plug}. The simulated EEG signal segments were generated with a response window of 35 time points, i.e., $T_0=35$. 
For cluster 0, we created a typical P300 pattern where the target ERP function reached its positive peak around the 10th time point post stimulus; for cluster 1, we reversed the sign of ERP functions of cluster 0; for cluster 2, we considered a delayed P300 pattern where the target ERP function reached its peak around the 25th time point post stimulus. 
We considered an autoregressive temporal structure of order 1 (i.e., $\mathrm{AR}(1)$) for the background noises, where the true parameters for three clusters were $(\rho_0=0.6, \sigma_0=3.0), (\rho_1=0.6, \sigma_1=4.0)$, and $(\rho_2=0.7, \sigma_2=3.0)$.
We designed the same two cases as the multi-channel scenario, i.e., a case with matched data among source participants (Case S1) and a case without matched data (Case S2). For Case S1, the cluster labels for source participants 1-6 were 0, 0, 1, 1, 2, and 2, respectively. For Case S2, the cluster labels for source participants 1-6 were 1, 1, 1, 2, 2, and 2, respectively. 
We performed 100 replications for each case. Within each replication, we assumed that each participant was spelling characters ``TTT'' with ten sequence replications for training, and we generated additional EEG data from the new participant with 19 characters, with ten sequence replications per character for testing.

\begin{table}[htbp]
\centering
\caption{\small \textbf{Upper Panel}: A summary of means and standard deviations of $Z_n=1$ for Case S1 across 100 replications. \textbf{Lower Panel}: A summary of means and standard deviations of $Z_n=1$ for Case 2 across 100 replications. The numerical values were multiplied by 100 for convenience and were reported with respect to the training sequence replications. For Case S1, all the values are consistently close to zero throughout the entire training sequence replications; for Case S2, the values associated with Participants 1-2 gradually increased to 80.0 and maintain after 4 sequence replications, while the values associated with Participants 3-6 were consistently close to zero.
}
\resizebox{0.95\columnwidth}{!}{
\begin{tabular}{c|cccccccccc}
\hline
\textbf{Partici} & \multicolumn{10}{c}{\textbf{Sequence Size}}                                  \\ \cline{2-11} 
\textbf{-pant ID}       & 1    & 2    & 3    & 4    & 5    & 6    & 7    & 8    & 9    & 10   \\ \hline
1 & 1.0, 0.3 & 1.0, 0.3 & 1.0, 0.3 & 1.0, 0.3 & 1.0, 0.2 & 1.0, 0.3 & 1.0, 0.2 & 1.0, 0.3 & 1.0, 0.3 & 1.0, 0.3 \\
2 & 1.0, 0.3 & 1.0, 0.3 & 1.0, 0.3 & 1.0, 0.2 & 1.0, 0.3 & 1.0, 0.3 & 1.0, 0.3 & 1.0, 0.3 & 1, 0.3 & 1.0, 0.3 \\
3 & 1.0, 0.2 & 1.0, 0.3 & 1.0, 0.3 & 1.0, 0.3 & 1.0, 0.3 & 1.0, 0.3 & 1.0, 0.3 & 1.0, 0.3 & 1.0, 0.2 & 1.1, 0.3 \\
4 & 1.0, 0.3 & 1.0, 0.2 & 1.0, 0.3 & 1.0, 0.3 & 1.1, 0.3 & 1.0, 0.3 & 1.0, 0.2 & 1.0, 0.2 & 1.0, 0.2 & 1.0, 0.2 \\
5 & 1.0, 0.3 & 1.0, 0.3 & 1.1, 0.3 & 1, 0.3 & 1.0, 0.3 & 1.0, 0.2 & 1.0, 0.3 & 1.0, 0.2 & 1, 0.3 & 1.0, 0.3 \\
6 & 1.0, 0.2 & 1.0, 0.3 & 1.0, 0.3 & 1.0, 0.3 & 1.0, 0.2 & 1.0, 0.3 & 1.0, 0.3 & 1.0, 0.3 & 1.0, 0.3 & 1.0, 0.3 \\\hline
\hline
\textbf{Partici} & \multicolumn{10}{c}{\textbf{Sequence Size}}                                  \\ \cline{2-11} 
\textbf{-pant ID} & 1    & 2    & 3    & 4    & 5    & 6    & 7    & 8    & 9    & 10   \\ \hline
1 & 19.9, 16.9 & 57.2, 21.6 & 70.8, 20.6 & 78.3, 19.7 & 80.6, 18.9 & 82.0, 19.0 & 81.4, 20.5 & 82.0, 20.7 & 81.6, 21.4 & 79.9, 23.0 \\
2 & 20.0, 17.6 & 57.7, 20.6 & 71.0, 20.1 & 78.6, 18.3 & 80.2, 17.4 & 81.0, 17.1 & 80.7, 18.1 & 80.7, 18.8 & 79.0, 21.3 & 77.4, 23.3 \\
3 & 1.0, 0.3 & 1.1, 0.3 & 1.0, 0.2 & 1.0, 0.2 & 1.0, 0.2 & 1.0, 0.2 & 1.0, 0.2 & 1.0, 0.3 & 1.0, 0.2 & 1.0, 0.2 \\
4 & 1.0, 0.3 & 1.0, 0.3 & 1.0, 0.3 & 1.0, 0.2 & 1.1, 0.3 & 1.0, 0.3 & 1.0, 0.2 & 1.0, 0.3 & 1.0, 0.2 & 1.0, 0.2 \\
5 & 1.1, 0.4 & 1.0, 0.3 & 1.0, 0.2 & 1.0, 0.3 & 1.0, 0.2 & 1.0, 0.2 & 1.1, 0.3 & 1.0, 0.2 & 1.0, 0.2 & 1.0, 0.2 \\
6 & 1.0, 0.3 & 1.0, 0.2 & 1.0, 0.3 & 1.0, 0.3 & 1.0, 0.3 & 1.0, 0.3 & 1.0, 0.2 & 1.0, 0.3 & 1.0, 0.2 & 1.0, 0.2 \\\hline
\end{tabular}}
\label{tab:sup_simulation_single_channel_cluster}
\end{table}

\noindent \textbf{Settings and Diagnostics} \quad All simulated datasets were fitted with equations in Section \ref{supsec:mcmc_single}. 
To simplify, we considered two $\gamma$-exponential covariance kernels $\kappa_{1}$ and $\kappa_{0}$ for target and non-target ERP functions, respectively. The length-scale and gamma hyper-parameters of $\kappa_{1}$ and $\kappa_{0}$ were $(0.3, 1.2)$ and $(0.4, 1.2)$, respectively. We ran the MCMC with three chains, with each chain containing 5,000 burn-ins and 1,000 MCMC samples. The Gelman-Rubin statistics were evaluated.

\begin{sidewaysfigure}[htbp]
     \centering
     \begin{subfigure}
     \centering
      \includegraphics[width=0.45\textwidth]{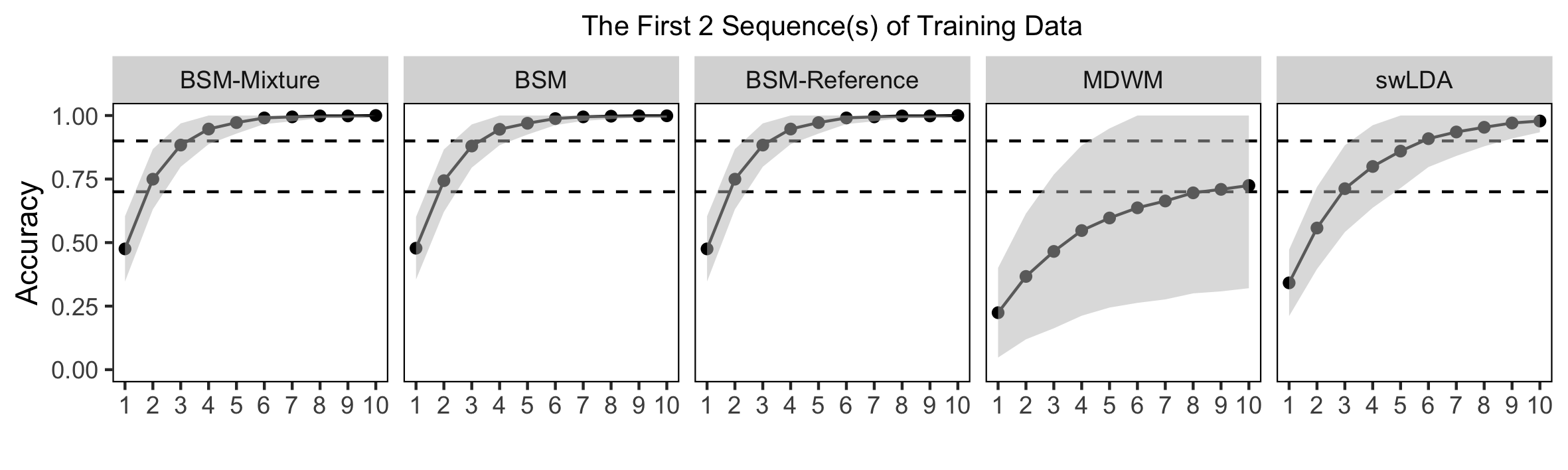}
     \end{subfigure}
     \hfill
     \begin{subfigure}
     \centering
      \includegraphics[width=0.45\textwidth]{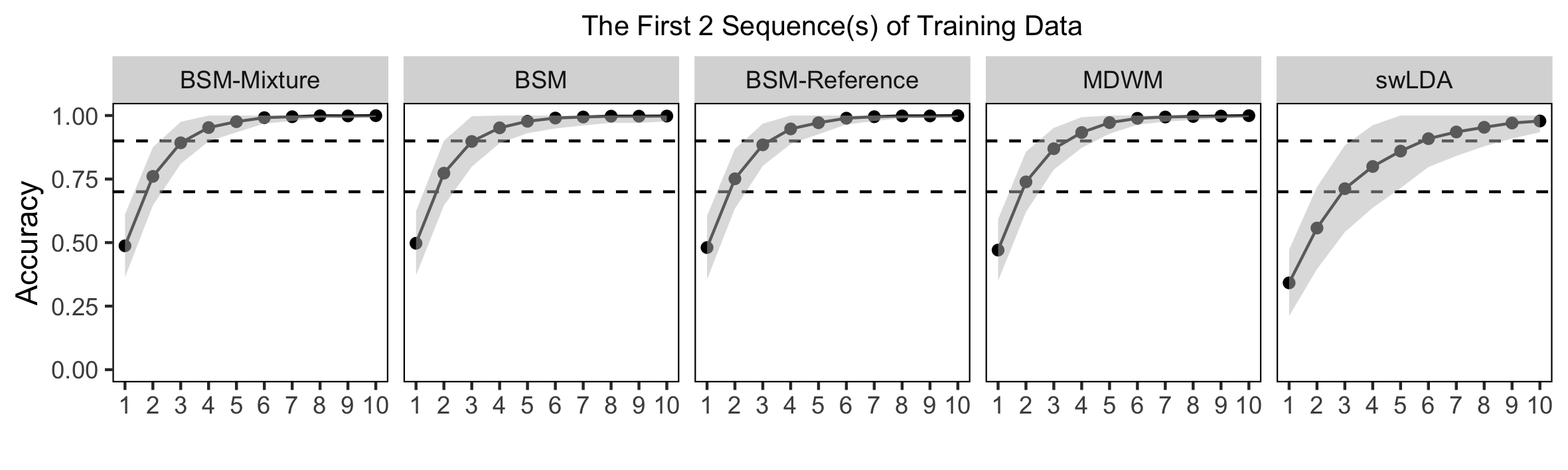}
    \end{subfigure}
     \vfill
     \begin{subfigure}
     \centering
      \includegraphics[width=0.45\textwidth]{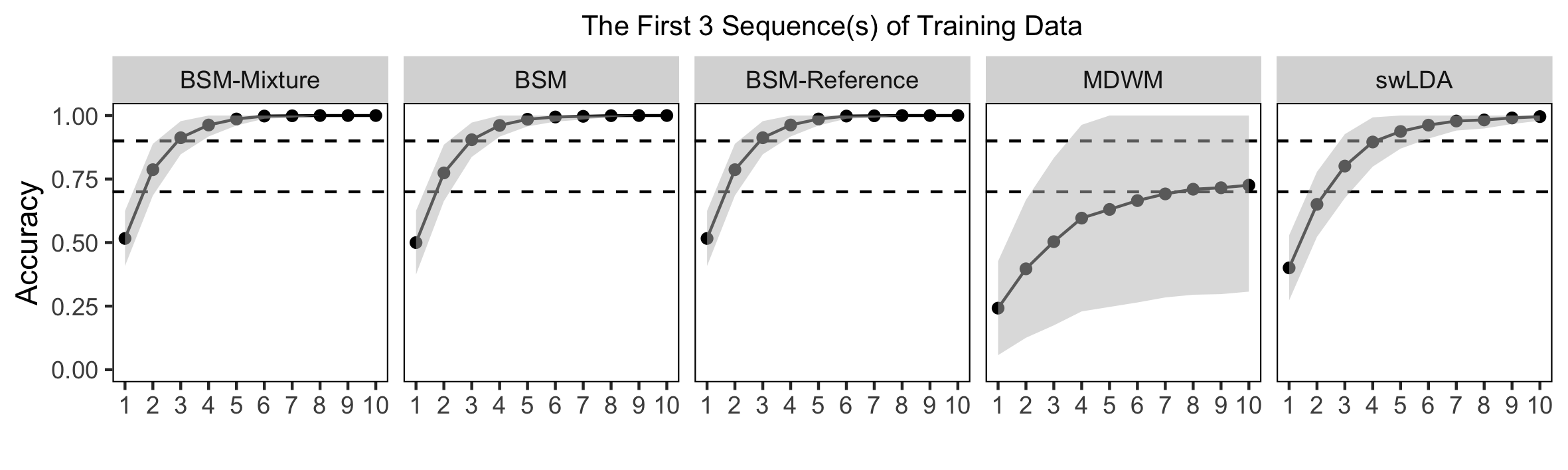}
     \end{subfigure}
     \hfill
     \begin{subfigure}
     \centering
      \includegraphics[width=0.45\textwidth]{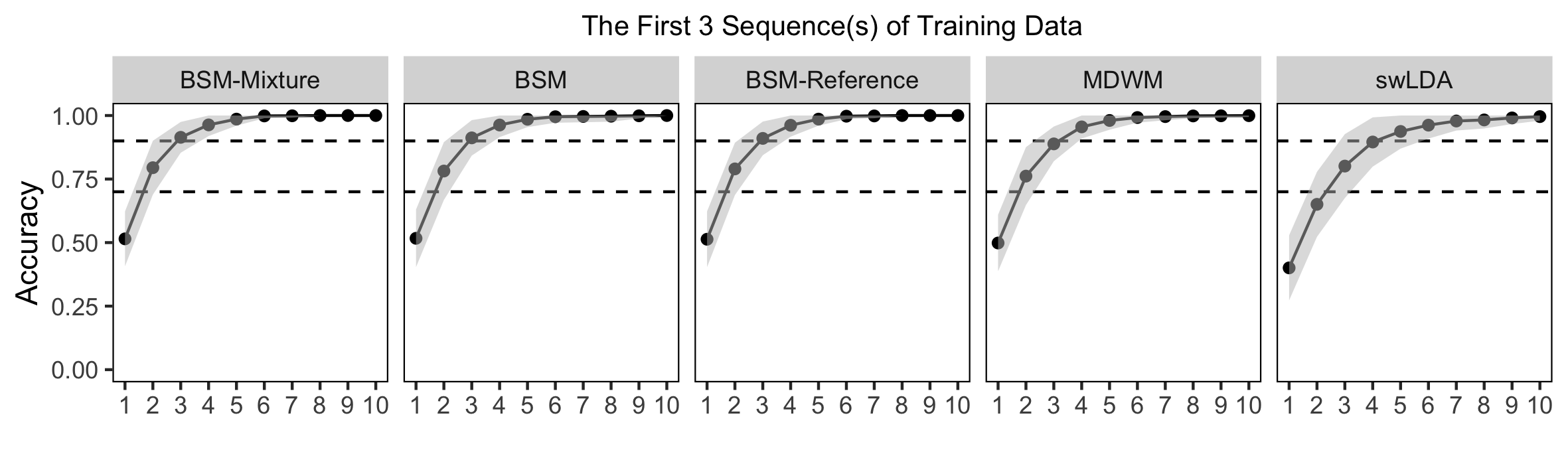}
    \end{subfigure}
     \vfill
     \begin{subfigure}
     \centering
      \includegraphics[width=0.45\textwidth]{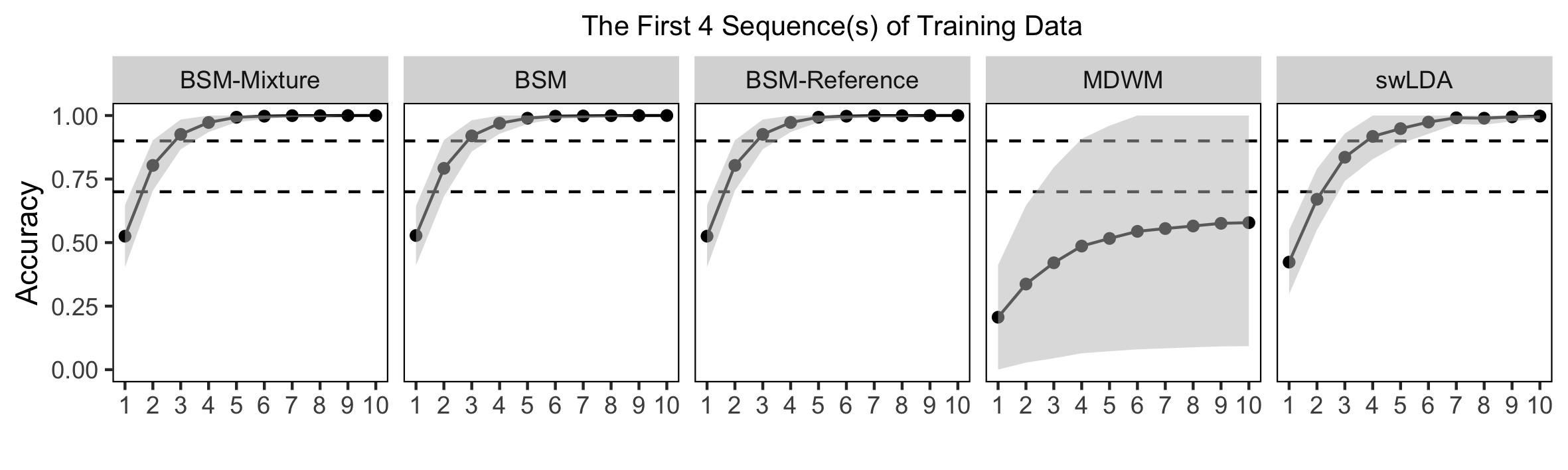}
     \end{subfigure}
     \hfill
     \begin{subfigure}
     \centering
      \includegraphics[width=0.45\textwidth]{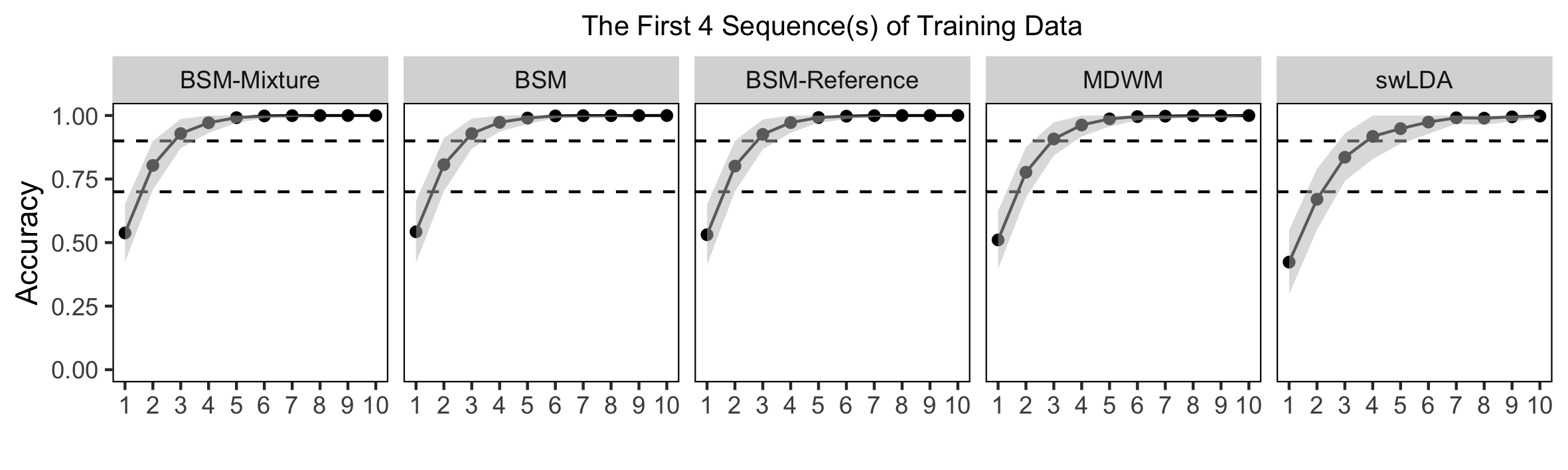}
    \end{subfigure}
    \vfill
    \caption{The \textbf{left} and \textbf{right} panels show the means and standard errors of testing prediction accuracy of Cases S1 and S2 by BSM-Mixture, BSM, BSM-Reference, MDWM, and swLDA, respectively, across 100 replications. The upper, middle, and lower panels show prediction accuracy with the first 2, 3, and 4 training sequence replications, respectively.}
    \label{fig:sup_simulation_single_channel_prediction}
\end{sidewaysfigure}

\noindent \textbf{Criteria} \quad 
We evaluated our method by matching and prediction. For matching, we reported the proportion that each source participant was matched to the new participant and produced the ERP function estimates with 95\% credible bands with respect to the training sequence size. 
For prediction, we reported the character-level prediction accuracy of the testing data, using Bayesian clustering with source participants' data (BSM), Bayesian generative methods with the new participant's data only (BSM-Reference), the hybrid model selection strategy (BSM-Mixture), an existing classification method with data borrowing using Riemannian Geometry (MWDM), and swLDA with new participant's data only (swLDA). Similarly, we did not include SMGP for comparison in the simulation studies because SMGP assumed that data were generated under sequences of stimuli, while our data generative mechanism was based on each stimulus. We did not include other conventional ML methods because we assumed that the prediction accuracy of swLDA was representative of those types of methods.

\noindent \textbf{Matching Results} \quad The upper and lower panels of Table \ref{tab:sup_simulation_single_channel_cluster} show the means and standard deviations of $Z_n=1$ across 100 replications for cases without and with matched data by BSM, respectively. The numerical values were multiplied by 100 for convenient reading and were reported with respect to the training sequence size of the new participant. For Case S1, the values were consistently close to zero throughout the entire training sequence replications; for Case S2, the values associated with Participants 1-2 gradually increased to 80.0 and maintain after 4 sequence replications, while the values associated with Participants 3-6 were consistently close to zero.

\begin{table}[htbp]
\small
\caption{\small \textbf{Upper Lanel}: The means and standard deviations of testing sequence size to achieve a 95\% testing accuracy given certain training sequence sizes for BSM-Mixture, BSM, MDWM, BSM-Reference, and swLDA for Cases S1 and S2 across 100 iterations. \textbf{Lower Panel}: The means and standard deviations of training sequence size to achieve a 95\% testing accuracy given certain testing sequence sizes for the same five methods across 100 iterations. 
}
\centering
\begin{tabular}{c|c|ccc|cc}\hline
\multirow{2}{*}{} & \multirow{2}{*}{\begin{tabular}[c]{@{}c@{}}\textbf{Train}\\  \textbf{Seq Size}\end{tabular}} & \multicolumn{5}{c}{\textbf{Test Sequence Size}} \\\cline{3-7}
                         &   & \textbf{BSM-Mixture}  & \textbf{BSM}          & \textbf{MDWM}         & \textbf{BSM-Reference} & \textbf{swLDA}        \\\hline
\multirow{3}{*}{\textbf{Case S1}} & 2 & 4.9, 1.5 & 5.1, 1.6 & 8.7, 1.8 & 4.9, 1.5  & 7.4, 2.0 \\
                         & 3 & 4.5, 1.1  & 4.6, 1.3 & 8.3, 1.9 & 4.5, 1.1  & 6.3, 1.9 \\
                         & 4 & 4.3, 1.0 & 4.4, 1.2 & 8.3, 2.1 & 4.3, 1.0  & 5.7, 1.8 \\\hline
\multirow{3}{*}{\textbf{Case S2}} & 2 & 4.8, 1.5  & 4.6, 1.5 & 5.2, 1.5 & 5.0, 1.5  & 7.4, 2.0 \\
                         & 3 & 4.4, 1.1 & 4.4, 1.3  & 4.9, 1.4 & 4.5, 1.2   & 6.3, 1.9 \\
                         & 4 & 4.3, 1.1 & 4.3, 1.1 & 4.6, 1.3 & 4.3, 1.1  & 5.7, 1.8\\\hline\hline
\multirow{2}{*}{} & \multirow{2}{*}{\begin{tabular}[c]{@{}c@{}}\textbf{Test} \\ \textbf{Seq Size}\end{tabular}} & \multicolumn{5}{c}{\textbf{Train Sequence Size}} \\\cline{3-7}
 &  & \textbf{BSM-Mixture} & \textbf{BSM} & \textbf{MDWM} & \textbf{BSM-Reference} & \textbf{swLDA} \\\hline
\multirow{3}{*}{\textbf{Case S1}} & 5 & 9.7, 1.3 & 9.5, 1.7 & 10.0, 0.0 & 9.7, 1.3 & 10.0, 0.2 \\
 & 6 & 7.6, 3.3 & 7.7, 3.1 & 10.0, 0.0 & 7.6, 3.3 & 8.8, 2.3 \\
 & 7 & 4.5, 3.6 & 4.7, 3.6 & 8.8, 2.6 & 4.5, 3.6 & 6.3, 3.2 \\\hline
\multirow{3}{*}{\textbf{Case S2}} & 5 & 9.7, 1.5 & 9.6, 1.5 & 9.9, 0.5 & 9.7, 1.3 & 10.0, 0.2 \\
 & 6 & 7.3, 3.5 & 7.2, 3.5 & 8.2, 3.1 & 7.5, 3.4 & 8.8, 2.3 \\
 & 7 & 4.4, 3.6 & 4.3, 3.6 & 5.3, 3.8 & 4.5, 3.6 & 6.3, 3.2 \\\hline\hline
\end{tabular}
\label{tab:sup_simulation_single_test_95_seq_size_forward_backward}
\end{table}

\begin{table}[htbp]
\centering
\caption{\small \textbf{Upper Panel}: The means and standard deviations of testing sequence sizes to achieve a 95\% testing accuracy given certain training sequence sizes for BSM-Mixture, BSM, MDWM, BSM-Reference, and swLDA for Cases 1 and 2 across 100 replications. 
\textbf{Lower Panel}: The mean and standard deviations of training sequence size to achieve a 95\% testing accuracy given certain testing sequence sizes for the same five methods across 100 iterations.
}
\resizebox{0.95\textwidth}{!}{%
\begin{tabular}{c|c|ccc|cc} \hline
\textbf{Upper} & \textbf{Train}    & \multicolumn{5}{c}{\textbf{Test Sequence Size}} \\ \cline{3-7}
 & \textbf{Seq Size} & \textbf{BSM-Mixture} & \textbf{BSM} & \textbf{MDWM} & \textbf{BSM-Reference} & \textbf{swLDA} \\\hline
\multirow{3}{*}{\textbf{Case 1}}
 & 2 & 8.4, 1.7 & 9.8, 0.7 & 7.9, 1.6 & 8.2, 1.7 & 10.0, 0.3 \\
 & 3 & 7.6, 1.8 & 9.6, 1.1 & 7.8, 1.6 & 7.6, 1.8 & 9.8, 0.6 \\
 & 4 & 7.2, 1.7 & 9.3, 1.1 & 7.8, 1.4 & 7.1, 1.7 & 9.6, 0.9 \\\hline
\multirow{3}{*}{\textbf{Case 2}}
 & 2 & 7.3, 1.6 & 7.3, 1.7 & 7.9, 1.6 & 8.2, 1.7 & 10.0, 0.3 \\
 & 3 & 7.2, 1.8 & 7.2, 1.8 & 7.9, 1.6 & 7.6, 1.8 & 9.8, 0.6 \\
 & 4 & 6.8, 1.7 & 6.8, 1.7 & 7.8, 1.5 & 7.1, 1.7 & 9.6, 0.9 \\\hline
 \hline
\textbf{Lower} & \textbf{Test}    & \multicolumn{5}{c}{\textbf{Train Sequence Size}} \\ \cline{3-7}
 & \textbf{Seq Size} & \textbf{BSM-Mixture} & \textbf{BSM} & \textbf{MDWM} & \textbf{BSM-Reference} & \textbf{swLDA} \\\hline
\multirow{3}{*}{\textbf{Case 1}} 
 & 5 & 8.0, 3.0 & 9.2, 1.7 & 9.0, 2.5 & 8.0, 3.1 & 9.7, 1.2 \\
 & 6 & 6.3, 3.4 & 8.3, 2.2 & 7.5, 3.5 & 6.3, 3.5 & 9.2, 1.6 \\
 & 7 & 4.9, 3.4 & 7.1, 2.9 & 5.2, 3.9 & 4.9, 3.5 & 8.7, 2.1 \\\hline
\multirow{3}{*}{\textbf{Case 2}}
 & 5 & 7.5, 3.5 & 7.4, 3.5 & 8.9, 2.6 & 8.0, 3.1 & 9.7, 1.2 \\
 & 6 & 5.7, 4.1 & 5.8, 4.1 & 7.5, 3.4 & 6.3, 3.5 & 9.2, 1.6 \\
 & 7 & 3.8, 3.7 & 3.9, 3.7 & 5.3, 4.0 & 4.9, 3.5 & 8.7, 2.1 \\\hline
\end{tabular}}
\label{tab:sup_simulation_multi_test_95_seq_size_forward_backward}
\end{table}

\noindent \textbf{Prediction Results} \quad The left panel of Figure \ref{fig:sup_simulation_single_channel_prediction} shows the mean and standard error of testing prediction accuracy of Case S1 by BSM-Mixture, BSM, BSM-Reference, MDWM, and swLDA. The testing prediction accuracy was further stratified by the first 2, 3, and 4 training sequence replications of the new participant. BSM-Mixture performed the best among all five methods, and MDWM surprisingly performed poorly in this scenario. Meanwhile, the right panel of
Figure \ref{fig:sup_simulation_single_channel_prediction} shows the means and standard errors of testing prediction accuracy of Case S2 by the same five methods. The testing prediction accuracy was further stratified by the first 2, 3, and 4 training sequence replications of the new participant. All five methods performed well, and BSM-Mixture achieved 100\% accuracy earliest and showed the narrowest error bars than the remaining four methods. The upper panel of Table \ref{tab:sup_simulation_single_test_95_seq_size_forward_backward} shows the means and standard deviations of testing sequences to achieve a 95\% testing acuracy given certain training sequences for BSM-Mixture, BSM, MDWM, BSM-Reference, and swLDA for Cases S1 and S2 across 100 iterations. Meanwhile, the lower panel of Table \ref{tab:sup_simulation_single_test_95_seq_size_forward_backward} shows the means and standard deviations of training sequence size to achieve a 95\% training accuracy given certain testing sequneces for the same five methods across 100 iterations. Similar results of the multi-channel scenario are shown in Table \ref{tab:sup_simulation_multi_test_95_seq_size_forward_backward}. 

\subsection{Real Data Based Simulation Scenario (Normal Noise)}
\label{supsubsec:multi_channel_sim_normal}
A real-data based multi-channel simulation scenario with normal noise is performed. The number of participants, sample sizes of training set, and true parameters are copied from the real data analysis. We follow the rule that participant 0 refers to the new participant, while participants 1 to 23 refer to source participants.
The participant-specific ERP functions and variance parameters are obtained from posterior means of each participant in the real data analysis. We multiply the variance estimates by 2 in this simulation scenario. The simulated EEG signals are also generated from multivariate normal distributions with a response window of 25 time points per channel, i.e., $T_0=25$. We perform 100 replications for this scenario. Within each replication, we assume that each participant is spelling the character ``T'' eight times with ten sequence replications per character for training and generate additional testing data of the same size as the testing data of the naive multi-channel scenario.

\begin{figure}[h]
    \centering
\includegraphics[width=\textwidth]{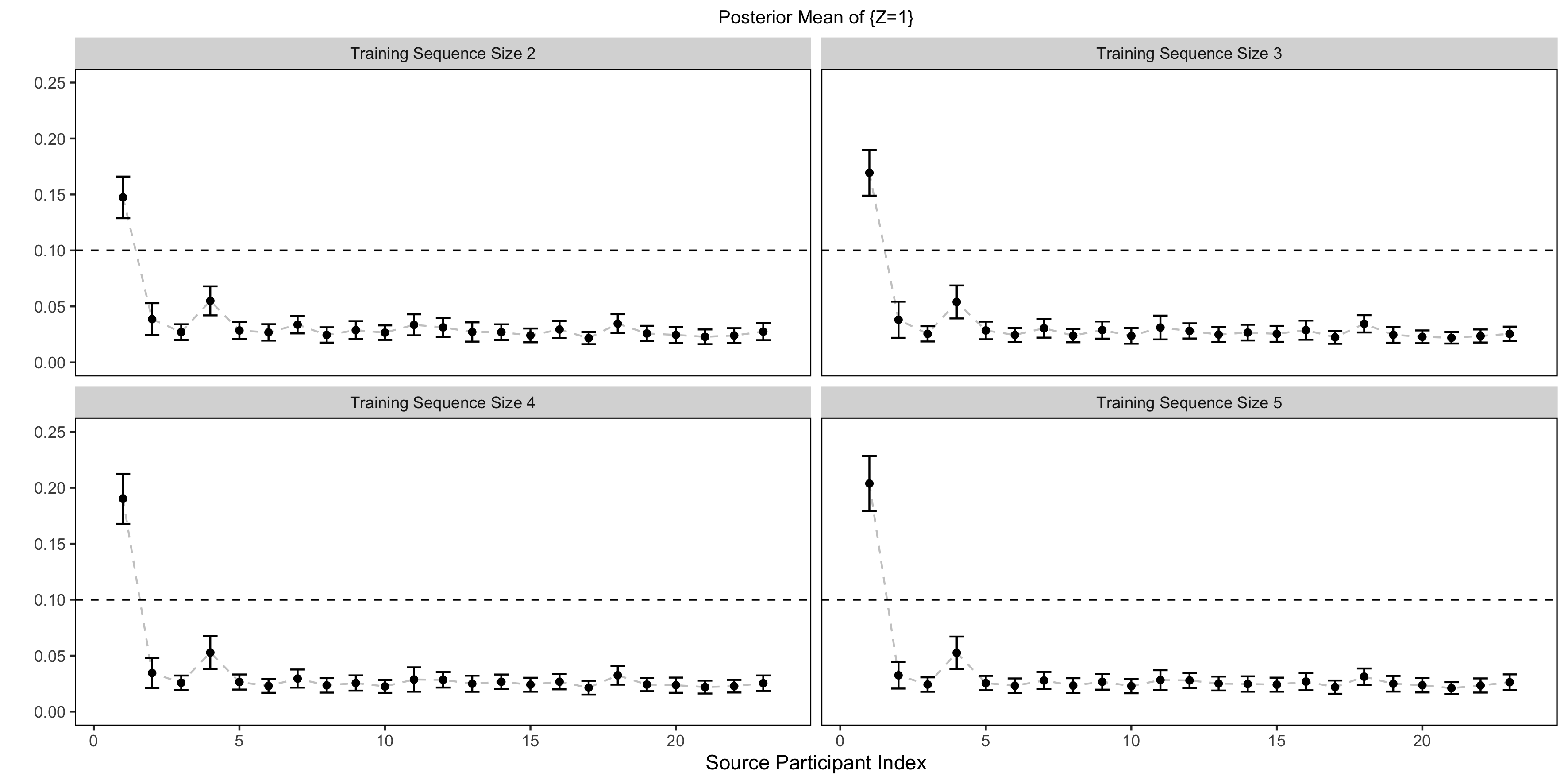}
\caption{A plot of posterior probability of $\{Z_n=1\}$ for 23 source participants with training sequence sizes ranging from 2 to 5. The means and one-standard error bars across 100 replications are shown.}
\label{fig:sup_sim_real_data_clustering}
\end{figure}

All simulated datasets are fitted with equations in Section \ref{supsec:mcmc_multi}. We used the same kernels and hyper-parameters as the real data analysis and set the pre-specified threshold $\delta_Z$ as 0.1. We run the MCMC with three chains, with each chain containing 5,000 burn-ins and 3,000 MCMC samples. Both matching and prediction results are reported, and we follow the same criteria as mentioned in Criteria part in Section \ref{supsubsec:single_channel_sim}.

\noindent \textbf{Matching Results} Figure \ref{fig:sup_sim_real_data_clustering} shows the posterior probability of $\Pr\{Z_n=1\},n=1,\cdots, 23$ with respect to different training sequence replications. The means and one-standard error bars across 100 replications are shown. A dashed horizontal line at the pre-specified $\delta_Z=0.1$ is plotted. We observe that source participant 1 gets selected with all four training sequence sizes, and the mean posterior probabilities increase from 0.15 to 0.20 when the training sequence size increases from 2 to 5. The remaining 22 source participants receive low posterior probability, which is similar to the results in the real data analysis section.

\begin{table}[ht]
\centering
\caption{\textbf{Upper Panel}: The means and standard deviations of testing sequences required to achieve a 85\% testing accuracy given training sequence sizes ranging from 2 to 5 for BSM-Mixture, BSM, MDWM, BSM-Reference, and swLDA across 100 replications.
\textbf{Lower Panel}: The means and standard deviations of training sequences required to achieve a 85\% testing accuracy given testing sequence sizes ranging from 7 to 10 for BSM-Mixture, BSM, MDWM, BSM-Reference, and swLDA across 100 replications.
}
\resizebox{0.95\textwidth}{!}{
\begin{tabular}{c|ccccc}
\hline
\multirow{2}{*}{\begin{tabular}[c]{@{}c@{}}\textbf{Train} \\ \textbf{Seq Size}\end{tabular}} & \multicolumn{5}{c}{\textbf{Test Sequence Size}} \\ \cline{2-6} 
  & \textbf{BSM-Mixture}   & \textbf{BSM}           & \multicolumn{1}{c|}{\textbf{MDWM}}          & \textbf{BSM-Reference} & \textbf{swLDA}         \\ \hline
2 & 9.49, (0.948) & 9.49, (0.948) & \multicolumn{1}{c|}{9.96, (0.281)} & 9.58, (0.912) & 9.85, (0.52)  \\
3 & 9.22, (1.08)  & 9.22, (1.08)  & \multicolumn{1}{c|}{9.93, (0.355)} & 9.46, (0.904) & 9.84, (0.581) \\
4 & 8.89, (1.16)  & 8.89, (1.16)  & \multicolumn{1}{c|}{9.91, (0.429)} & 9.1, (1.15)   & 9.69, (0.706) \\
5 & 8.59, (1.34)  & 8.59, (1.34)  & \multicolumn{1}{c|}{9.89, (0.447)} & 8.81, (1.29)  & 9.61, (0.84)  \\ \hline
\hline
\multirow{2}{*}{\textbf{\begin{tabular}[c]{@{}c@{}}Test\\ Seq Size\end{tabular}}} &
  \multicolumn{5}{c}{\textbf{Train Sequence Size}} \\ \cline{2-6} 
 &
  \textbf{BSM-Mixture} &
  \textbf{BSM} &
  \multicolumn{1}{c|}{\textbf{MDWM}} &
  \textbf{BSM-Reference} &
  \textbf{swLDA} \\ \hline
7  & 9.34, (2.14) & 9.34, (2.14) & \multicolumn{1}{c|}{10, (0)}      & 9.51, (1.83) & 9.85, (1.06) \\
8  & 8.47, (3.03) & 8.47, (3.03) & \multicolumn{1}{c|}{9.93, (0.7)}  & 8.42, (3.12) & 9.64, (1.61) \\
9  & 6.42, (3.9)  & 6.42, (3.9)  & \multicolumn{1}{c|}{9.59, (1.8)}  & 7.06, (3.68) & 8.73, (2.76) \\
10 & 5.04, (3.97) & 5.04, (3.97) & \multicolumn{1}{c|}{9.41, (2.04)} & 5.18, (3.81) & 7.45, (3.64) \\ \hline
\end{tabular}}
\label{tab:supsec_sim_real_data_based}
\end{table}

\noindent \textbf{Prediction Results} The upper panel of Table \ref{tab:supsec_sim_real_data_based} shows the numbers of testing sequences required to achieve a 85\% testing prediction accuracy given training sequence size ranging from 2 to 5, while the lower panel of Table \ref{tab:supsec_sim_real_data_based} shows the numbers of training sequences required to achieve a 85\% testing prediction accuracy given testing sequences ranging from 7 to 10. The means and standard deviations across 100 replications are reported. Overall, we observe that BSM-Mixture performs the best, although the number of sequence saved is moderate compared to BSM-Reference, possibly because we only find one partial match from source participants' pool.

\subsection{Real Data Based Simulation Scenario (Student-t Noise)}
\label{supsubsec:multi_channel_sim_t}

To examine the robustness of our method with respect to non-normal noises, a real-data based multi-channel simulation scenario with a student-t noise is performed. We choose the student-t distribution with a moderate tail to generate a subset of EEG signals with ``extreme'' values. Since the extremeness is not expected to be severe with task-based EEG, we choose a student-t distribution with a degree of freedom 5 and we use the exact variances estimated from the real data. Other than those two differences, the remaining components are exactly the same as Section \ref{supsubsec:multi_channel_sim_normal}.

\begin{figure}[ht]
    \centering
\includegraphics[width=\textwidth]{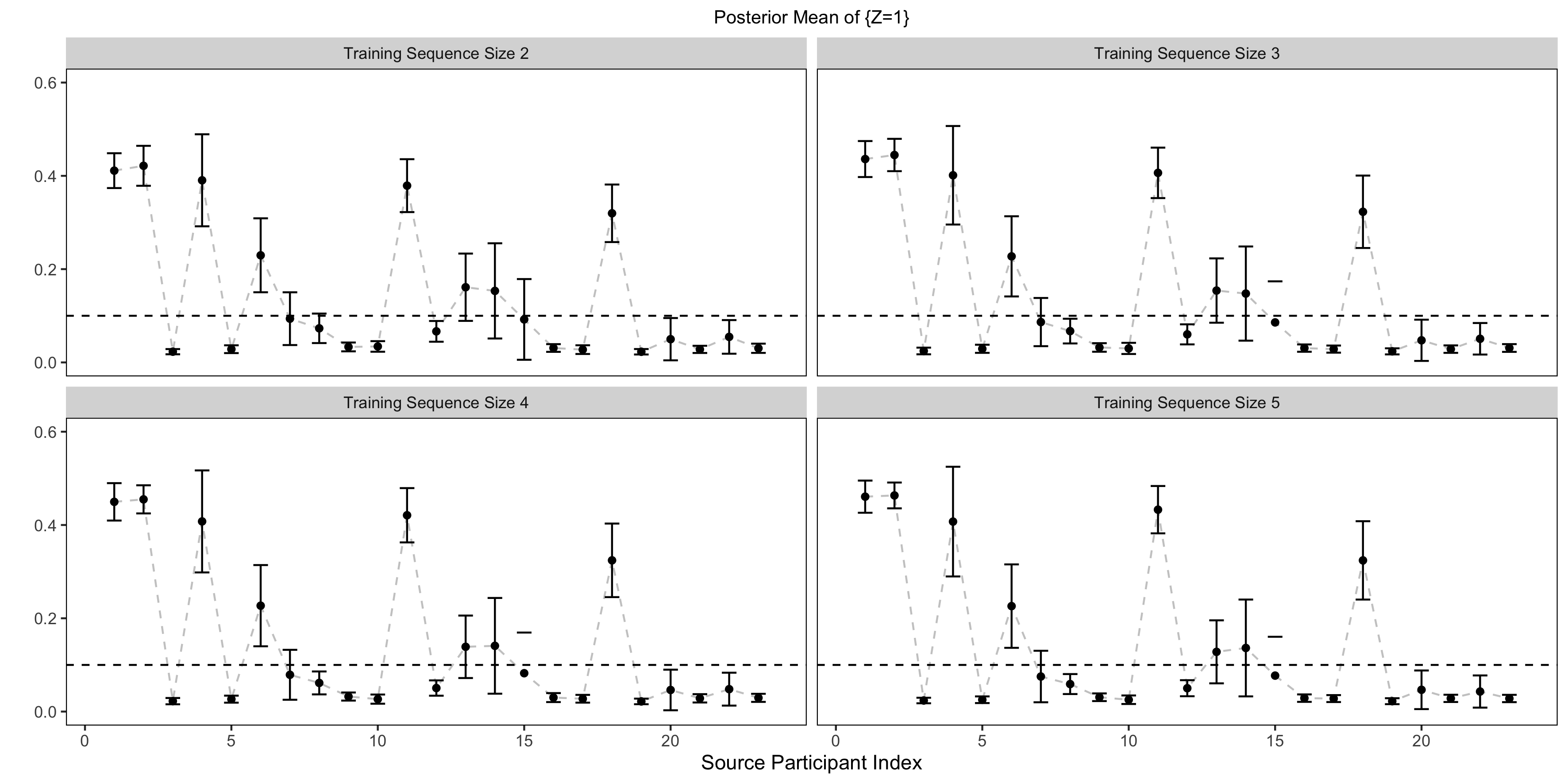}
\caption{A plot of posterior probability of $\{Z_n=1\}$ for 23 source participants with training sequence sizes ranging from 2 to 5. The noise is generated from a multivariate student-t distribution with pre-specified autoregressive temporal and compound symmetry spatial covariance matrices and a degree of freedom 5. The means and one-standard error bars across 100 replications are shown.}
\label{fig:sup_sim_real_data_clustering_t}
\end{figure}

All simulated datasets are fitted with equations in Section \ref{supsec:mcmc_multi}. The hyper-parameters for MCMC are the same as \ref{supsubsec:multi_channel_sim_normal}. Both matching and prediction results are shown with the same criteria as mentioned before.

\noindent \textbf{Matching Results} Figure \ref{fig:sup_sim_real_data_clustering_t} shows the posterior probability of $\Pr\{Z_n=1\},n=1,\cdots, 23$ with respect to various training sequence replications. The means and one-standard error bars across 100 replications are shown. A dashed horizontal line at the pre-specified $\delta_Z=0.1$ is plotted. We observe that source participants 1, 2, 4, 6, 11, and 18 consistently are selected with all four training sequence sizes.

\begin{figure}[ht]
     \centering
     \begin{subfigure}
     \centering
      \includegraphics[width=0.95\textwidth]{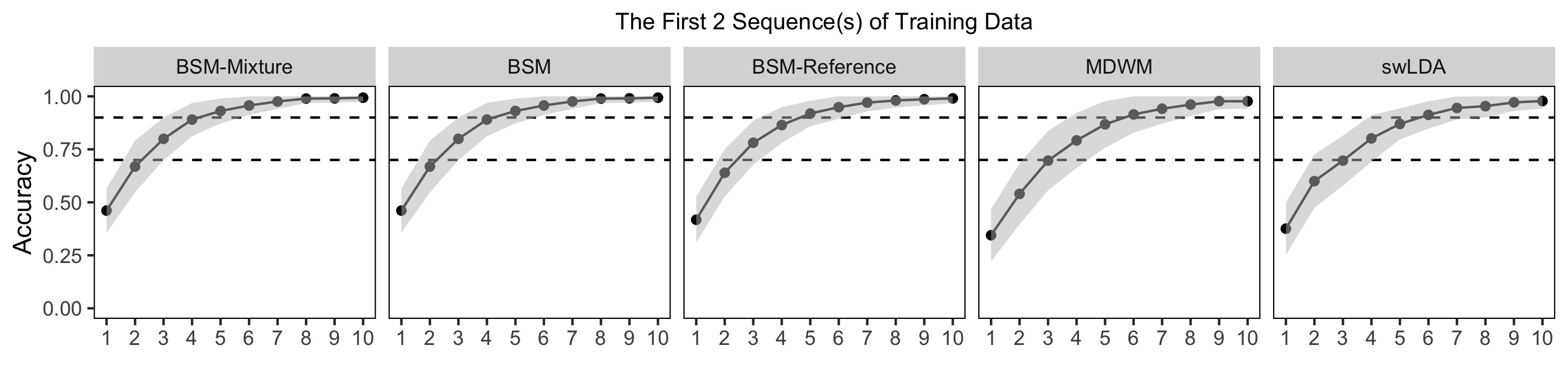}
     \end{subfigure}
     \vfill
     \begin{subfigure}
     \centering
      \includegraphics[width=0.95\textwidth]{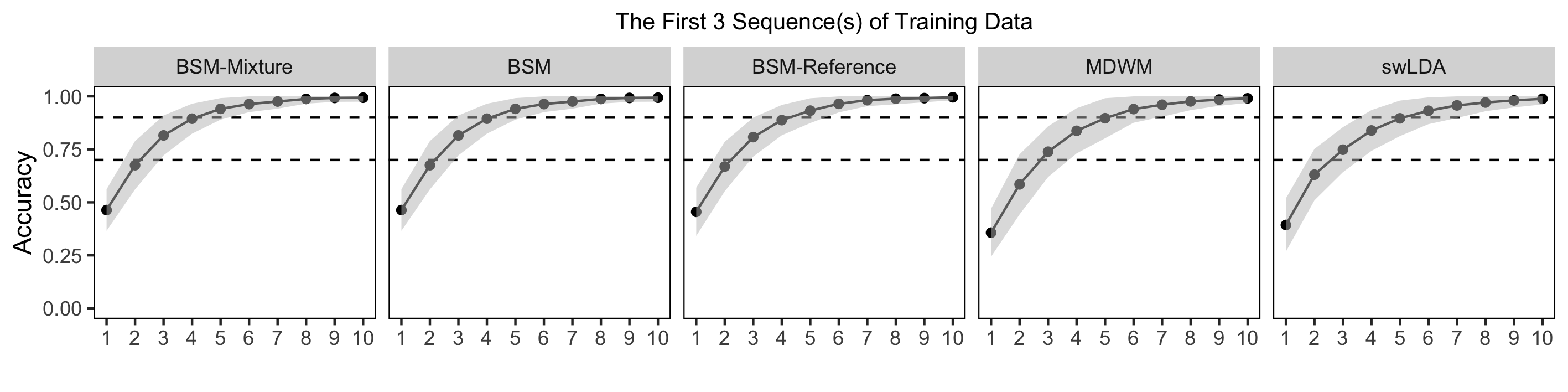}
     \end{subfigure}
     \vfill
     \begin{subfigure}
     \centering
      \includegraphics[width=0.95\textwidth]{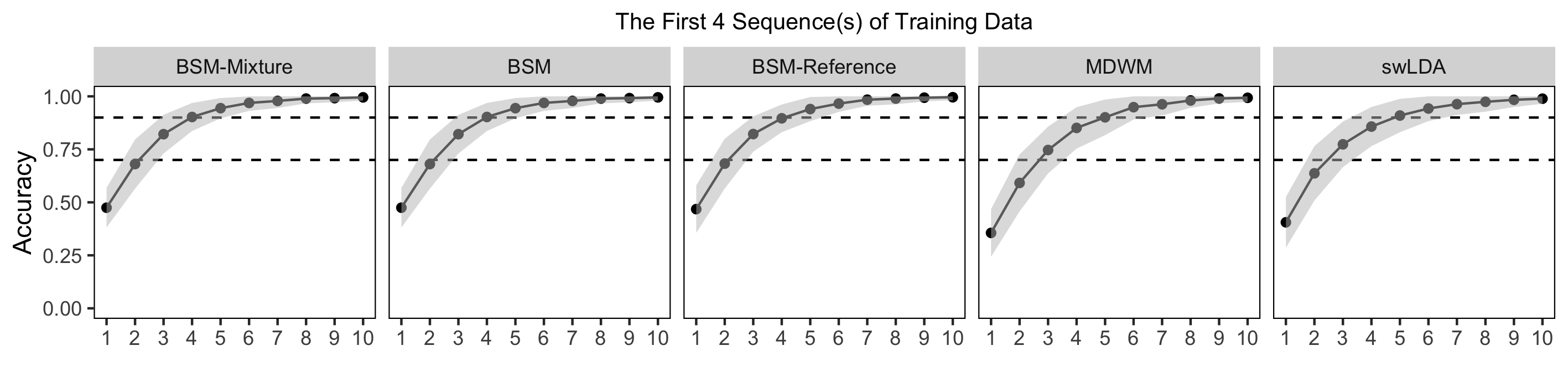}
     \end{subfigure}
     \vfill
    \caption{A graphic of the testing prediction accuracy. The upper, middle, and lower panel corresponds to the prediction accuracy with the first 2, 3, and 4 sequences of training data, respectively. All results are evaluated across 100 replications. The noise is generated from a multivariate student-t distribution with a pre-specified covariance matrix and a degree of freedom 5.}
    \label{fig:sup_simulation_multi_channel_real_data_t_prediction}
\end{figure}

\begin{table}[ht]
\centering
\caption{\textbf{Upper Panel}: The means and standard deviations of testing sequences required to achieve a 90\% testing accuracy given training sequence sizes ranging from 2 to 5 for BSM-Mixture, BSM, MDWM, BSM-Reference, and swLDA across 100 replications.
\textbf{Lower Panel}: The means and standard deviations of training sequences required to achieve a 90\% testing accuracy given testing sequence sizes ranging from 5 to 10 for BSM-Mixture, BSM, MDWM, BSM-Reference, and swLDA across 100 replications.
The noise is generated from a multivariate student-t distribution with a pre-specified covariance matrix and a degrees of freedom 5.
}
\resizebox{0.95\textwidth}{!}{
\begin{tabular}{c|ccccc}
\hline
\textbf{Train} & \multicolumn{5}{c}{\textbf{Test Sequence Size}}                                               \\ \cline{2-6}
\textbf{Seq Size} & \textbf{BSM-Mix} & \textbf{BSM-Cluster} & \multicolumn{1}{c|}{\textbf{MDWM}} & \textbf{BSM-Ref} & \textbf{swLDA} \\ \hline
2              & 4.84, (1.34) & 4.84, (1.34) & \multicolumn{1}{c|}{6.23, (1.87)} & 5.34, (1.33) & 6.4, (1.68)  \\
3              & 4.71, (1.23) & 4.71, (1.23) & \multicolumn{1}{c|}{5.79, (1.75)} & 4.93, (1.28) & 5.81, (1.64) \\
4              & 4.63, (1.23) & 4.63, (1.23) & \multicolumn{1}{c|}{5.6, (1.59)}  & 4.84, (1.2)  & 5.54, (1.68) \\
5              & 4.56, (1.17) & 4.56, (1.17) & \multicolumn{1}{c|}{5.28, (1.33)} & 4.7, (1.25)  & 5.27, (1.52) \\\hline
\hline
\textbf{Test} & \multicolumn{5}{c}{\textbf{Train Sequence Size}}                                                \\ \cline{2-6} 
\textbf{Seq Size} & \textbf{BSM-Mix} & \textbf{BSM-Cluster} & \multicolumn{1}{c|}{\textbf{MDWM}} & \textbf{BSM-Ref} & \textbf{swLDA} \\ \hline
5             & 5.04, (4.3)  & 5.04, (4.3)  & \multicolumn{1}{c|}{6.81, (3.99)} & 5.69, (4.16)  & 6.46, (3.86)  \\
6             & 3.08, (3.54) & 3.08, (3.54) & \multicolumn{1}{c|}{4.92, (4.14)} & 3.43, (3.61)  & 4.45, (3.75)  \\
7             & 2.02, (2.7)  & 2.02, (2.7)  & \multicolumn{1}{c|}{2.91, (3.23)} & 1.98, (2.45)  & 2.58, (2.75)  \\
8             & 1.47, (1.82) & 1.47, (1.82) & \multicolumn{1}{c|}{1.75, (1.73)} & 1.29, (1.31)  & 1.68, (1.66)  \\
9             & 1, (0)       & 1, (0)       & \multicolumn{1}{c|}{1.66, (1.85)} & 1.21, (1.27)  & 1.31, (0.677) \\
10            & 1, (0)       & 1, (0)       & \multicolumn{1}{c|}{1.22, (1)}    & 1.03, (0.171) & 1.11, (0.424) \\ \hline
\end{tabular}}
\label{tab:supsec_sim_real_data_t}
\end{table}

\noindent \textbf{Prediction Results} Each row of Figure \ref{fig:sup_simulation_multi_channel_real_data_t_prediction}  shows the testing prediction accuracy (mean and one-standard error bars) by the same five methods across 100 replications. The upper, middle, and lower panels are the prediction accuracy associated with 2, 3, and 4 training sequences of the pseudo new participant, respectively. Overall, we observe that BSM-Mixture outperforms BSM-Reference, MDWM and swLDA. (In this case, BSM-Mixture is essentially BSM because at least one source participant is identified.) Given first 2 training sequences, BSM-Mixture can already achieve the 0.90 threshold with 4 testing sequences, while BSM-Reference, MDWM, and swLDA require 5, 6, and 6 testing sequences, respectively. The lag is reduced as the number of training sequences increases. Although the mean number of testing sequences to achieve the 0.90 threshold for BSM-Mixture is 4 regardless of training sequences, the error bars are gradually narrower, indicating the potential gain in precision of data borrowing with more training sequences. Details can be found in Table \ref{tab:supsec_sim_real_data_t}.

\section{Additional EEG-BCI Results}
\label{supsec:real_data_analysis}

\subsection{Experimental Design and Model Fitting}
\label{subsec:real_data_overall}
We showed the results of real data analysis for the remaining participants. A total of forty-one participants attended the experiment. Each participant was asked to complete one calibration session (TRN) and up to three free-typing sessions (FRT). During the calibration session, each participant was asked to copy a 19-character phrase ``THE\_QUICK\_BROWN\_FOX'' including three spaces while wearing an EEG cap with 16 channels and sitting close to a monitor with a virtual keyboard (See Figure 1). For all TRN files, it contained 15 sequences per character, while for FRT files, the numbers of testing sequences depended on the calibration accuracy of each participant. Details were available in \citeauthor{thompson2014plug} in 2014. 

As mentioned in the main text, we applied the same bandpass filter and down-sampled technique and extracted a fixed response window about 800 ms after each stimulus to obtain the EEG signal matrix. Since the stimulus-to-stimulus interval is around 150 ms, overlapping components in the temporal domain existed in the EEG signal matrix. Finally, we applied the participant-specific xDAWN spatial filter to the EEG signal matrix with all 16 channels \citep{rivet2009xdawn} and selected the first two components as the transformed EEG signals. For each participant, we fitted their data with BSM-Mixture model in equation 6, used the same covariance kernels as for K151's, and ran the MCMC with three chains, with each chain containing 8,000 burn-ins and 1,000 MCMC samples.

\begin{figure}[ht]
\centering
\includegraphics[width=0.95\textwidth]{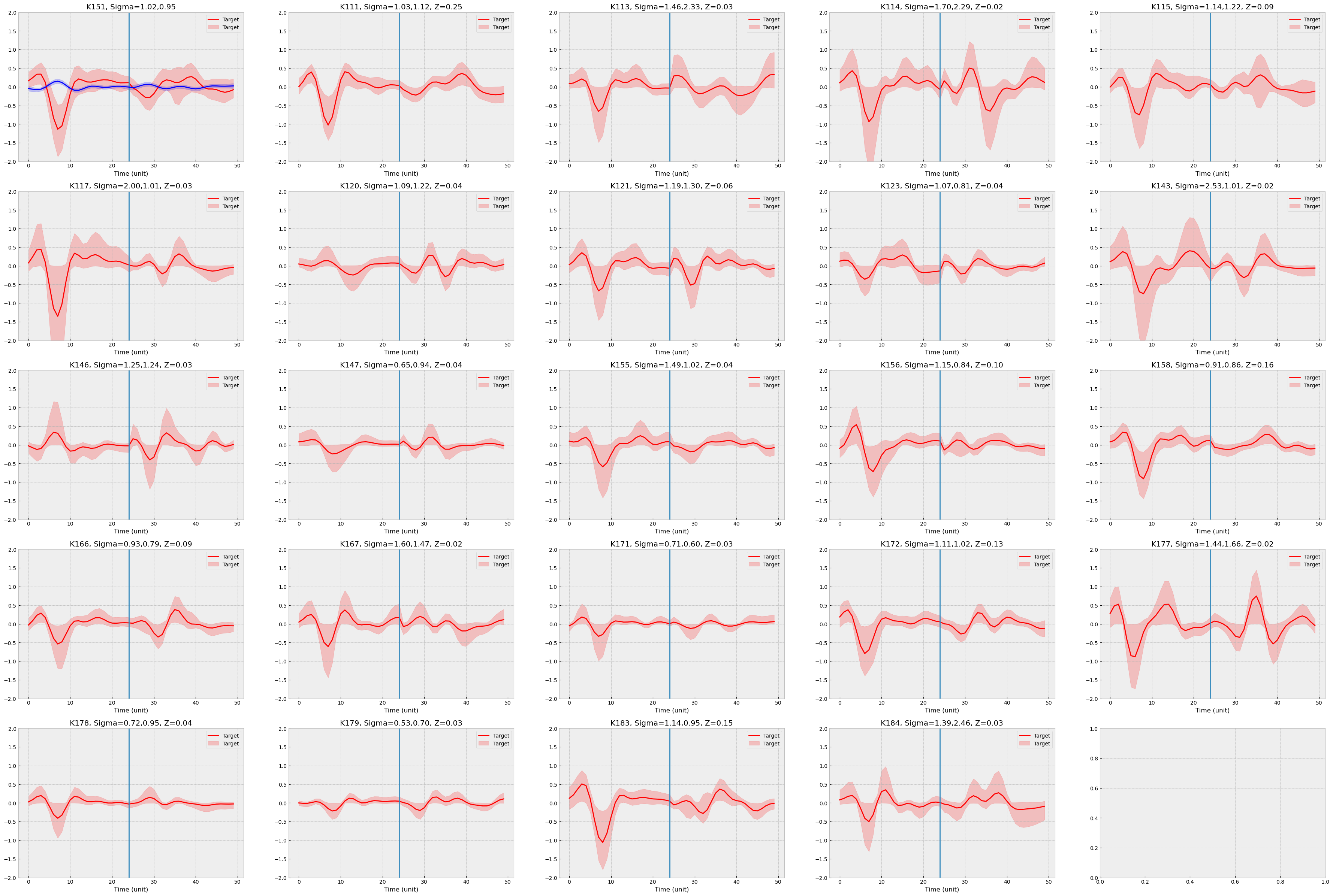}
    \caption{A plot that summarizes parameter estimates of transformed ERP functions, variance nuisance parameters, and posterior probability of $Z_n=1$. Here, K151 is the new participant, while the remaining 23 ones are source participants. K111, K158, K183, and K172 are source participants with $\Pr\{Z_n=1\}>0.1$.}
    \label{fig:sup_K151_matching}
\end{figure}

\subsection{Matching Results of K151}
\label{subsec:K151_matching}

Figure \ref{fig:sup_K151_matching} shows the parameter estimates of transformed ERP functions, variance nuisance parameters, and posterior probability of $Z_n=1$. The top left plot refers to the new participant, K151, while the remaining plots refer to the 23 source participants. For participant-specific plot, posterior means and 95\% credible bands of transformed ERP functions are shown. The first 25 and right 25 points correspond to two components, respectively. Since we only borrow target transformed ERP data, we only plot parameter estimates of non-target transformed ERP functions for K151. The posterior means of component-specific variance parameters are reported. The posterior means of $Z_n=1$ for 23 source participants are also shown. Among 23 source participants, we observe four participants have $\Pr\{Z_n=1\}>\delta_Z$, i.e., K111 with 24\%, K158 with 16\%, K183 with 15\%, and K172 with 13\%.

\subsection{Sensitivity Analysis of K151}
\label{subsec:K151_sensitivity}

\begin{figure}[ht]
    \centering
    \includegraphics[width=0.45\textwidth]{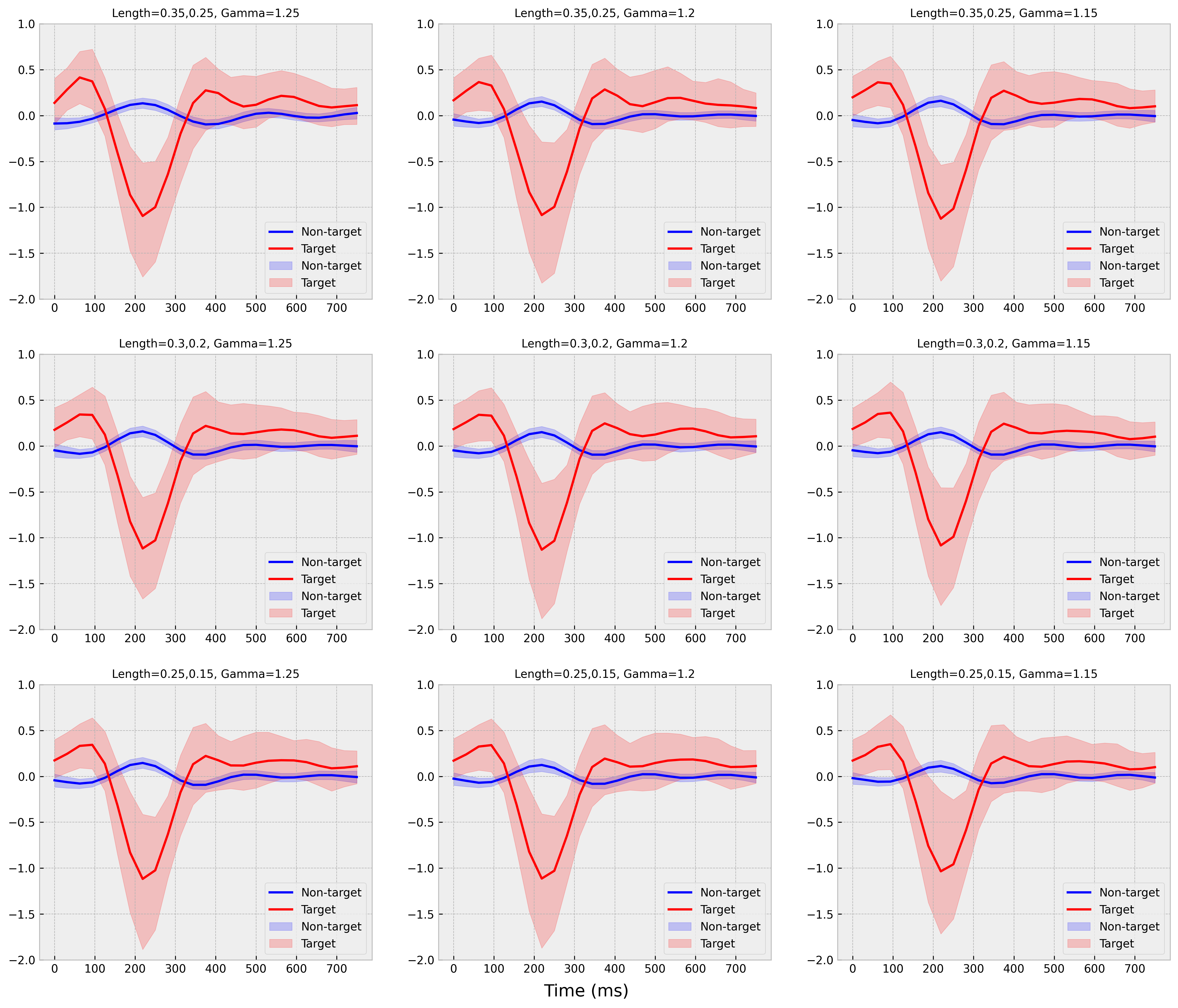}
    \hfill
    \includegraphics[width=0.45\textwidth]{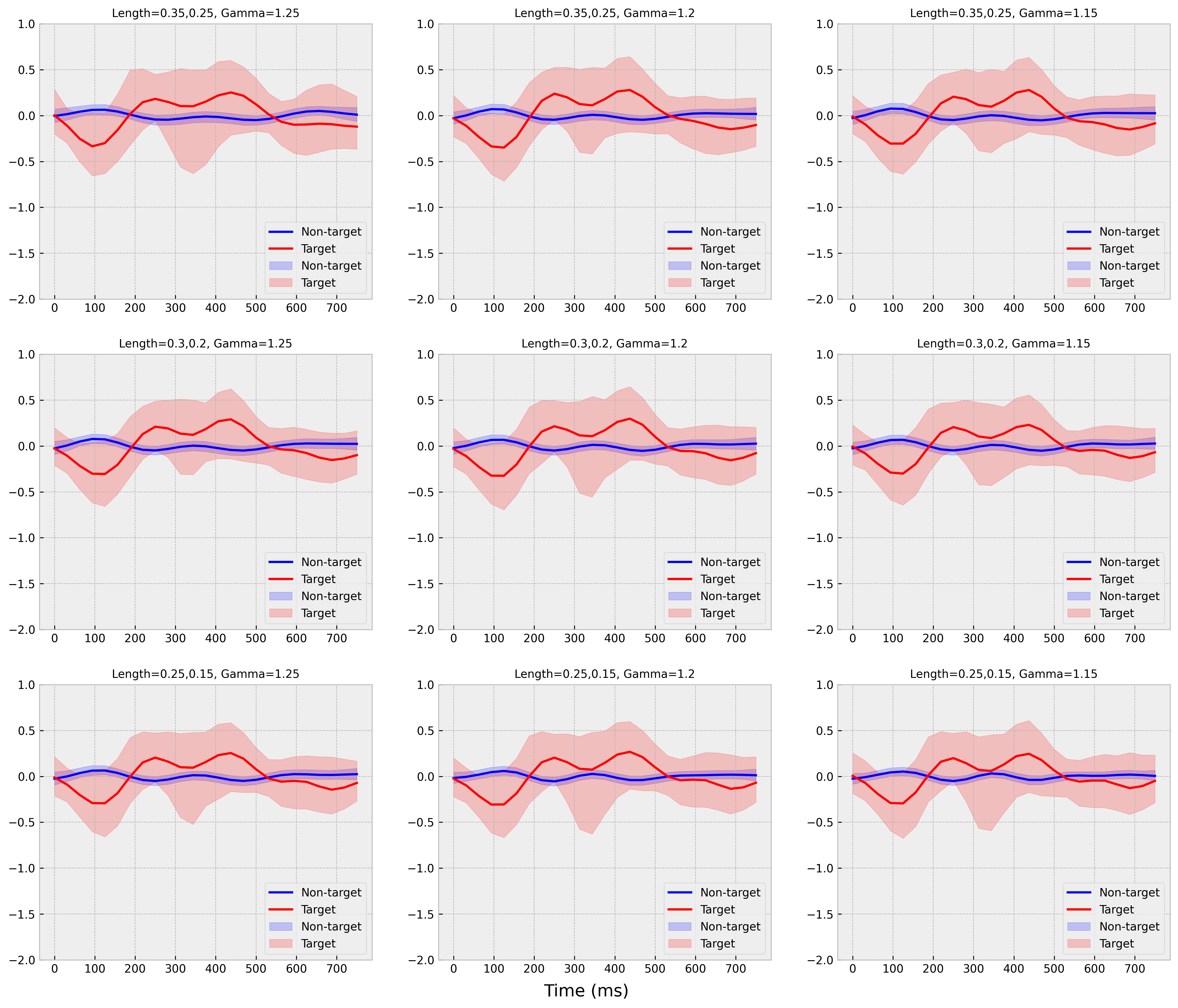}
    \caption{\small The mean estimates with 95\% credible bands of the first (Left) and second (Right) transformed ERP functions by BSM-Mixture stratified by nine combinations of kernel hyper-parameters length $s_0$ and gamma $\gamma_0$. We restricted $s_0$ of target ERP functions to be smaller than the non-target ones by $0.1$.}
    \label{fig:sup_K151_sensitivity}
\end{figure}

To evaluate the robustness of our method, we performed the sensitivity analysis on K151 by varying the hyper-parameters of $\gamma$-exponential kernels that fit the target and non-target transformed ERP function estimates. We evaluated the length parameter $s_0$ of target transformed ERP function estimates with values $0.35, 0.5,$ and $0.25$, the length parameter $s_0$ of non-target transformed ERP function estimates with values $0.25, 0.2,$ and $0.15$, and gamma parameter $\gamma_0$ with values $1.25, 1.2,$ and $1.15$. To simplify, we assigned every participant (source or new) with the same combination of kernel hyper-parameters. We produced the plots of transformed ERP function estimates and reported the testing prediction accuracy under nine combinations of kernel hyper-parameters. The left and right panels of Figure \ref{fig:sup_K151_sensitivity} show the mean estimates with 95\% credible bands of the first two transformed ERP functions by BSM-Mixture under nine combinations of kernel hyper-parameters length $s_0$ and gamma $\gamma_0$. Among nine plots, we found few differences among target transformed ERP function estimates. The plot of the mean estimates of the second transformed ERP functions is available in the supplementary material. Table \ref{tab:sup_K151_sensitivity} shows the testing prediction accuracy percentage of BSM-Mixture with up to five testing sequences under the same nine combinations of kernel hyper-parameters. We found few differences across different kernel hyper-parameters, and the prediction accuracy was consistently higher than $90.0\%$ with five sequences. 

\begin{table}[htbp]
\centering
\caption{Testing prediction accuracy (percentage) of BSM-Mixture with testing sequences up to five stratified by nine combinations of kernel hyper-parameters length $s_0$ and gamma $\gamma_0$. We restricted $s_0$ of target ERP functions to be smaller than the non-target ones by $0.1$. The prediction accuracy was consistently over 90.0\% with five testing sequence replications.}
\begin{tabular}{c|ccc|ccc|ccc}\hline
\backslashbox{$\gamma_0$}{$s_0$} &
  \multicolumn{3}{c|}{$0.35, 0.25$} &
  \multicolumn{3}{c|}{$0.3, 0.2$} &
  \multicolumn{3}{c}{$0.25, 0.15$} \\\hline 
  \textbf{\begin{tabular}[c]{@{}c@{}}Testing \\Seq Size\end{tabular}}& 
 1.25 &
  $1.2$ &
  $1.15$ &
  $1.25$ &
  $1.2$ &
  $1.15$ &
  $1.25$ &
  $1.2$ &
  $1.15$ \\\hline
1 & 56.8 & 56.8 & 55.4 & 55.4 & 54.1 & 52.7 & 54.1 & 55.4 & 52.7 \\
2 & 78.4 & 81.1 & 81.1 & 81.1 & 82.4 & 82.4 & 82.4 & 82.4 & 83.8 \\
3 & 85.1 & 83.8 & 83.8 & 83.8 & 83.8 & 83.8 & 83.8 & 83.8 & 83.8 \\
4 & 90.5 & 89.2 & 89.2 & 89.2 & 89.2 & 89.2 & 89.2 & 89.2 & 89.2 \\
5 & 91.9 & 90.5 & 90.5 & 90.5 & 90.5 & 90.5 & 90.5 & 90.5 & 90.5 \\
6 & 94.6 & 94.6 & 93.2 & 93.2 & 94.6 & 94.6 & 93.2 & 94.6 & 93.2 \\
\hline
\end{tabular}
\label{tab:sup_K151_sensitivity}
\end{table}

\begin{figure}[ht]
     \centering
      \begin{subfigure}
          \centering
          \includegraphics[width=0.48\textwidth]{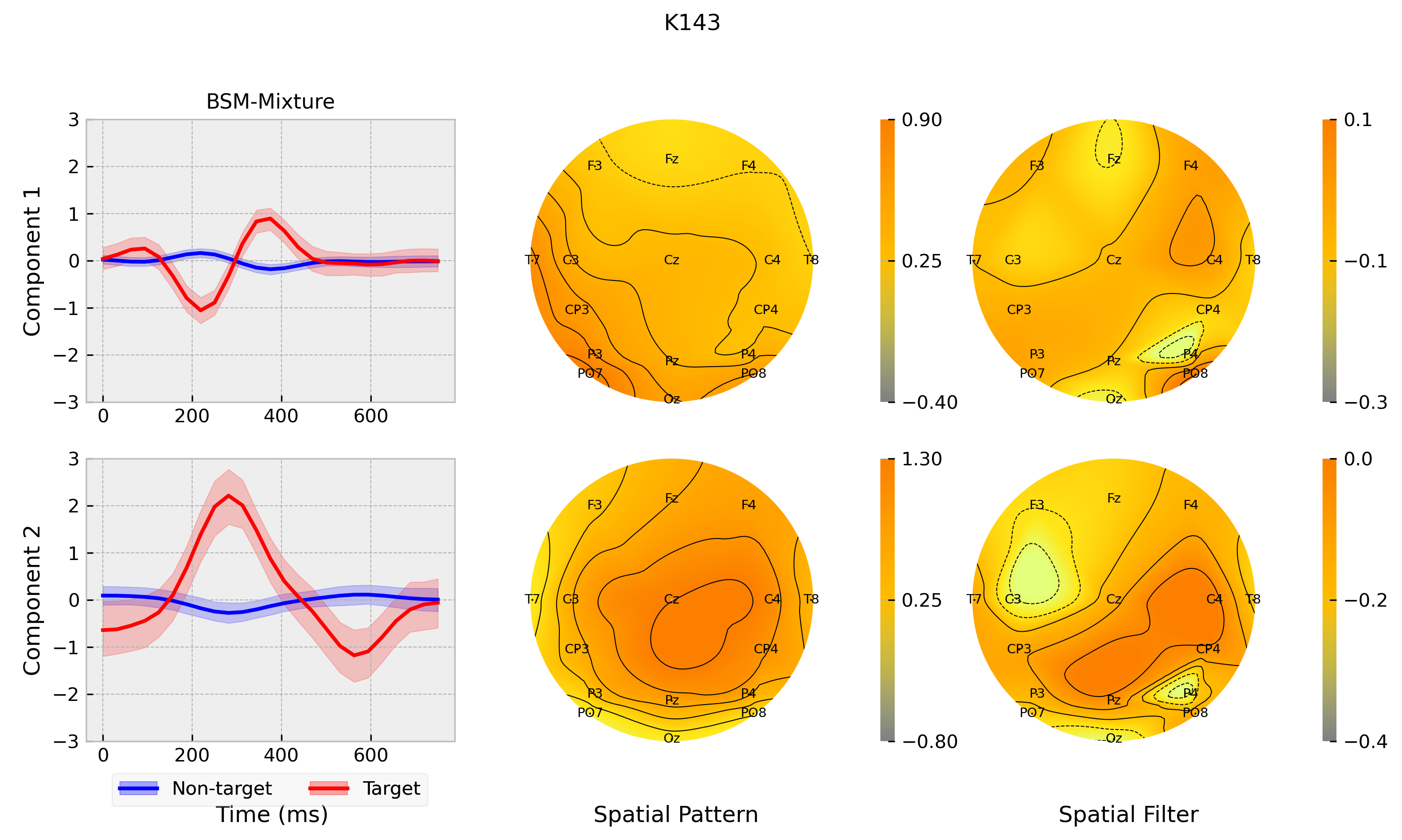}
      \end{subfigure}
      \hfill
      \begin{subfigure}
          \centering
          \includegraphics[width=0.48\textwidth]{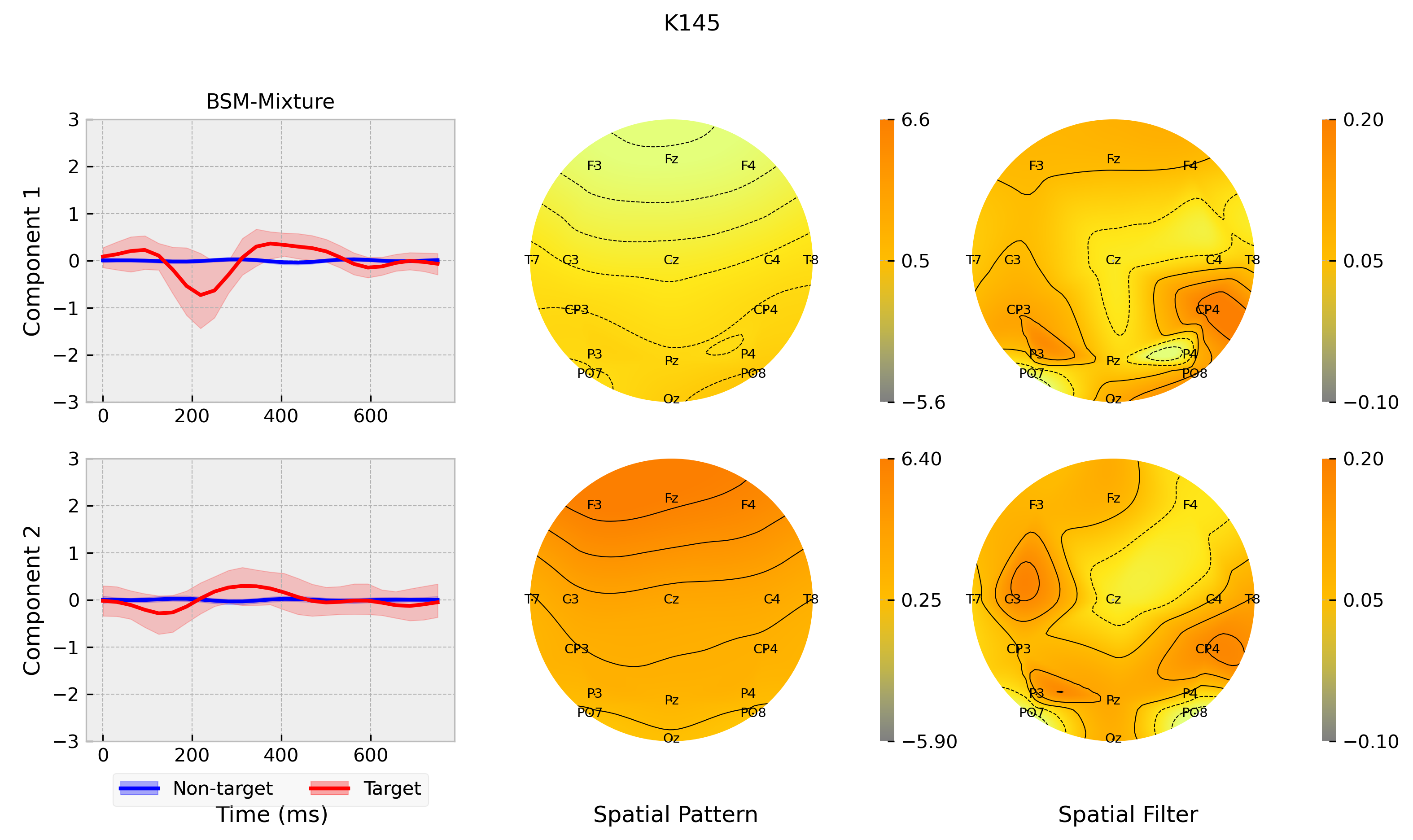}
      \end{subfigure}
      \vfill
      \begin{subfigure}
          \centering
          \includegraphics[width=0.48\textwidth]{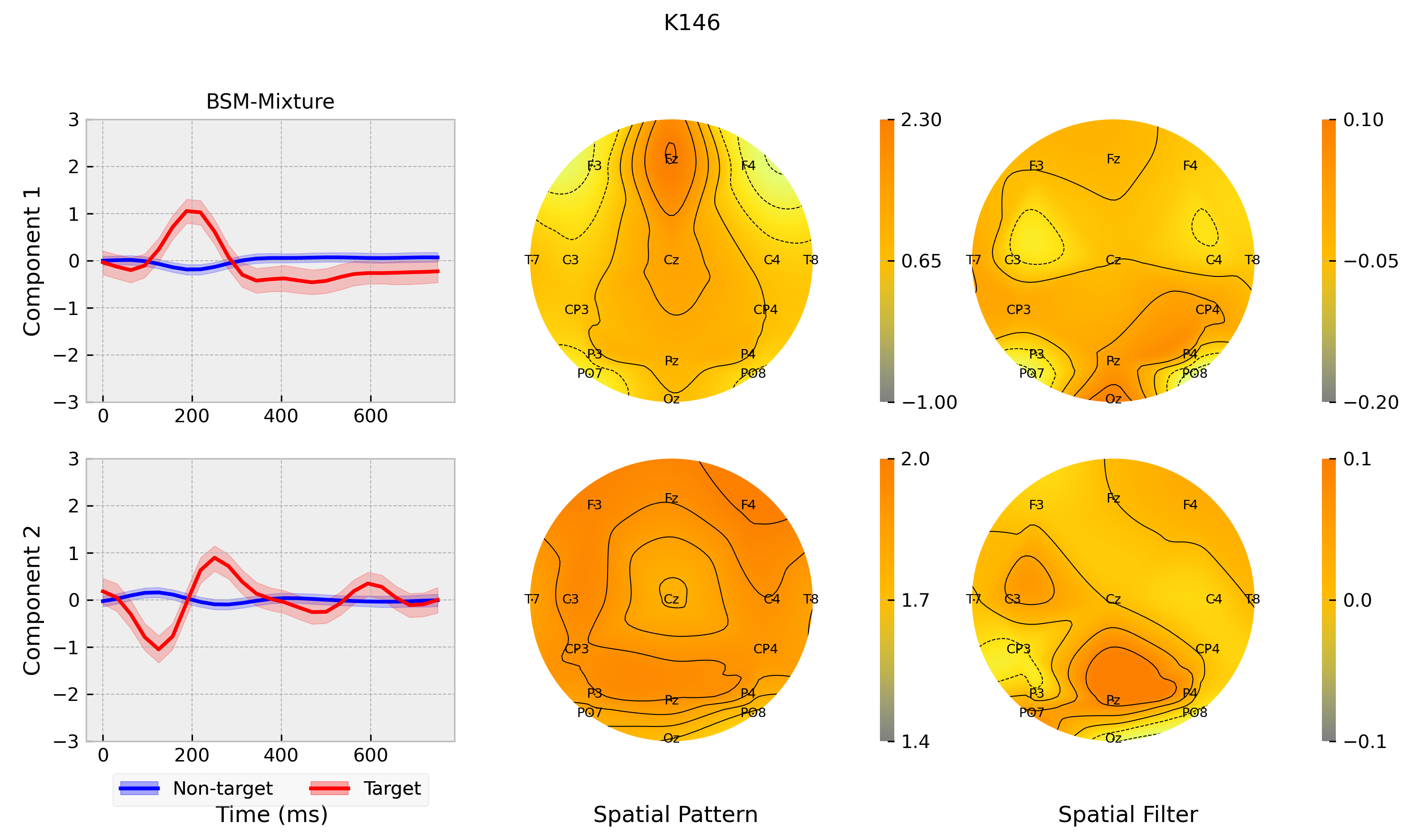}
      \end{subfigure}
      \hfill
      \begin{subfigure}
          \centering
          \includegraphics[width=0.48\textwidth]{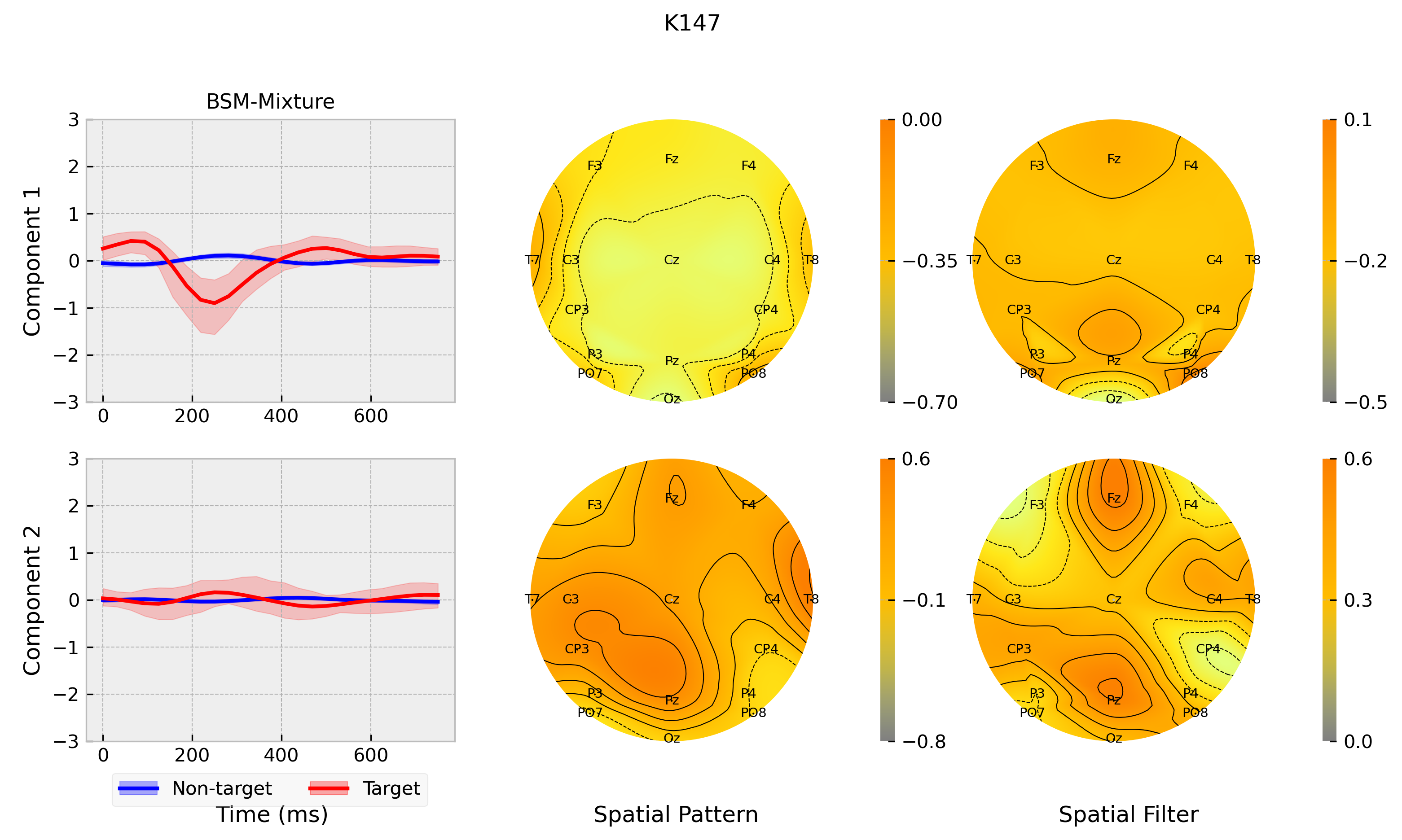}
      \end{subfigure}
      \vfill
      \begin{subfigure}
          \centering
          \includegraphics[width=0.48\textwidth]{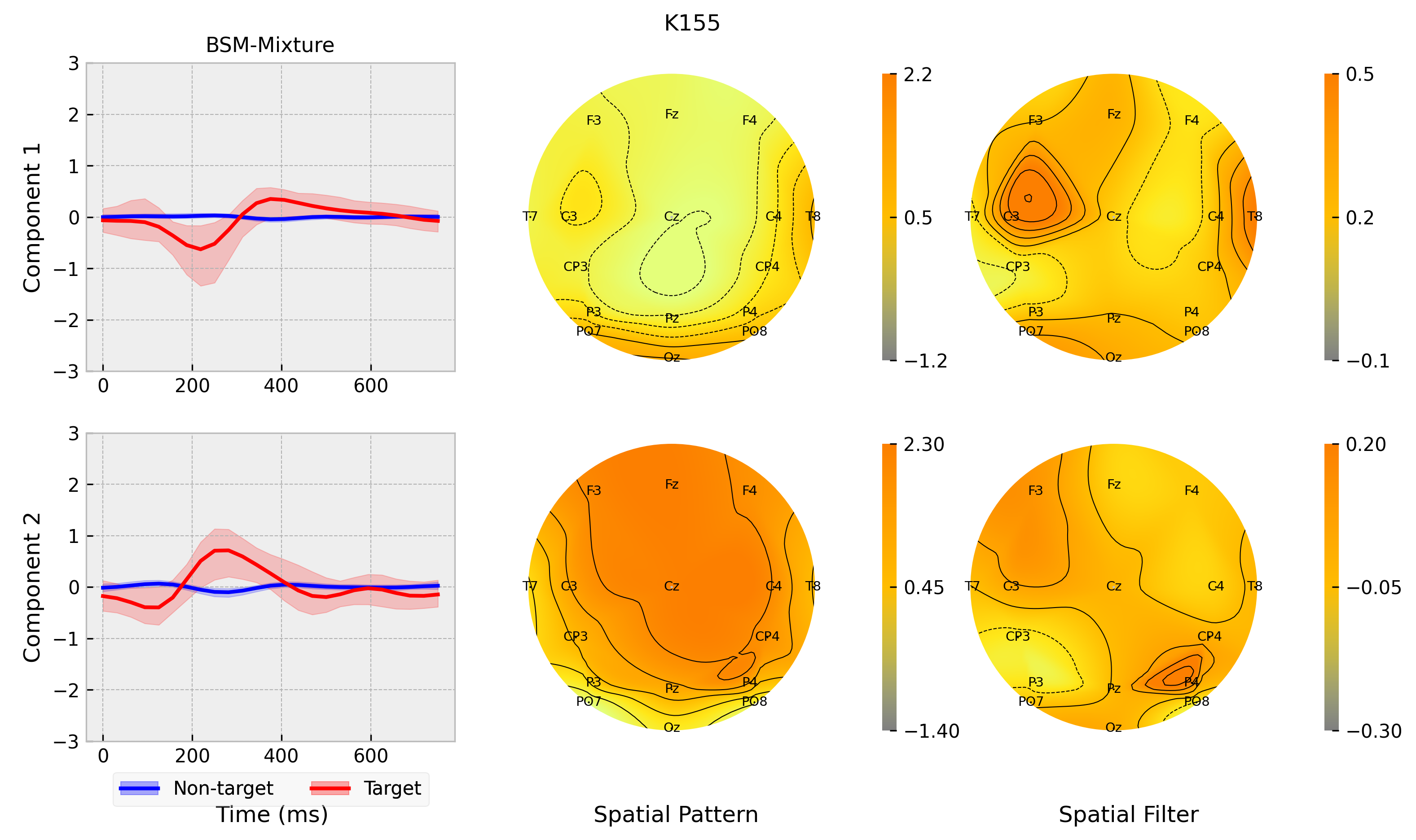}
      \end{subfigure}
      \hfill
      \begin{subfigure}
          \centering
          \includegraphics[width=0.48\textwidth]{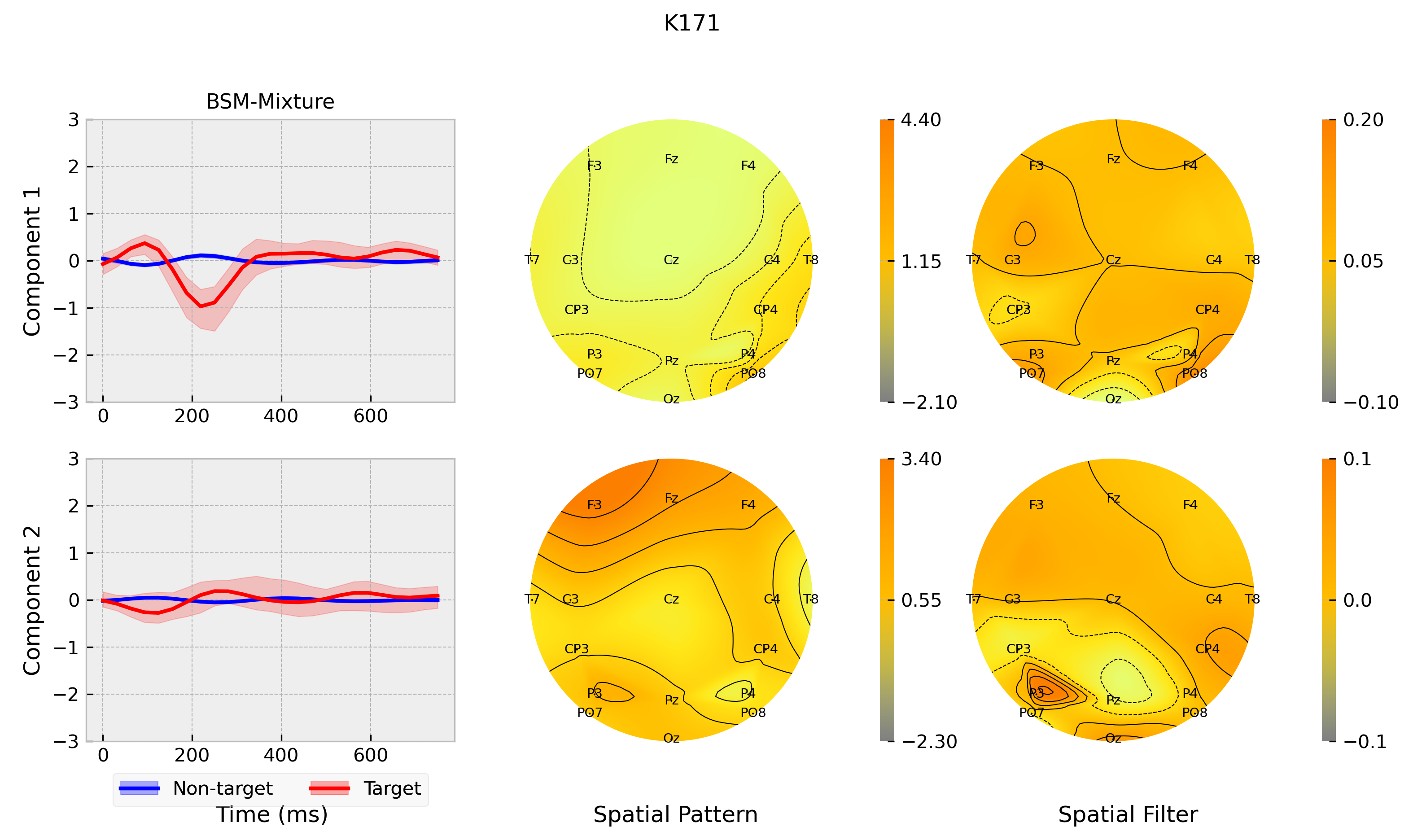}
      \end{subfigure}
      \vfill
      \begin{subfigure}
          \centering
          \includegraphics[width=0.48\textwidth]{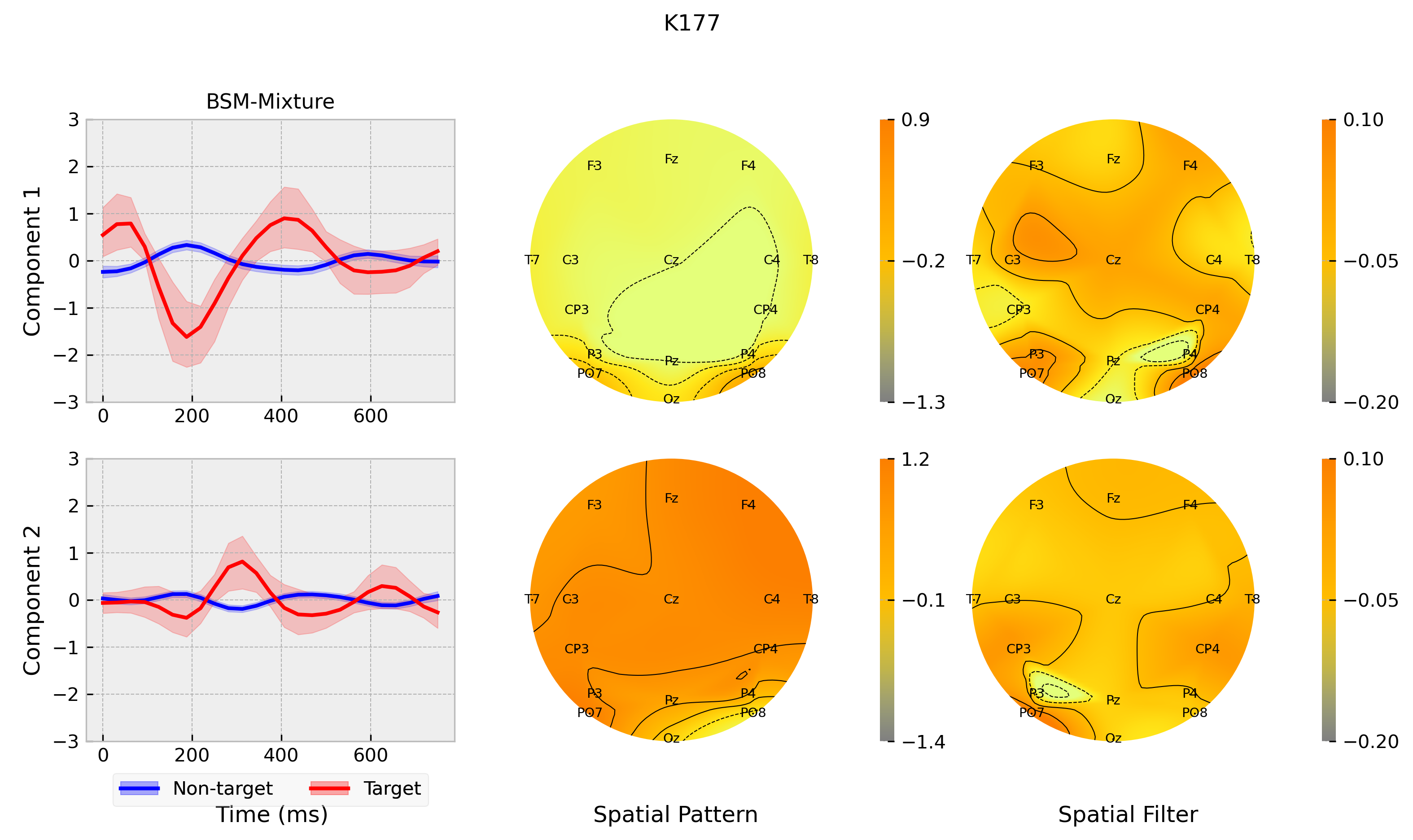}
      \end{subfigure}
      \hfill
      \begin{subfigure}
          \centering
          \includegraphics[width=0.48\textwidth]{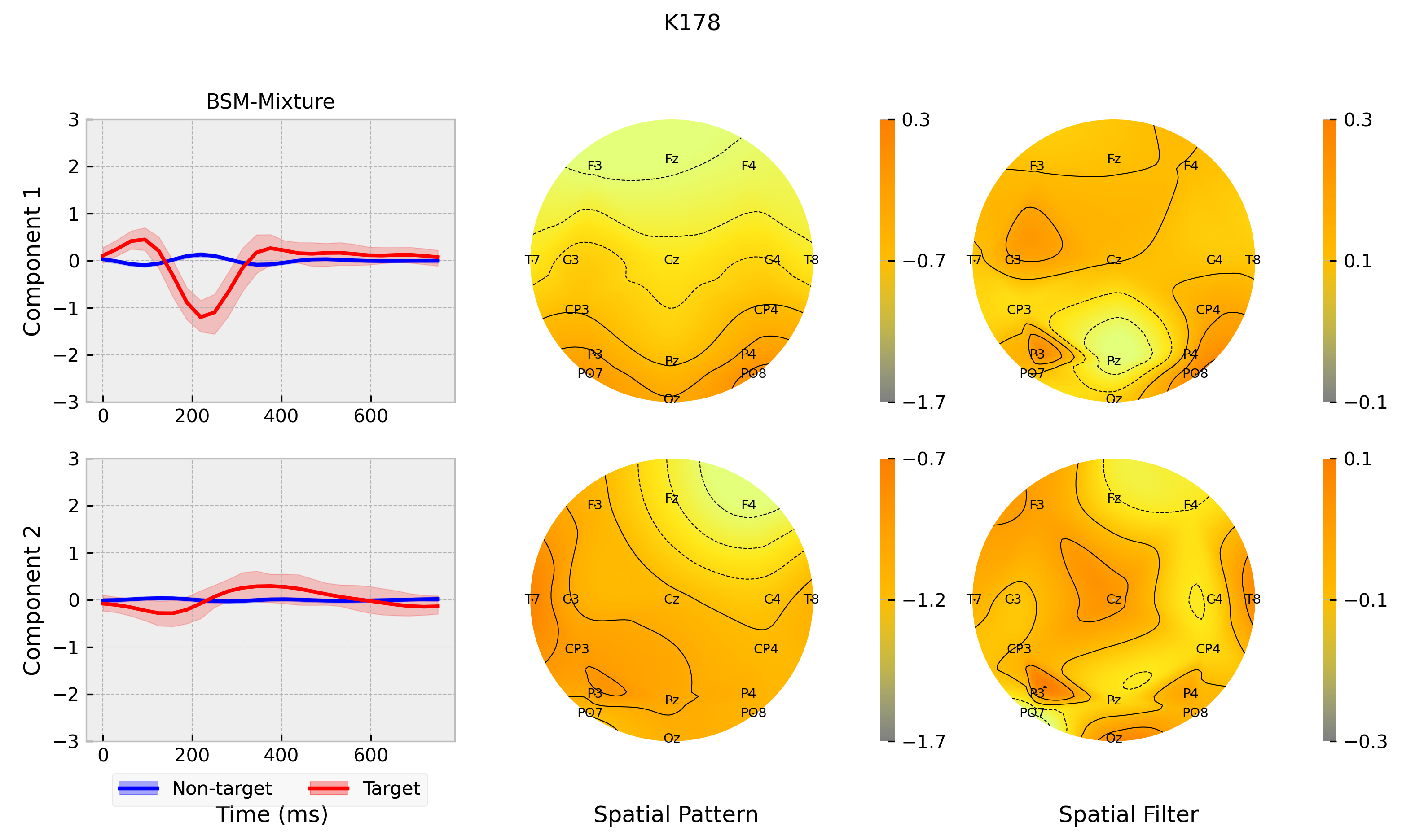}
      \end{subfigure}
      \vfill
    \caption{The transformed ERP function estimates by BSM-Mixture, spatial pattern, and spatial filters for the remaining eight participants of interests. 
    }\label{fig:sup_real_data_remaining_8}
\end{figure}

\subsection{Results of Other Participants}
\label{subsec:real_data_cluster_predict}

Results of all other participants showed that the average selection indicators for source participants were typically small and the overall magnitude and shape were similar between BSM and BSM-Reference methods. In particular, BSM-Mixture tended to attenuate the magnitude of peaks for transformed ERP function estimates, especially for the second component. The credible bands for BSM-Mixture were slightly wider due to the larger uncertainty in the mixture model. 

\begin{figure}[htbp]
\centering
\includegraphics[width=0.95\textwidth]{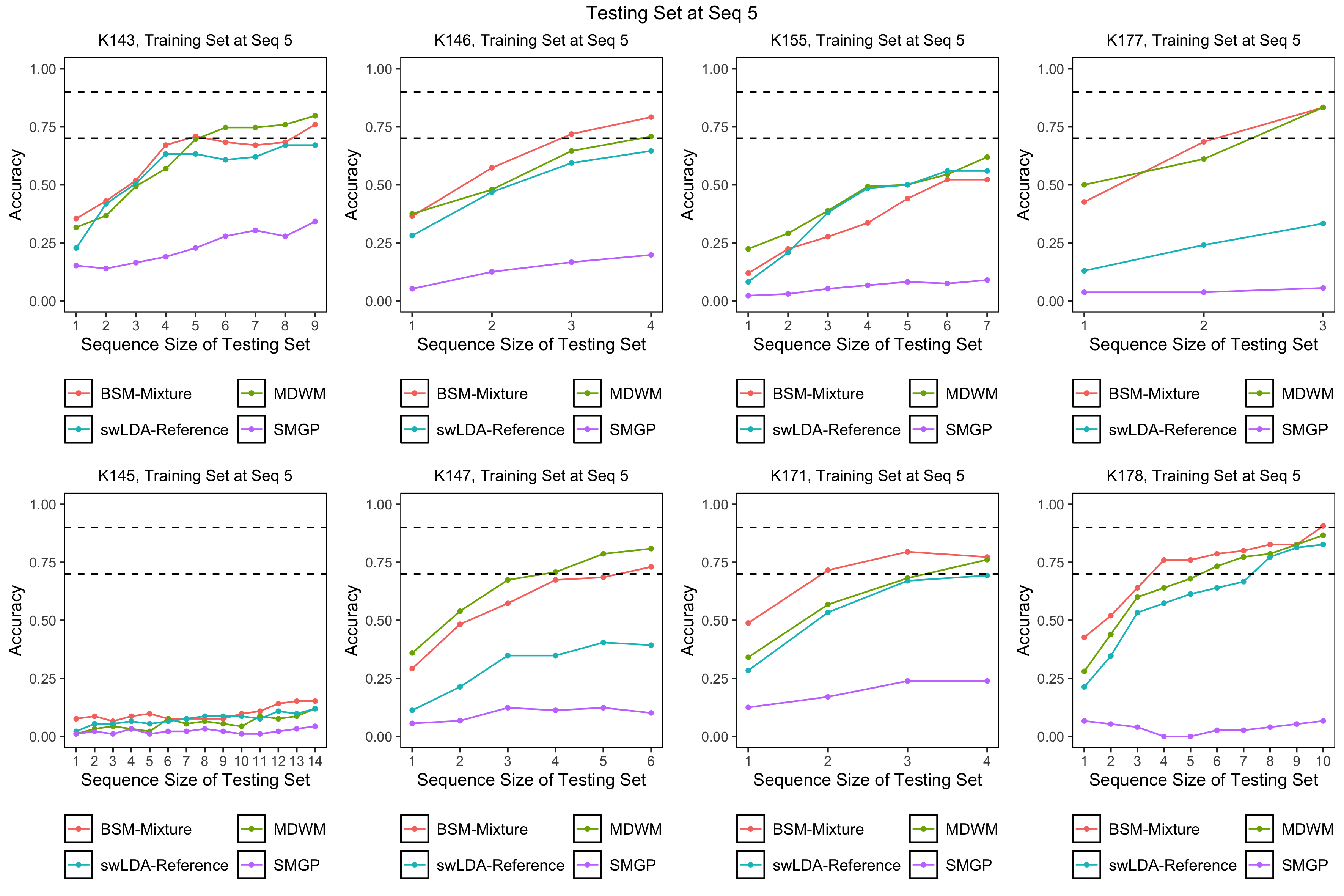}
    \caption{\small A plot to show testing prediction accuracy with respect by BSM-Mixture, MDWM, swLDA, and SMGP of remaining eight participants. Two horizontal dashed lines corresponding to 0.7 and 0.9 were shown.}
    \label{fig:sup_real_data_remaining_8_prediction}
\end{figure}

Figure \ref{fig:sup_real_data_remaining_8} shows the transformed ERP function estimates by BSM, spatial pattern, and spatial filters for participants of which BSM outperformed BSM-Reference and swLDA-Reference. Figure \ref{fig:sup_real_data_remaining_8_prediction} 
shows the total prediction accuracy of all FRT files by BSM-Mixture, MDWM, swLDA, and SMGP with five training sequences of remaining seven participants with brain damage and neuro-degenerative diseases. Overall, BSM-Mixture performed better than reference methods and similarly to MDWM.

\clearpage
\newpage
\bibliographystyle{apalike}
\bibliography{supplement_1}